\documentclass[twoside,twocolumn,9pt]{article}
\usepackage{extsizes}
\usepackage[super,sort&compress,comma]{natbib} 
\usepackage[version=3]{mhchem}
\usepackage[left=1.5cm, right=1.5cm, top=1.785cm, bottom=2.0cm]{geometry}
\usepackage{balance}
\usepackage{mathptmx}
\usepackage{sectsty}
\usepackage{graphicx} 
\usepackage{lastpage}
\usepackage[format=plain,justification=justified,singlelinecheck=false,font={stretch=1.125,small,sf},labelfont=bf,labelsep=space]{caption}
\usepackage{float}
\usepackage{fancyhdr}
\usepackage{fnpos}
\usepackage[english]{babel}
\addto{\captionsenglish}{%
  
}
\usepackage{array}
\usepackage{droidsans}
\usepackage{charter}
\usepackage[T1]{fontenc}
\usepackage[usenames,dvipsnames]{xcolor}
\usepackage{setspace}
\usepackage[compact]{titlesec}
\usepackage{hyperref}

\usepackage{balance}
\usepackage{times,mathptmx}
\usepackage{sectsty}
\usepackage{graphicx} 
\usepackage{lastpage}
\usepackage[format=plain,justification=justified,singlelinecheck=false,font={stretch=1.125,small,sf},labelfont=bf,labelsep=space]{caption}
\usepackage{float}
\usepackage{fancyhdr}
\usepackage{fnpos}
\usepackage[english]{babel}
\addto{\captionsenglish}{%
  
}
\usepackage{array}
\usepackage{droidsans}
\usepackage{charter}
\usepackage[T1]{fontenc}
\usepackage[usenames,dvipsnames]{xcolor}
\usepackage{setspace}
\usepackage[compact]{titlesec}
\usepackage{hyperref}

\usepackage{abstract}
\usepackage{graphicx}
\usepackage{gensymb}
\usepackage{caption}
\usepackage{amsmath}
\usepackage{amsthm}
\usepackage{amsfonts}
\usepackage{sidecap}
\usepackage{mathtools}
\usepackage{adjustbox}
\usepackage{ upgreek }

\usepackage{epstopdf}

\definecolor{cream}{RGB}{222,217,201}

\begin{document}

\pagestyle{fancy}
\thispagestyle{plain}
\fancypagestyle{plain}{
\renewcommand{\headrulewidth}{0pt}
}

\makeFNbottom
\makeatletter
\renewcommand\LARGE{\@setfontsize\LARGE{15pt}{17}}
\renewcommand\Large{\@setfontsize\Large{12pt}{14}}
\renewcommand\large{\@setfontsize\large{10pt}{12}}
\renewcommand\footnotesize{\@setfontsize\footnotesize{7pt}{10}}
\makeatother

\renewcommand{\thefootnote}{\fnsymbol{footnote}}
\renewcommand\footnoterule{\vspace*{1pt}%
\color{cream}\hrule width 3.5in height 0.4pt \color{black}\vspace*{5pt}} 
\setcounter{secnumdepth}{5}

\makeatletter 
\renewcommand\@biblabel[1]{#1}            
\renewcommand\@makefntext[1]%
{\noindent\makebox[0pt][r]{\@thefnmark\,}#1}
\makeatother 
\renewcommand{\figurename}{\small{Fig.}~}
\sectionfont{\sffamily\Large}
\subsectionfont{\normalsize}
\subsubsectionfont{\bf}
\setstretch{1.125} 
\setlength{\skip\footins}{0.8cm}
\setlength{\footnotesep}{0.25cm}
\setlength{\jot}{10pt}
\titlespacing*{\section}{0pt}{4pt}{4pt}
\titlespacing*{\subsection}{0pt}{15pt}{1pt}

\fancyfoot{}
\fancyfoot[LO,RE]{\vspace{-7.1pt}\includegraphics[height=9pt]{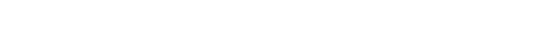}}
\fancyfoot[CO]{\vspace{-7.1pt}\hspace{13.2cm}\includegraphics{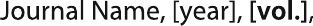}}
\fancyfoot[CE]{\vspace{-7.2pt}\hspace{-14.2cm}\includegraphics{head_foot/RF}}
\fancyfoot[RO]{\footnotesize{\sffamily{1--\pageref{LastPage} ~\textbar  \hspace{2pt}\thepage}}}
\fancyfoot[LE]{\footnotesize{\sffamily{\thepage~\textbar\hspace{3.45cm} 1--\pageref{LastPage}}}}
\fancyhead{}
\renewcommand{\headrulewidth}{0pt} 
\renewcommand{\footrulewidth}{0pt}
\setlength{\arrayrulewidth}{1pt}
\setlength{\columnsep}{6.5mm}
\setlength\bibsep{1pt}

\makeatletter 
\newlength{\figrulesep} 
\setlength{\figrulesep}{0.5\textfloatsep} 

\newcommand{\topfigrule}{\vspace*{-1pt}%
\noindent{\color{cream}\rule[-\figrulesep]{\columnwidth}{1.5pt}} }

\newcommand{\botfigrule}{\vspace*{-2pt}%
\noindent{\color{cream}\rule[\figrulesep]{\columnwidth}{1.5pt}} }

\newcommand{\dblfigrule}{\vspace*{-1pt}%
\noindent{\color{cream}\rule[-\figrulesep]{\textwidth}{1.5pt}} }

\makeatother

\twocolumn[
  \begin{@twocolumnfalse}
{\includegraphics[height=30pt]{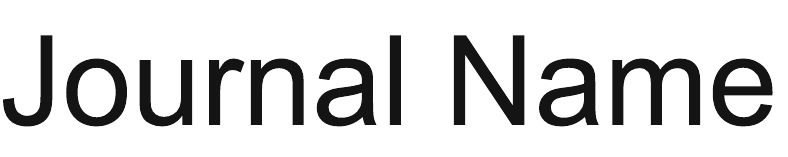}\hfill\raisebox{0pt}[0pt][0pt]{\includegraphics[height=55pt]{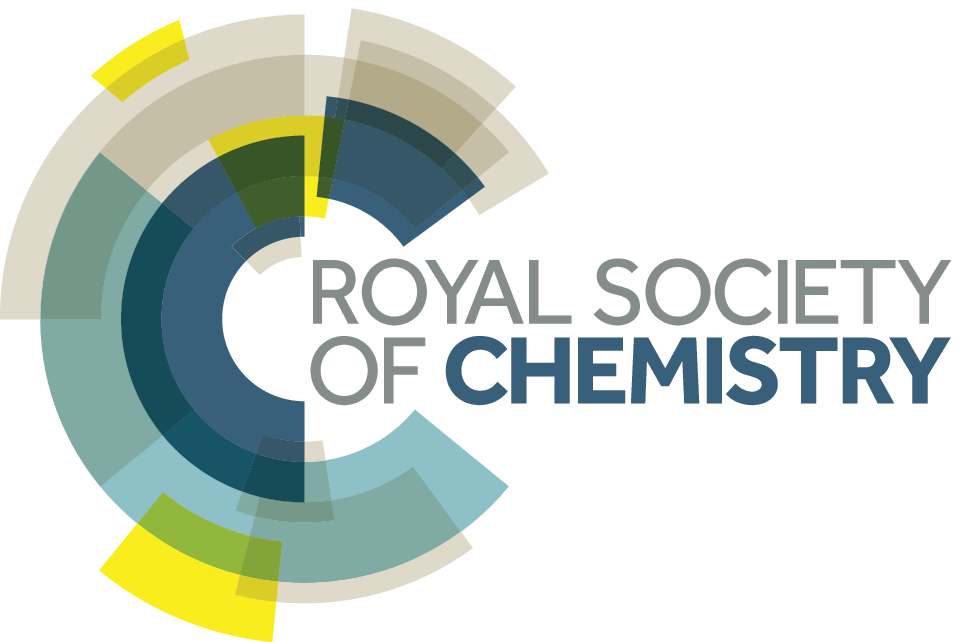}}\\[1ex]
\includegraphics[width=18.5cm]{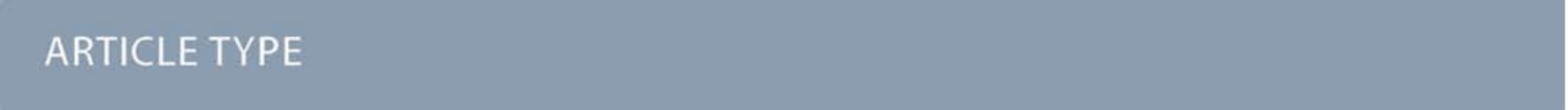}}\par
\vspace{1em}
\sffamily
\begin{tabular}{m{4.5cm} p{13.5cm} }

\includegraphics{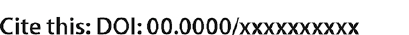} & \noindent\LARGE{\textbf{Data-Driven Design of Novel Halide Perovskite Alloys$^\dag$}} \\
\vspace{0.3cm} & \vspace{0.3cm} \\

 & \noindent\large{Arun Mannodi-Kanakkithodi,$^{\ast}$\textit{$^{a,b}$} and Maria K.Y. Chan\textit{$^{b}$}} \\

\includegraphics{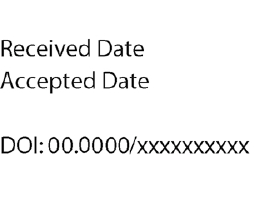} & \noindent\normalsize{The great tunability of the properties of halide perovskites presents new opportunities for optoelectronic applications as well as significant challenges associated with exploring combinatorial chemical spaces. In this work, we develop a framework powered by high-throughput computations and machine learning for the design and prediction of mixed cation halide perovskite alloys. In a chemical space of ABX$_{3}$ perovskites with a selected set of options for A, B, and X atoms, pseudo-cubic structures of compounds with B-site mixing are simulated using density functional theory (DFT) and several properties are computed, including stability, lattice constant, band gap, vacancy formation energy, refractive index, and optical absorption spectrum, using both semi-local and hybrid functionals. Neural networks (NN) are used to train predictive models for every property using tabulated elemental properties of A, B, and X site atoms as descriptors. Starting from the DFT dataset of 229 points, we use the trained NN models to predict the structural, energetic, electronic and optical properties of a complete dataset of 17,955 compounds, and perform high-throughput screening in terms of stability, band gap and defect tolerance, to obtain 574 promising compounds that are ranked as potential absorbers according to their photovoltaic figure of merit. Compositional trends in the screened set of attractive mixed cation halide perovskites are revealed and additional computations are performed on selected compounds. The data-driven design framework developed here is promising for designing novel mixed compositions and can be extended to a wider perovskite chemical space in terms of A, B, and X atoms, different kinds of mixing at the A, B, or X sites, non-cubic phases, and other properties of interest.} \\

\end{tabular}

 \end{@twocolumnfalse} \vspace{0.6cm}

  ]

\renewcommand*\rmdefault{bch}\normalfont\upshape
\rmfamily
\section*{}
\vspace{-1cm}


\footnotetext{\textit{$^{a}$~School of Materials Engineering, Purdue University, West Lafayette, IN 47907, USA. }}
\footnotetext{\textit{$^{b}$~Center for Nanoscale Materials, Argonne National Laboratory, Argonne, IL 60439, USA; E-mail: amannodi@purdue.edu, mchan@anl.gov}}

\footnotetext{\dag~Electronic Supplementary Information (ESI) available: [details of any supplementary information available should be included here]. See DOI: 00.0000/00000000.}




\section*{Introduction}

The confluence of high-throughput computation, machine learning, and targeted synthesis has been key to the accelerated discovery of new materials in recent times. This nexus has made it possible to create better electronics \cite{ML2}, dielectrics \cite{ML3,MK6}, catalysts \cite{ML1}, and photovoltaic absorbers \cite{ML5,ML6}, and has increasingly become a necessary component of every materials design endeavor \cite{ML4,MK5}. The thriving research area of halide perovskites has also benefited from data-driven studies. There are many examples of computational screening being effectively performed for exploring the compositional space \cite{MLHP1,MLHP2,MLHP3}, replacing Pb with similar divalent cations \cite{MLHP4,MLHP5,MLHP6}, exploring defects and impurities \cite{MK1,MK2}, and predicting electronic and optical properties \cite{MLHP7,MLHP8} of halide perovskites. Machine learning approaches applied on computational or experimental data have been successful in identifying stable halide perovskite compositions, predicting new properties, and elucidating synthesis and stabilization pathways \cite{MLHP9,MLHP10}. \\

New halide perovskite compositions with tailored optical, electronic and defect properties can lead to improved performances in applications ranging from solar cells to light emitting diodes to electronics to qubits. Indeed, the engineering of known semiconductors in terms of their compositions and in terms of functional defects and impurities has been a common methodology for obtaining properties that are better than known pure compositions \cite{Semi}. Halide perovskites provide a very fertile space for exploration and improvements based on composition, impurities, interfaces, and dimensionality. The general consensus is that as a class of materials, halide perovskites have a high ceiling of performances that can be achieved, and we are still only scratching the surface. Of special interest in this regard is the engineering of the perovskite composition via mixing at the cation or anion sites. \\

A quick survey of more than 50 journal articles \cite{C1,C2,C3,C4,C5,C6,C7,C8,C9,C10,C11,C12,C13,C14,C15,C16,C17,C18,C19,E1,E2,E3,E4,E5,E6,E7,E8,E9,E10,E11,E12,E13,E14,E15,E16,E17,E18,E19,E20,E21,E22,E23,E24,E25,E26,E27,E28,E29,E30,E31,Comput_mixed_1,Sn_Pb_perovs_1,Sn_Pb_perovs_2,Sn_Pb_perovs_3,Sn_Pb_perovs_4,Sn_Pb_perovs_5,Sn_Pb_perovs_6,Sn_Pb_perovs_7} on ABX$_{3}$ halide perovskites with mixed compositions (i.e., alloying at A, B, and/or X sites) revealed that (a) 66$\%$ reported experimental data and 44$\%$ reported computational properties; (b) 44$\%$ involved A-site mixing, 44$\%$ involved B-site mixing, and 52$\%$ involved X-site mixing; and (c) a wide variety of properties such as band gaps, effective masses, exciton binding energies, photoluminescence spectra, and photon conversion efficiencies are studied and reported. It is clear there is tremendous interest in optimizing halide perovskite alloys, and quick predictions and screening of new compositions may be key to continuous improvements in properties and performance. A few important observations from this literature survey are explained below:

\begin{enumerate}

    \item One of the primary motivations for exploring mixed compositions originates from the seemingly inherent instability of some of the most commonly used compounds --- MAPbI$_{3}$, MAPbBr$_{3}$, CsPbI$_{3}$, etc. --- that manifests in the form of thermal degradation, phase segregation, and photo-induced halide segregation \cite{C19,E8,E21,E30}. On the A-site, organic monovalent molecular cations such as methylammonium (MA) and formamidinium (FA) serve to enhance the structural stability of the perovskite lattice, and it has been suggested that using two or three cations simultaneously (e.g. MA$_{x}$FA$_{y}$Cs$_{1-x-y}$) makes the perovskite more robust and thermally stable \cite{C8,C13,C17,E8,E15,E21,E24,E25,E28,E30}. Many compounds also have a propensity to decompose into their constituent halides, that is, ABX$_{3}$ $\rightarrow$ AX + BX$_{2}$, or oxidize into a double perovskite variant, that is, 2ABX$_{3}$ $\rightarrow$ A$_{2}$BX$_{6}$ \cite{C19,E30}.

    \item Most ABX$_{3}$ perovskites with standard choices of A, B, and X atoms display a narrow range of band gap values. A key advantage of mixing at A, B or X sites is the existence of band gap or stability ``bowing", wherein the properties may drastically rise or drop for intermediate compositions as compared to end points \cite{C2,C4,C10,C13,C19,E3,E15,E22}.

    \item Although the organic cation MA$^+$ enhances the perovskite's structural robustness, it also increases the risk of chemical instability when encountering moisture, oxidizing environment, and heat. Mixed cations, particularly when using cesium cation (Cs$^+$), prove to be an effective way of improving both stability and optoelectronic performances of hybrid perovskite films applied in solar cells \cite{C19,E4,E30}.

    \item Pb is the most commonly used divalent cation at the B-site, but there is tremendous effort focused on the reduction or complete removal of Pb from the perovskite lattice owing to its toxicity. Pb$^{2+}$ also shows a tendency to be oxidized to Pb$^{4+}$ in the presence of air and moisture, which accelerates PV degradation \cite{Comput_mixed_1}. However, the caveat remains that removing Pb entirely has so far reduced the stability and worsened the PV performance \cite{Sn_Pb_perovs_1,Sn_Pb_perovs_2}.

    \item There is evidence that ASnX$_{3}$ perovskites are more resilient to decomposition than APbX$_{3}$, and adding Sn to Pb can certainly increase the stability \cite{Comput_mixed_1,Sn_Pb_perovs_1,Sn_Pb_perovs_2,Sn_Pb_perovs_3,Sn_Pb_perovs_4,Sn_Pb_perovs_5,Sn_Pb_perovs_6,Sn_Pb_perovs_7}. However, it is also suggested that combining Pb-Sn mixing at the B-site with A-site mixing, such as adding Cs to FA, may be necessary for further stability improvement \cite{C19}.

\end{enumerate}

The wealth of halide perovskite literature emerging every single day, and the analysis presented here of a small subset of this literature, suggests that ways of efficient exploration of mixed perovskite compositions would be beneficial to the semiconductor and halide perovskite communities. Although unequivocal predictions of multiple properties of all perovskite compositions and the ability to design novel structures with desired properties would be invaluable, they are inevitably complicated by the combinatorial nature of the chemical space. There are dozens to hundreds of possibilities for ions at the A, B, and X sites, and there is a veritable infinitude of fractions in which they could combine to create new compositions, not even accounting for competing phases, octahedral rearrangements, etc. A complete understanding of the structure and properties of halide perovskite chemical spaces both broad (choices of A, B, and X) and deep (mixing several components in tiny or large fractions at any site) is missing, and not conducive to brute-force experimentation or even computation. \\

\begin{figure*}[h]
\centering
\includegraphics[width=0.80\linewidth]{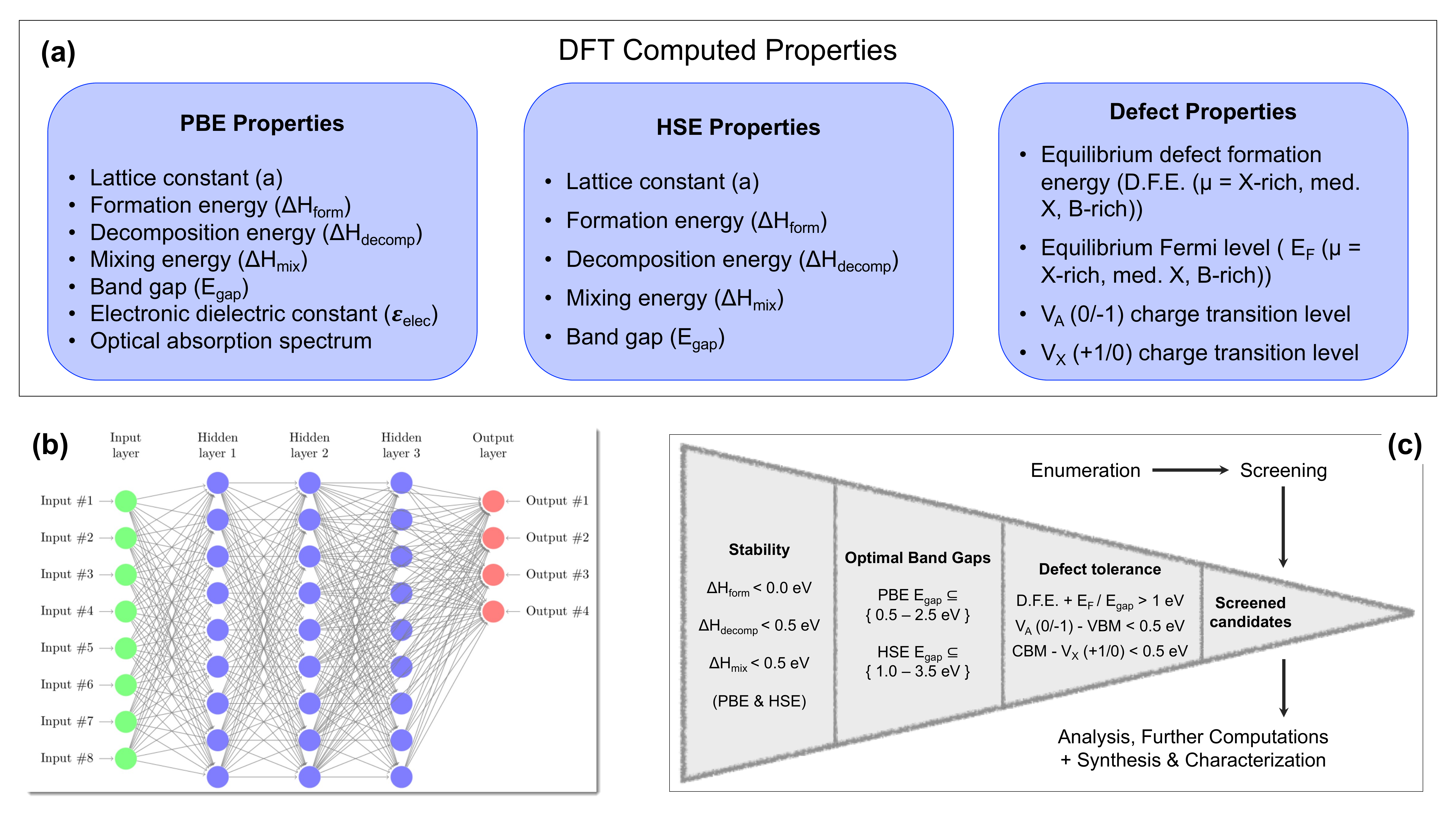}
\caption{\label{Fig:outline} 
(a) DFT properties computed for 229 perovskite compounds at the PBE and HSE06 levels of theory. (b) Typical neural network architecture used to train predtictive models from DFT data. (c) Screening performed on ML predicted dataset of 17,955 perovskite compounds in terms of their stability, band gaps and defect tolerance.}
\end{figure*}

Density functional theory (DFT) computations are widely applied towards the in-depth studies of halide perovskites from first principles \cite{MK4}, serving both as validation or explanation of experimental observations and in screening procedures alluded to earlier. DFT can reliably predict the low energy crystal structures, electronic band gaps and band edges, optical absorption coefficients, dielectric constants, and defect formation energies of halide perovskites, among many other properties. One of the biggest challenges with DFT computations for halide perovskites is the absence of universal benchmarking of DFT-computed properties, and the identification of an acceptable level of theory that can be applied. Standard DFT using semi-local Generalized Gradient Approximation (GGA) functionals like Perdew–Burke–Ernzerhof (PBE) are often insufficient for electronic and defect properties, but can reasonably estimate relative energies and aid in initial screening of properties prior to the application of more accurate but expensive functionals \cite{ChanBandGap,GGA,PBE}. The use of the Heyd–Scuseria–Ernzerhof (HSE) functionals like HSE06 or HSE03, or Green’s function-based GW approximation, along with the incorporation of spin-orbit coupling (SOC) has been suggested to be important for accurate band gaps and charge-dependent defect formation energies \cite{MK1,SOC1,SOC2,SOC3}. It has also been reported that for Sn-containing halide perovskites, HSE06 may be necessary for formation energies and HSE+SOC needed for correct band edges \cite{Comput_mixed_1}. However, PBE calculations have also been shown to work well, generally via a cancellation of errors, for band gaps and defect charge transition levels \cite{MK1,Yanfa}. \\

Since computing time, resources and manpower are finite in nature, DFT computations on halide perovskites, at a desirable level of theory, cannot be endlessly applied to all possible compositions. Approaches rooted in artificial intelligence (AI), machine learning (ML), or data science can function as conduits for information transfer from a small portion of the chemical space to the entire region or dataset. DFT data generated on a subset of compositions that are representative of the entire chemical space can be coupled with ML techniques to train predictive models that take as input \textit{descriptors} that uniquely identify every data point, or every composition, and give as output the properties of interest with acceptable error bars. DFT+ML frameworks have frequently been applied for quick prediction and screening over large chemical spaces \cite{MK4}. The identification of appropriate descriptors, the generation of sufficient quantity of accurate DFT data, and the adoption of a suitable ML algorithm (such as neural networks) following standard data science practices are key to the development of accurate, generalizable models that can significantly accelerate materials discovery. \\

Motivated by the challenges and prospects of halide perovskites and a DFT+ML materials design methodology, we develop a data-driven framework for the prediction of structural, electronic, optical and defect properties of a selected chemical space of halide perovskite alloys with B-site mixing. This framework is powered by high-throughput DFT computations performed using both the PBE and HSE06 functionals, and neural network regression models trained upon the resulting datasets using mean elemental properties of various constituent atoms as input descriptors. The ABX$_{3}$ perovskite chemical space is restricted to the following choices of A, B and X atoms: A $\in$ [FA, MA, Cs, Rb, K], B $\in$ [Ca, Sr, Ba, Ge, Sn, Pb] (with mixing), and X $\in$ [I, Br, Cl]. This leads to 5*6*3 = 90 pure ABX$_{3}$ compounds; we allow mixing at the B-site, for any number of atoms, in fractions of [0, 1/8, 2/8, 3/8, ... 1], which results in a total number of 17,955 unique compositions in this space. A set of DFT computations are performed to predict around a dozen different properties for 229 compounds, selected such that every A, B, and X atom appears roughly equally frequently, forming just about 1.3$\%$ of the total chemical space. Neural network (NN) models are trained upon this data with rigorous tuning of hyperparameters, training-test splits, and cross-validation, following which the models are deployed for predictions upon the dataset of 17,955 compounds, leading to screening of stable materials with appropriate band gaps, optical absorption, and defect tolerance. The general outline of this work is captured in Fig. \ref{Fig:outline}. Each computed property is listed in Fig. \ref{Fig:outline}(a), divided into the type of DFT functional used and type of property. A typical NN architecture used to train models from the DFT data is shown in Fig. \ref{Fig:outline}(b), and the ML-based screening funnel is presented in Fig. \ref{Fig:outline}(c), showing how quick enumeration and predictions over the 17,955 compounds can be used to obtain stable materials with suitable band gaps and defect tolerance. \\

The perovskite design framework demonstrated here is, to our knowledge, one of the most comprehensive computation-driven studies of halide perovskite alloys thus far. The strategy can easily be extended to additional properties, choices of A, B, and X atoms, and mixing in fractions other than factors of 1/8, as the ML models in theory apply to any type of mixing within the considered chemical space, as explained later. With the consideration of multiple functionalities including the electronic structure, optical behavior, and defect formation, the data-driven design framework promises to benefit a variety of optoelectronic applications. Although screening has been demonstrated for the specific example of solar cell absorbers here, this framework can help identify stable materials with suitable band gaps for IR sensors and UV lasers, suitable defect levels which can be used as qubits for quantum computing, suitable electronic structure for white light emission, and wide band gap semiconductors for power electronics. In the following sections, we describe in detail the methods applied for DFT computations and training neural networks, the different sets of properties predicted, a visualization of the atom-composition-property space, best predictive performances for various properties, and screening of promising materials followed by a visualization of the screened chemical space. We end with a discussion of the predictive power and possible limitations of the current framework, and the extensions being pursued in the near future. \\

\section*{Methods}

\subsection*{DFT Details}

DFT computations were performed on 2$\times$2$\times$2 supercells of pseudo-cubic ABX$_{3}$ structures, such that there are 8 ABX$_{3}$ formula units, which facilitates alloying at the B-site in factors of 1/8. We begin with the optimized cubic structure of an end-point composition (that is, a pure ABX$_{3}$ compound with no alloying) and perform varying degrees of B-site substitution in the 2$\times$2$\times$2 supercell to simulate alloys with special quasi-random structures \cite{SQS1,SQS2}. A total of 229 compositions are selected from the entire space of 17,955 possible compounds, where 90 systems correspond to pure compositions and there are 139 alloy compositions. For every compound, geometry optimization is first performed to obtain the low energy pseudo-cubic lattice constant and the formation, decomposition, and mixing energies, following which a series of additional computations are performed, namely: a static calculation to obtain band edges and band gap, an optical absorption calculation to obtain absorption coefficients as a function of incident photon energy, density functional perturbation theory (DFPT) calculations to obtain static dielectric constant, and finally, optimization of A-site or X-site vacancy containing systems to obtain defect formation energies and charge transition levels. A comprehensive list of every property computed in this work is presented in Fig. \ref{Fig:outline}(a). \\

As mentioned earlier, key questions remain about the choice of DFT functional that provides reasonably accurate estimates of the structures and properties of halide perovskites. To explore the effect of the level of theory, we performed calculations using both the GGA-PBE functional as well as the hybrid HSE06 functional. All computations were performed using the Vienna Ab initio Simulation Package (VASP) \cite{vasp1,vasp2}, applying the generalized gradient approximation (GGA) parametrized by Perdew, Burke and Ernzerhof (PBE) \cite{PBE} and using the projector-augmented wave (PAW) pseudopotentials \cite{PAW}. For PBE geometry optimization calculations, the plane wave energy cut-off was set at 500 eV and all atomic structures were fully relaxed until forces on all atoms were less than 0.05 eV/{\AA}. Brillouin zone integration was performed using a 3$\times$3$\times$3 Monkhorst-Pack mesh, and later expanded to a 4$\times$4$\times$4 mesh for band structure, DFPT, and optical absorption calculations. DFPT computations were performed in a static manner to generate the electronic components of the dielectric tensor, and the electronic dielectric constant is reported as the trace of the tensor. The optical absorption spectra are obtained from calculations of the frequency-dependent dielectric function, which yield coefficients of absorption as a function of incident energy. We further calculate the degree of overlap between the computed absorption spectrum and the AM1.5 solar spectral irradiance data reported by NREL \cite{AM1.5} to determine a PV figure of merit, higher values of which indicate better suitability for single-junction solar cell absorption. \\

Vacancy containing systems are simulated by removing one A-site (V$_{A}$) or X-site (V$_{X}$) atom in the supercell, and performing DFT optimization (using the same convergence conditions as before, keeping cell shape and size fixed) in the neutral and charged states (-1 for V$_{A}$ and +1 for V$_{X}$), to obtain both the neutral state formation energy of vacancies and the charge transition levels created by them relative to the valence band edge. Multiple symmetry-inequivalent vacancy sites are explored for alloyed systems and the lowest energy ones are chosen. Finally, HSE06 calculations (which are between one and two orders of magnitude more expensive than corresponding PBE calculations) are performed for complete structure optimization of the 229 compounds, initially employing Gamma-point only unconstrained relaxation, followed by a static calculation using a 2$\times$2$\times$2 Monkhorst-Pack mesh. The mixing parameter $\alpha$ is chosen to be 0.25, and the same convergence conditions are used as listed above for PBE calculations. Thus, lattice constants, energetics, and band gaps were calculated from both PBE and HSE, whereas dielectric constants, absorption spectra, and vacancy energetics were obtained from PBE only. Additional PBE and HSE calculations with the same convergence criteria are performed on the elemental standard states and known halide compounds of all A, B, and X species (all such starting structures taken from the Materials Project database \cite{MP}), in order to calculate the formation and decomposition energies. \\

\begin{figure}[h]
\centering
\includegraphics[width=\columnwidth]{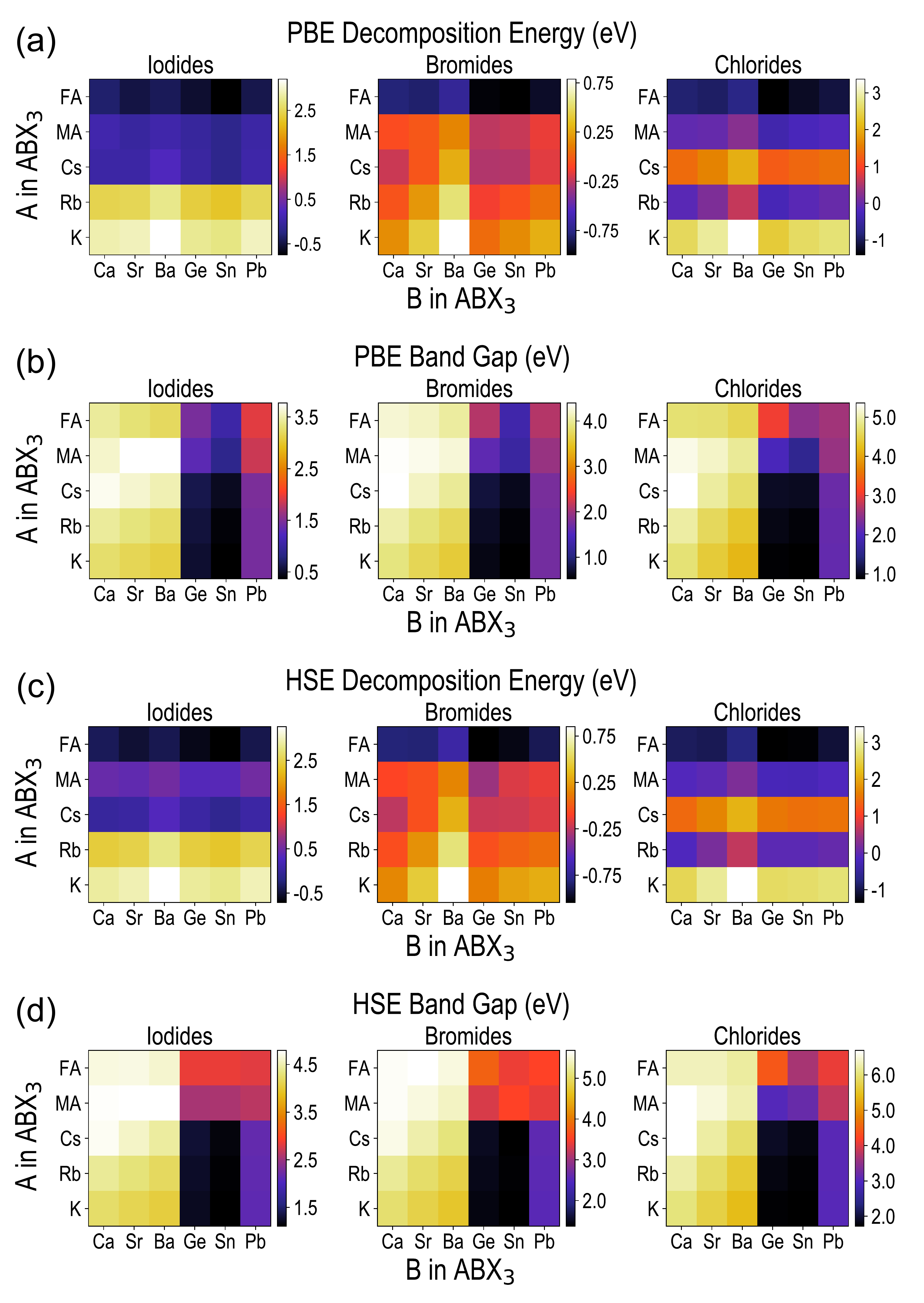}
\caption{\label{Fig:Decomp_gap_heatmaps} 
Heatmaps for properties of 90 pure ABX$_{3}$ perovskites: (a) PBE computed $\Delta$H$_{decomp}$, (b) PBE computed E$_{gap}$, (a) HSE06 computed $\Delta$H$_{decomp}$ and (b) HSE06 computed E$_{gap}$.}
\end{figure}

\subsection*{Training Neural Network Models}

NNs find utility in a wide variety of applications that deal with large datasets and demands of high accuracy, and are a staple in materials informatics studies today, with examples including accelerating molecular dynamics simulations, mapping the structure-synthesis-property spaces of materials, and on-demand prediction of properties \cite{MLHP8,NN1,NN2}. NN-based models can be trained on materials datasets to obtain regression models, classification boundaries, and generative design of novel structures \cite{NN1,NN2,NN3,NN4,NN5}. The key aspects of an NN framework (example pictured in Fig. \ref{Fig:outline}(b)) include an input layer with multiple nodes corresponding to different descriptors, hidden layers with multiple nodes each, and an output layer with one or more nodes corresponding to the properties of interest that need to be predicted. Using the DFT dataset of 229 compounds, NN models are trained for every property listed in Fig. \ref{Fig:outline}(a), using a common set of descriptors emerging from an averaging of different tabulated elemental/molecular properties of the ions that constitute the A, B, and X sites, leading to instant predictions of the same properties for the entire set of 17,955 compounds. \\

The 15 elemental/molecular properties used in this work are listed for every A, B, and X species in Table \ref{Fig:SI_elem_prop}. Each ABX$_{3}$ compound is thus represented as a 45-dimensional vector, with 15 dimensions each for the properties of A, B, and X; the dimensions corresponding to B are averaged over each species that is mixed at the B-site. For example, for a compound CsSn$_{0.5}$Sr$_{0.25}$Ba$_{0.25}$Cl$_{3}$, the first 15 dimensions will be the elemental properties of Cs (Elem$_{A}$ = Elem$_{Cs}$), the next 15 dimensions will be a weighted average of the elemental properties of Sn, Sr and Ba (that is, Elem$_{B}$ = 0.5*Elem$_{Sn}$ + 0.25*Elem$_{Sr}$ + 0.25*Elem$_{Ba}$), and the final 15 dimensions will be the elemental properties of Cl (Elem$_{X}$ = Elem$_{Cl}$). The NN input layer thus contains 45 nodes corresponding to the 45 descriptor dimensions. The number of hidden layers are restricted to be two in this work, making the NN a shallow network. Different types of output layers are used, corresponding to multiple combinations of properties as output nodes; however, it is observed that best performances are obtained by training separate NN models for every property which is a single node in the output layer, and the results presented in this work reflect that. \\

\begin{figure*}[h]
\centering
\includegraphics[width=\linewidth]{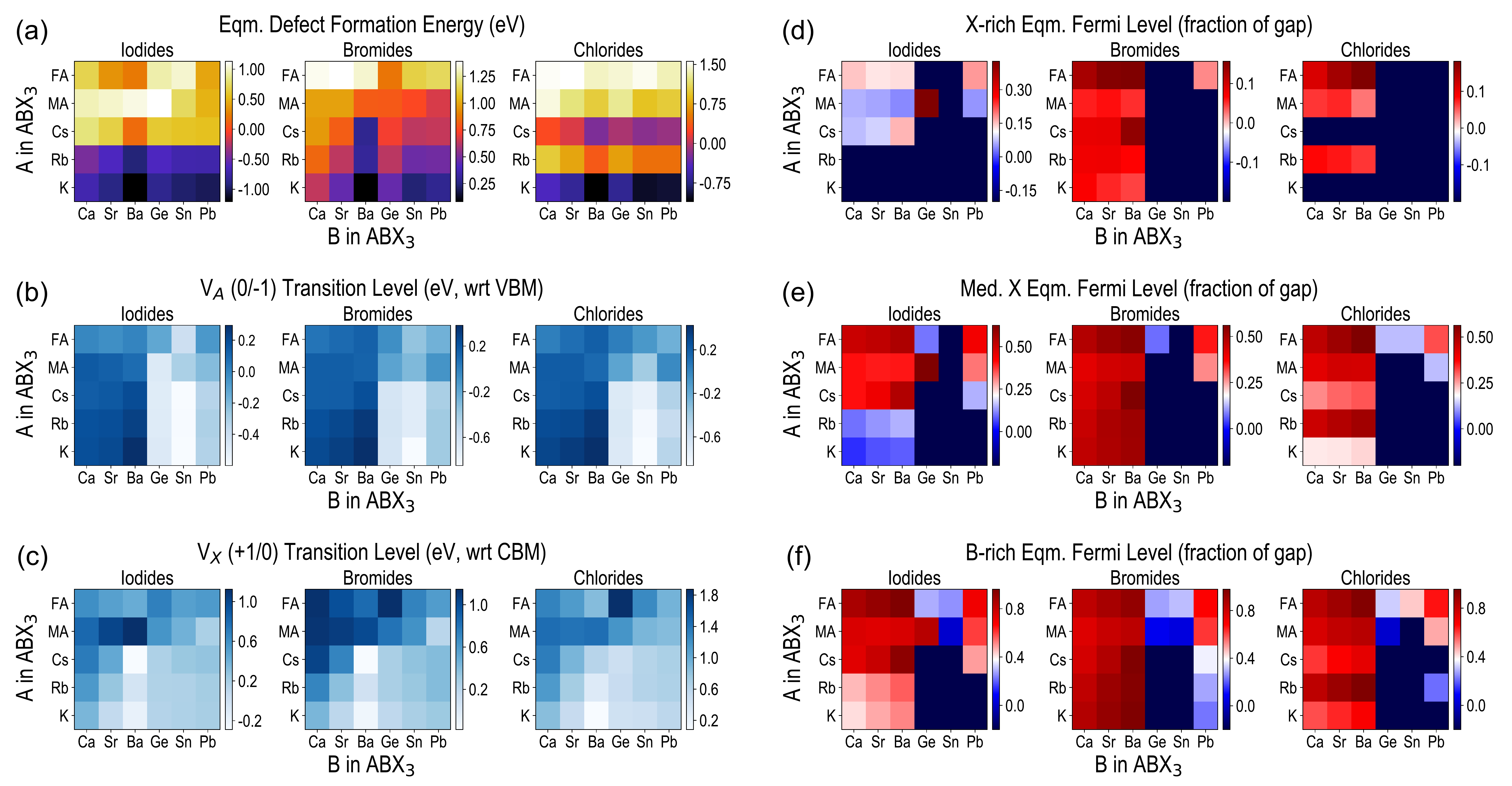}
\caption{\label{Fig:Defect_heatmaps} 
Heatmaps for defect properties of 90 pure ABX$_{3}$ perovskites: (a) equilibrium D.F.E. under medium halide chemical potential conditions, (b) V$_{A}$ (0/-1) charge transition level relative to VBM, (c) V$_{X}$ (+1/0) charge transition level relative to CBM, and equilibrium E$_{F}$ as a fraction of the PBE band gap at (d) halide rich, (e) medium halide, and (f) B-rich chemical potential conditions.}
\end{figure*}

We follow standard ML practices in training each predictive model. The DFT dataset is first divided into training (80$\%$) and test (20$\%$) sets, and the test set is kept aside while the NN model is trained using the training set data. The training-test split is fixed at 80-20 for every model in this work. Five-fold cross-validation is applied to tackle over-fitting. The error metric used in this work is the root mean square error (RMSE) between the DFT values and NN model predictions, calculated separately on the training and test sets. Hyperparameter optimization is accomplished by applying a grid-search strategy to six important hyperparameters, namely the number of neurons in the first (n1) and second (n2) hidden layers, the learning rate (lr, which defines how quickly a network updates its parameters), the dropout value (dp, which is a regularization technique used to handle overfitting), the number of epochs (ep, defining the number of times training is performed on the entire dataset), and the batch size (bs, the number of training samples propagated through the network) \cite{NN_hp1,NN_hp2}. The final models for all properties after complete optimization are reported as ML vs DFT parity plots, making distinctions between training and test set points. Other regression algorithms, including random forests and Gaussian processes, were also tested on the data, but did not improve on the performance of the best NN models. \\

\section*{Results and discussion}

\subsection*{All Computed Properties}

The following properties are predicted from PBE or HSE:

\begin{enumerate}

    \item Lattice constant: Computed with both PBE and HSE, the optimized simulation cell volume (V) is used to determine the pseudocubic lattice constant (a) for one ABX$_3$ formula unit using the formula a = (1/2)*V$^{1/3}$ .
    
    \item Formation energy: This quantity, computed from both PBE ($\Delta$H$_{form}$$^{PBE}$) and HSE ($\Delta$H$_{form}$$^{HSE}$), determines the stability of the ABX$_3$ compound w.r.t. decomposition to elemental or molecular standard states of A, B, and X. For a general perovskite alloy represented as AB$^*$X$_3$ (where B$^{*}$ represents the mixed B-site species such as Sn$_{0.5}$Pb$_{0.5}$, Ba$_{0.25}$Ge$_{0.25}$Sn$_{0.5}$, etc.), the formation energy is calculated as $\Delta$H$_{form}$ (AB$^*$X$_3$) = E(AB$^*$X$_3$) - E(A) - $\sum$$_i$w$_{i}$*E(B$_{i}$) - 1.5*E(X$_2$), where w$_{i}$ is the mixed fraction of element B$_{i}$.
    
    \item Decomposition energy: This quantity, computed from both PBE ($\Delta$H$_{decomp}$$^{PBE}$) and HSE ($\Delta$H$_{decomp}$$^{HSE}$), determines the stability of the ABX$_3$ compound w.r.t. decomposition to the halide compounds of A and all B atoms, and is given by the formula: $\Delta$H$_{decomp}$ (AB$^*$X$_3$) = E(AB$^*$X$_3$) - E(AX) - $\sum$$_i$w$_{i}$*E(B$_{i}$X$_2$).
    
    \item Mixing energy: This quantity, computed from both PBE ($\Delta$H$_{mix}$$^{PBE}$) and HSE ($\Delta$H$_{mix}$$^{HSE}$), determines the stability of the mixed-cation ABX$_3$ compound w.r.t. decomposition to pure ABX$_3$ perovskites (or end-point compositions), and is given by the formula: $\Delta$H$_{mix}$ (AB$^*$X$_3$) = E(AB$^*$X$_3$) - $\sum$$_i$w$_{i}$*E(AB$_{i}$X$_3$).
    
    \item Band gap: This is computed in eV from both PBE (E$_{gap}$$^{PBE}$) and HSE (E$_{gap}$$^{HSE}$).
    
    \item Dielectric constant and refractive index: The static electronic component of the dielectric constant ($\epsilon$$_{elec}$) is computed using DFPT, and the refractive index ($\eta$) is reported as a square root of $\epsilon$$_{elec}$.
    
    \item Optical absorption spectrum and PV Figure of Merit (FOM): Absorption coefficients ($\alpha$) are calculated as a function of incident photon energy, and used to determine the FOM based on the degree of overlap with the AM1.5 standard solar spectrum. Computed absorption and solar irradiance intensity are expressed as functions of wavelength $\lambda$ as $\alpha$($\lambda$) and I$_{s}$($\lambda$), respectively, and the FOM is calculated as follows:

\begin{equation}\label{eqn-0}
\begin{multlined}
FOM = \sum_{\lambda_i} \alpha(\lambda_i) * I_s(\lambda_i) * (\lambda_{i+1} - \lambda_i) / \sum_{\lambda_i} I_s(\lambda_i) * (\lambda_{i+1} - \lambda_i)
\end{multlined}
\end{equation}

    I$_{s}$ values are used in units of Wm$^{-2}$nm$^{-1}$ and $\alpha$ in units of cm$^{-1}$, leading to FOM values in ($\sim$ 10$^{3}$ to 10$^{6}$) cm$^{-1}$, expressed ultimately in log$_{10}$ scale.
    
    \item Vacancy defect formation energy: For V$_{A}$ and V$_{X}$ defects in the AB$^{*}$X$_{3}$ perovskite, DFT energies obtained from neutral and charged calculations are used in the equations below to estimate their charge and Fermi level dependent formation energies, and their charge transition levels:
    
\begin{equation}\label{eqn-1}
\begin{multlined}
E^f(vac^q) = E(AB^*X_3-vac^q) - E(AB^*X_3) + \mu + \\ q(E_F + E_{vbm}) + E_{corr}
\end{multlined}
\end{equation}
    
\begin{equation}\label{eqn-2}
\begin{multlined}
{\epsilon}(q_1/q_2) = \frac{E(vac^{q_1}) - E(vac^{q_2})}{q_{2}-q_{1}} - E_{vbm}
\end{multlined}
\end{equation}
    
    In equation \ref{eqn-1}, E(AB$^{*}$X$_{3}$-vac$^{q}$) is the total DFT energy of the AB$^{*}$X$_{3}$ supercell with a vacancy at the A or X sites and a total system charge = q, $\mu$ is the chemical potential of A ($\mu$$_{A}$) or X ($\mu$$_{X}$), depending on whether we are considering V$_{A}$ or V$_{X}$, E$_{vbm}$ is the valence band maximum (VBM) of AB$^{*}$X$_{3}$, E$_{F}$ is the Fermi level as it goes from the VBM to the conduction band minimum (CBM), and E$_{corr}$ is the correction energy necessary to account for the periodic interaction between charges \cite{Corr1,Corr2}. Equation \ref{eqn-2} will yield the Fermi level where V$_{A}$ displays a (0/-1) charge transition or where V$_{X}$ displays a (+1/0) transition \cite{MK1,MK2,MK3,MK7}. 

\end{enumerate}

To identify stable compounds with attractive optoelectronic properties, we will look for low $\Delta$H$_{form}$, $\Delta$H$_{decomp}$, and $\Delta$H$_{mix}$ values, E$_{gap}$ values that lie in suitable ranges, and high FOM values. Assuming vacancies are the most likely point defects to exist in halide perovskites (a reasonable assumption from earlier studies of intrinsic point defects across many halide perovskites \cite{MK1,MK2}), defect tolerance can be tested based on the formation energies and transition levels of V$_{A}$ and V$_{X}$. A low vacancy formation energy with a transition level deep in the band gap must be avoided to prevent the spontaneous creation of active defects that can cause nonradiative recombination of charge carriers and a potential reduction in PV efficiency. \\

\begin{figure}[!b]
\centering
\includegraphics[width=\columnwidth]{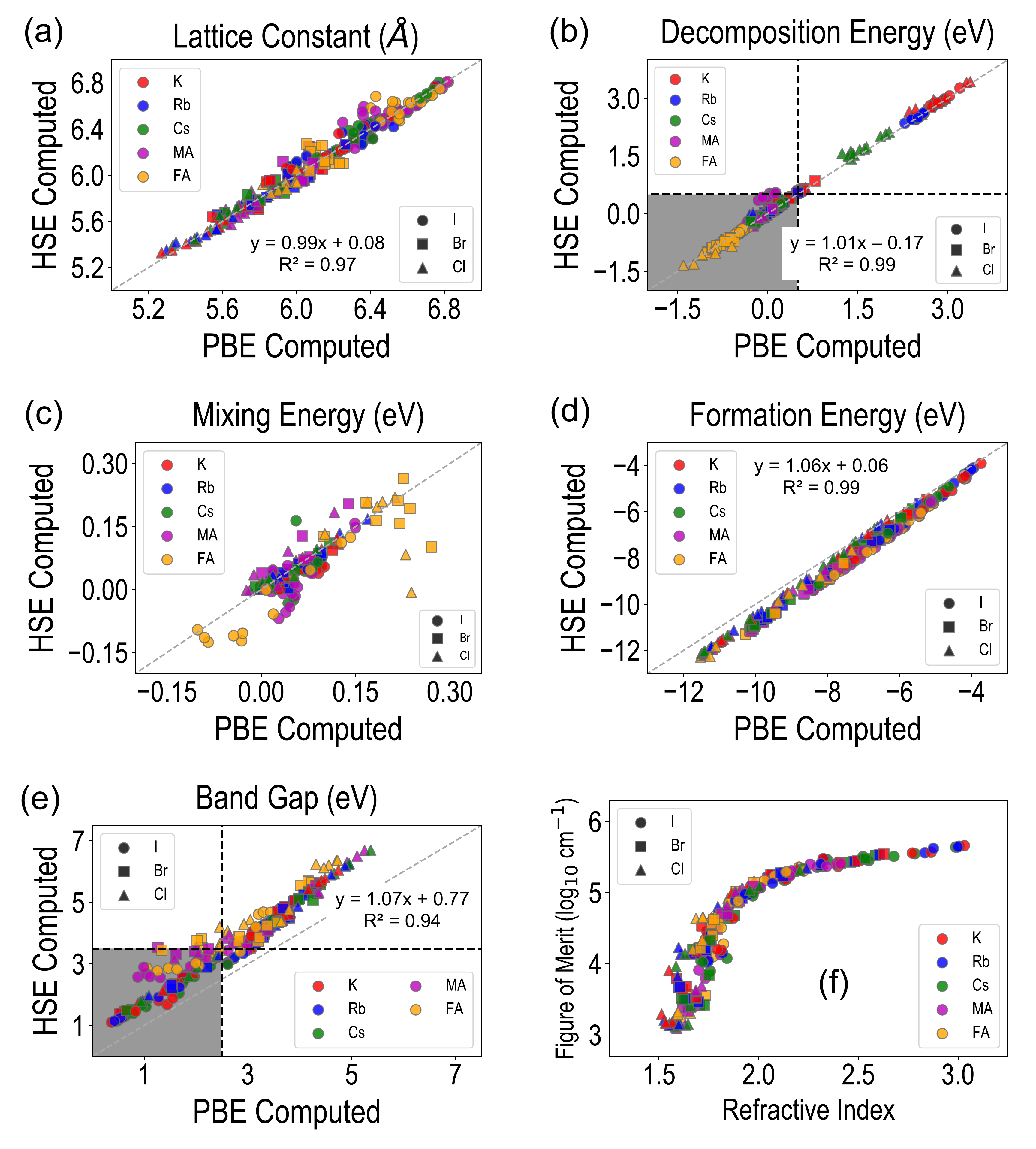}
\caption{\label{Fig:PBE_vs_HSE} 
A comparison between the PBE and HSE computed lattice constants (a), $\Delta$H$_{decomp}$ (b), $\Delta$H$_{mix}$ (c), $\Delta$H$_{form}$ (d) and E$_{gap}$ (e) for the DFT dataset of 229 compounds. PBE vs HSE linear regression equations and R$^2$ values are also pictured. Further, the computed optical absorption properties are shown as a refractive index vs Figure of Merit (as a log value of the computed FOM in cm$^{-1}$) plot in (f).}
\end{figure}

\subsection*{Visualization of DFT Data}

We begin with a visualization of the DFT dataset through a series of plots, highlighting some critical trends and correlations. Even before venturing into the behavior of perovskite alloys, it is important to consider the stability and trends in properties of the 90 pure ABX$_{3}$ compositions. The most commonly used measures for perovskite stability are the octahedral ($o$) and tolerance ($t$) factors \cite{Sampson}, defined below in terms of the ionic radii of A (r$_{A}$), B (r$_{B}$), and X (r$_{X}$):

\begin{equation}\label{eqn-3}
\begin{multlined}
o = r_B / r_X
\end{multlined}
\end{equation}

\begin{equation}\label{eqn-4}
\begin{multlined}
\textit{t} = (r_A + r_X) / (\sqrt2(r_B + r_X))
\end{multlined}
\end{equation}

For robust stability of perovskite compounds, $o$ is suggested to lie in the range 0.442–-0.895 while $t$ should be in the range 0.813–1.107. The $o$ and $t$ values of the 90 ABX$_{3}$ compounds are visualized using heatmaps in Fig. \ref{Fig:SI_oct_fact} and Fig. \ref{Fig:SI_tol_fact}, respectively. Nearly all B-X combinations satisfy the octahedral factor requirement, except for Ge-iodides. The tolerance factor falls just outside the stability range for certain combinations of A and B, such as FA-Ge and K-Ba, clearly because r$_{A}$ and r$_{B}$ need to proportionally change with each other to maintain stability, something that works well with combinations such as MA-Pb-I, MA-Sn-Br, Cs-Ge-Br, etc. The PBE and HSE computed $\Delta$H$_{decomp}$ values are plotted against $o$ and $t$ for the entire DFT dataset of 229 compounds (calculated using weighted averages for the B site when there is mixing), in Fig. \ref{Fig:SI_pbe_tol} and Fig. \ref{Fig:SI_hse_tol} respectively. It is seen that a majority of the MA and FA containing compounds, and some Cs bromides and Rb chlorides, lie in the range of low $\Delta$H$_{decomp}$ and suitable $\mu$ and $t$. However, a number of compounds that are stable w.r.t. decomposition fall outside the acceptable ranges for $o$ or $t$, indicating that perovskite stability measures ought to be captured using more complex relationships as are being pursued in recent literature \cite{Tol1,Tol2,Tol3,Sampson}. One such tolerance factor proposed by Bartel et al. \cite{Tol1}, represented as $\tau$, was estimated for all 229 compounds using eqn. \ref{eqn-5}. $\Delta$H$_{decomp}$$^{PBE}$ and $\Delta$H$_{decomp}$$^{HSE}$ are plotted against $\tau$ in Fig. \ref{Fig:SI_pbe_tau} and Fig. \ref{Fig:SI_hse_tau}, respectively. Using the suggested stability condition of $\tau$ < 4.18, it is seen that nearly all FA and MA compounds are stable w.r.t. both $\tau$ and $\Delta$H$_{decomp}$, while several Cs and Rb compounds may be included by relaxing the $\tau$ condition. Almost all the compounds satisfying the $\tau$ condition have $\Delta$H$_{decomp}$ < 0.5 eV, except for a few Cs-chlorides, but a number of compounds with $\Delta$H$_{decomp}$ $\in$ (0.0, 0.5) eV fall outside the desired $\tau$ range.

\begin{equation}\label{eqn-5}
\begin{multlined}
\tau = r_B/r_X - [ 1 - (r_A/r_X)/ln(r_A/r_X) ]
\end{multlined}
\end{equation}

Next, a series of properties, namely $\Delta$H$_{decomp}$$^{PBE}$, E$_{gap}$$^{PBE}$, $\Delta$H$_{decomp}$$^{HSE}$, and E$_{gap}$$^{HSE}$, are plotted for all pure ABX$_{3}$ compounds as heatmaps in Fig. \ref{Fig:Decomp_gap_heatmaps}. The $\Delta$H$_{decomp}$ values reveal large instabilities for all K and Rb iodides, but stability increases for the corresponding bromides and some chlorides. On the other hand, nearly all Cs, MA, and FA based iodides are stable, as are most of the corresponding bromides and chlorides, except for Cs chlorides. In order to account for the level of accuracy of DFT energies and the possible realization of metastable phases, we adopt a relaxed $\Delta$H$_{decomp}$ screening criterion, shown in Fig. \ref{Fig:outline}(c), of < 0.5 eV, from both PBE and HSE. The formation energies are found to be large negative values throughout the dataset, while the mixing energies of perovskite alloys are discussed later.

\begin{figure}[!b]
\centering
\includegraphics[width=\columnwidth]{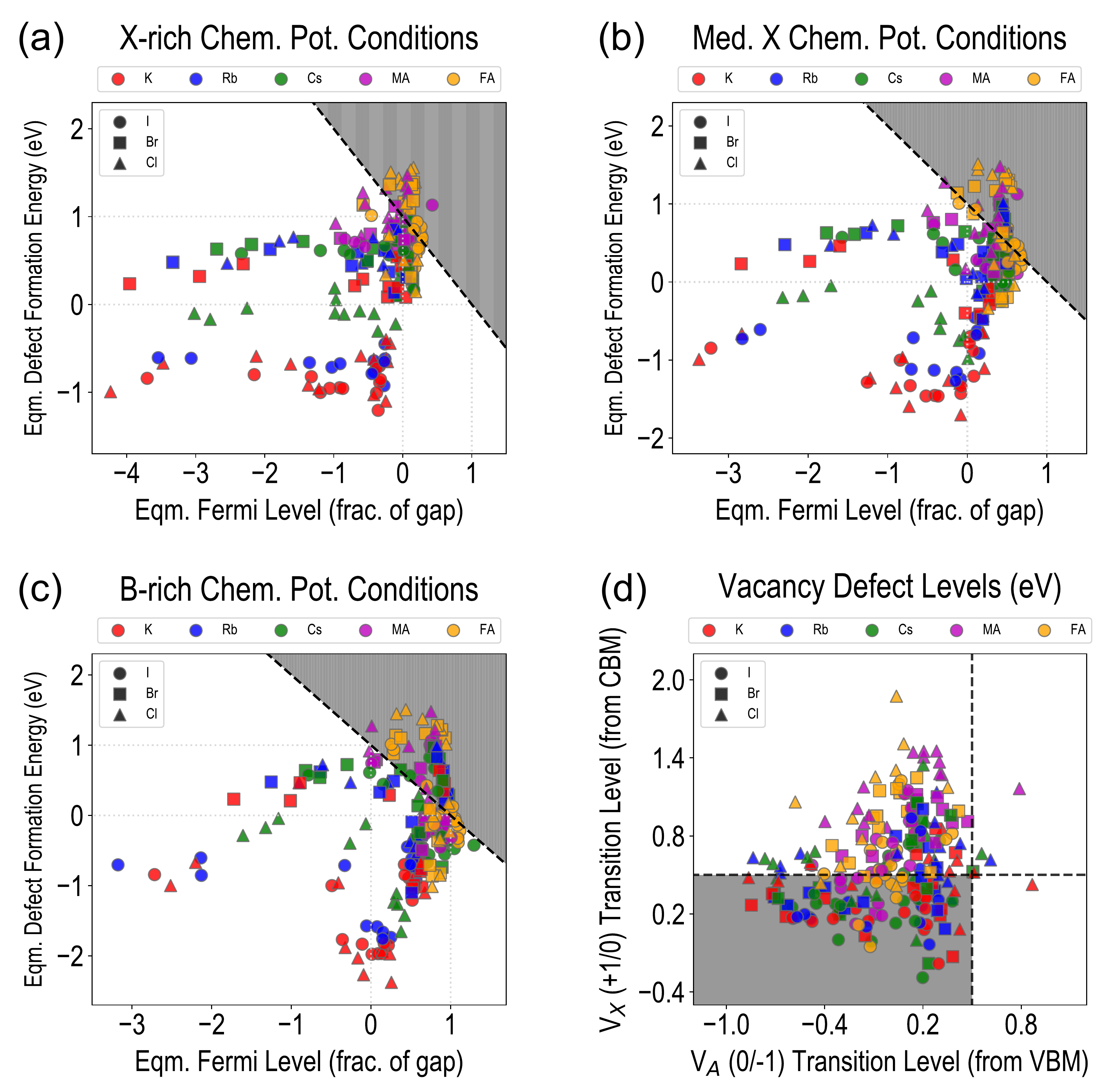}
\caption{\label{Fig:Defect_data} 
Defect computed properties for dataset of 229 perovskite compounds: Equilibrium D.F.E. vs E$_{F}$/E$_{gap}$ plots are shown for (a) halide rich, (b) medium halide, and (c) B-rich chemical potential conditions. The shaded region represents materials that satisfy the criterion D.F.E. + E$_{F}$/E$_{gap}$ > 1.0 eV and are thus defect tolerant; for high-throughput screening, this threshold is relaxed to 0.8 eV. The V$_{A}$ (0/-1) and V$_{X}$ (+1/0) charge transition levels are plotted in (d), with respect to the VBM and CBM respectively, and the shaded region represents cases with defect levels < 0.2 eV from band edges while the remaining materials have deep vacancy defect levels.}
\end{figure}

When it comes to band gap, it is seen that Ca, Sr, and Ba containing compounds show the largest E$_{gap}$$^{PBE}$ (E$_{gap}$$^{HSE}$) values for a given halogen atom, lying around 3.5 eV (4.5 eV) for iodides, $\sim$ 4 eV (5 eV) for bromides and $\sim$ 5 eV (6 eV) for chlorides. Completely inorganic perovskites with (Cs,Rb,K)-(Ge,Sn) A-B combinations show the lowest band gap values, with the iodides showing E$_{gap}$$^{PBE}$ (E$_{gap}$$^{HSE}$) $\sim$ 0.5 eV (1.5 eV), bromides $\sim$ 1 eV (2 eV) and chlorides $\sim$ 1.5 eV (2.5 eV). The MA and FA perovskites with Ge and Sn show larger band gaps for each halogen atom. All the Pb-based perovskites have higher band gaps than corresponding Ge and Sn compounds; MA and FA halides with Pb show E$_{gap}$$^{PBE}$ (E$_{gap}$$^{HSE}$) $\sim$ 2 eV (3 eV) for iodides, $\sim$ 2.5 eV (3.5 eV) for bromides and $\sim$ 3 eV (4 eV) for chlorides. The Cs, Rb and K halides with Pb show band gaps that are approximately lower by $\sim$ 1 eV than their MA and FA counterparts. It should be noted here that due to general uncertainty about the exact accuracies of PBE and HSE band gaps, we adopt a screening criterion as shown in Fig. \ref{Fig:outline}(c) of E$_{gap}$$^{PBE}$ $\in$ (0, 2.5) eV and E$_{gap}$$^{HSE}$ $\in$ (0, 3.5) eV for suitable solar cell absorbers. It is observed here that many Pb-based perovskites do occupy this acceptable range of band gaps, and their band gaps could potentially be reduced further by adding Sn or Ge, or increased by adding Ba, Sr, or Ca. Incorporating spin-orbit coupling with HSE calculations will bring down band gaps but trends will largely remain the same. \\

The A-site and X-site vacancy defect properties---namely, the equilibrium defect formation energy (D.F.E.) and Fermi level (E$_{F}$) for different chemical potential conditions, and the charge transition levels---are visualized next for the set of 90 pure ABX$_{3}$ compounds in Fig. \ref{Fig:Defect_heatmaps}. This allows the evaluation of the perovskite dataset in terms of defect tolerance, another important stability metric. For any given compound, using eqn. \ref{eqn-3}, the charge-dependent formation energies of V$_{A}$ and V$_{X}$ defects are plotted against E$_{F}$ as it goes from the VBM to the CBM, and the equilibrium E$_{F}$ is determined by applying charge neutrality conditions. The equilibrium D.F.E. and E$_{F}$ change as the chemical potential changes from halide-rich to medium halide to B-rich conditions, such that the equilibrium conductivity may transition between n-type, p-type, or intrinsic. Further, the relevant defect charge transition levels, V$_{A}$ (0/-1) and V$_{X}$ (+1/0), are calculated using eqn. \ref{eqn-4}. The condition for restricting the spontaneous formation of vacancies with transition levels deep in the perovskite band gap is captured in Fig. \ref{Fig:outline}(c) as D.F.E + E$_{F}$/E$_{gap}$$^{PBE}$ > 1 eV (which prevents D.F.E. from being negative anywhere in the band gap region) under any chemical potential condition, V$_{A}$ (0/-1) being less than 0.5 eV from the VBM, and V$_{X}$ (+1/0) being less than 0.5 eV from the CBM. \\

It can be seen from Fig. \ref{Fig:Defect_heatmaps}(a) that the lowest D.F.E. values (pictured here for medium-X conditions) are shown by Rb and K compounds while Cs, MA, and FA generally show higher D.F.E. values. The transition levels, plotted as a fraction of E$_{gap}$$^{PBE}$ in Fig. \ref{Fig:Defect_heatmaps}(b) and (c), appear to generally be < 0.2 for V$_{A}$ (0/-1) for all halides except for (Rb,K)-Ba compounds, while for V$_{X}$ (+1/0), there is a much higher tendency for deeper values (between 0.2 and 0.8), with (MA,Cs)-(Ge,Sn) compounds being primary culprits. Several Ca, Sr, and Ba containing perovskites show V$_{X}$ (+1/0) levels inside the VB or CB and would thus be safer w.r.t deeper defect levels. Finally, the equilibrium E$_{F}$, plotted as a fraction of E$_{gap}$$^{PBE}$ for X-rich, medium-X, and B-rich conditions in Fig. \ref{Fig:Defect_heatmaps}(d), (e) and (f) respectively, shows a general tendency for p-type conductivity under X-rich conditions that gradually becomes more n-type under B-rich conditions. All K and Rb iodides, and Ge, Sn, and Pb based bromides and chlorides, are strongly p-type under X-rich conditions and only become slightly less p-type under medium-X or B-rich conditions. All Cs, MA, and FA containing halides are p-type under X-rich, intrinsic under medium-X, and n-type under B-rich conditions, which is consistent with our previous reports on MAPbX$_{3}$ perovskites and other published computational literature \cite{MK1,MK2,Yanfa}. \\

Next, the entire DFT dataset of 229 compounds, containing both pure and mixed cation compositions, is visualized in the form of plots between different properties of interest in Fig. \ref{Fig:PBE_vs_HSE} and Fig. \ref{Fig:Defect_data}. It can be seen from Fig. \ref{Fig:PBE_vs_HSE}(a), (b) and (d) respectively that the PBE and HSE computed lattice constant, $\Delta$H$_{decomp}$, and $\Delta$H$_{form}$ values respectively correlate highly and often match very well with each other, indicating that the structure and energetics from the PBE functional provide sufficient accuracy. Fig. \ref{Fig:PBE_vs_HSE} also shows PBE vs HSE linear regression equations which could be used for predicting the expensive HSE level information using the moderate PBE data, with R$_2$ values of 0.97, 0.99, 0.99 and 0.94 respectively for lattice constant, $\Delta$H$_{decomp}$, $\Delta$H$_{form}$, and E$_{gap}$. $\Delta$H$_{mix}$ is more sensitive to the functional as can be seen from Fig. \ref{Fig:PBE_vs_HSE}(c), with some of the FA-based compounds showing a difference in the relative energetics of pure and mixed composition compounds from PBE and HSE energies. While the entire DFT dataset falls within an acceptable stability range for $\Delta$H$_{form}$ and $\Delta$H$_{mix}$, results are much more interesting for $\Delta$H$_{decomp}$, where the shaded region of stability in Fig. \ref{Fig:PBE_vs_HSE}(b) shows that while nearly all FA and MA compounds are resistant to decomposition, many Cs, Rb, and K compounds, mostly chlorides, are not. Further, a plot between E$_{gap}$$^{PBE}$ and E$_{gap}$$^{HSE}$ in Fig. \ref{Fig:PBE_vs_HSE}(e) shows a clear correlation between the two, with the HSE band gaps generally $\sim$ 1 eV higher than their PBE counterparts, and the shaded region of favorability showing the existence of many iodides, some bromides, and few chlorides. Fig. \ref{Fig:SI_gap_eps_dft} shows $\epsilon$$_{elec}$ plotted against E$_{gap}$$^{PBE}$ and E$_{gap}$$^{HSE}$, showing an inverse relationship between the electronic dielectric constant and band gap as expected. \\

Finally, the PV FOM (in log$_{10}$ scale) estimated from the optical absorption spectrum is plotted against the refractive index $\eta$ in Fig. \ref{Fig:PBE_vs_HSE}(f) to complete the list of the electronic/dielectric/optical properties being studied here. Some examples of calculated absorption coefficients are presented in Fig. \ref{Fig:SI_abs_spectra} along with the solar irradiance intensities, as a function of wavelength. We performed additional computations for some well known semiconductors, namely Si, SiC, GaAs, CdTe and CdSe (all in the zincblende structure), and their absorption spectra are shown in Fig. \ref{Fig:SI_abs_spectra}(a) to obtain an idea of how their overlap looks like with the AM1.5 spectrum. Calculated spectra for some selected halide perovskite alloys are shown in Fig. \ref{Fig:SI_abs_spectra}(b). Rather than a region of favorability, what we seek as a screening criterion here is a high value of FOM to maximize absorption in the visible range. FOM values values for the dataset lie between 3 and 6 in the log scale; in Table \ref{table:SI_FOM}, we listed the calculated FOM values of known semiconductors as well as some example perovskite compounds from our dataset. The FOM values of Si and GaAs are close to 6, similar to peak values in our dataset. \\

The defect properties for the 229 compounds are visualized in Fig. \ref{Fig:Defect_data}, in the form of plots between equilibrium D.F.E. and E$_{F}$ (a--c) and between V$_{A}$ (0/-1) and V$_{X}$ (+1/0) (d). In the D.F.E. vs E$_{F}$ (as a fraction of E$_{gap}$$^{PBE}$) plots, the shaded region indicates where D.F.E + E$_{F}$/E$_{gap}$$^{PBE}$ > 1 eV, showing the compounds which have sufficiently large defect formation energies. It can be seen from Fig. \ref{Fig:Defect_data}(a), (b) and (c) that very few compounds satisfy this condition, and are dominated by FA and MA containing perovskites. The number of defect tolerant compounds increases from X-rich to medium-X to B-rich chemical potential conditions, with the number of defect-tolerant Cs and Rb based compounds also consequently increasing. From Fig. \ref{Fig:Defect_data}(d), it is seen that a lot more compounds satisfy the charge transition level criterion, and in particular, many K, Rb, and Cs containing perovskites have V$_{A}$ and V$_{X}$ defects that are shallow in nature. On the other hand, many FA and MA compounds may create deeper defect levels, which could be a problem if they do not exist in the shaded region in Fig. \ref{Fig:Defect_data}(a), (b) or (c). It should be noted that there isn't one accepted definition of what constitutes a deep defect level and the criterion we apply here is slightly relaxed (< 0.5 eV from the band edges) by considering the level of accuracy expected in DFT (PBE) computed defect levels. \\ 

\begin{figure}[!h]
\centering
\includegraphics[width=0.90\columnwidth]{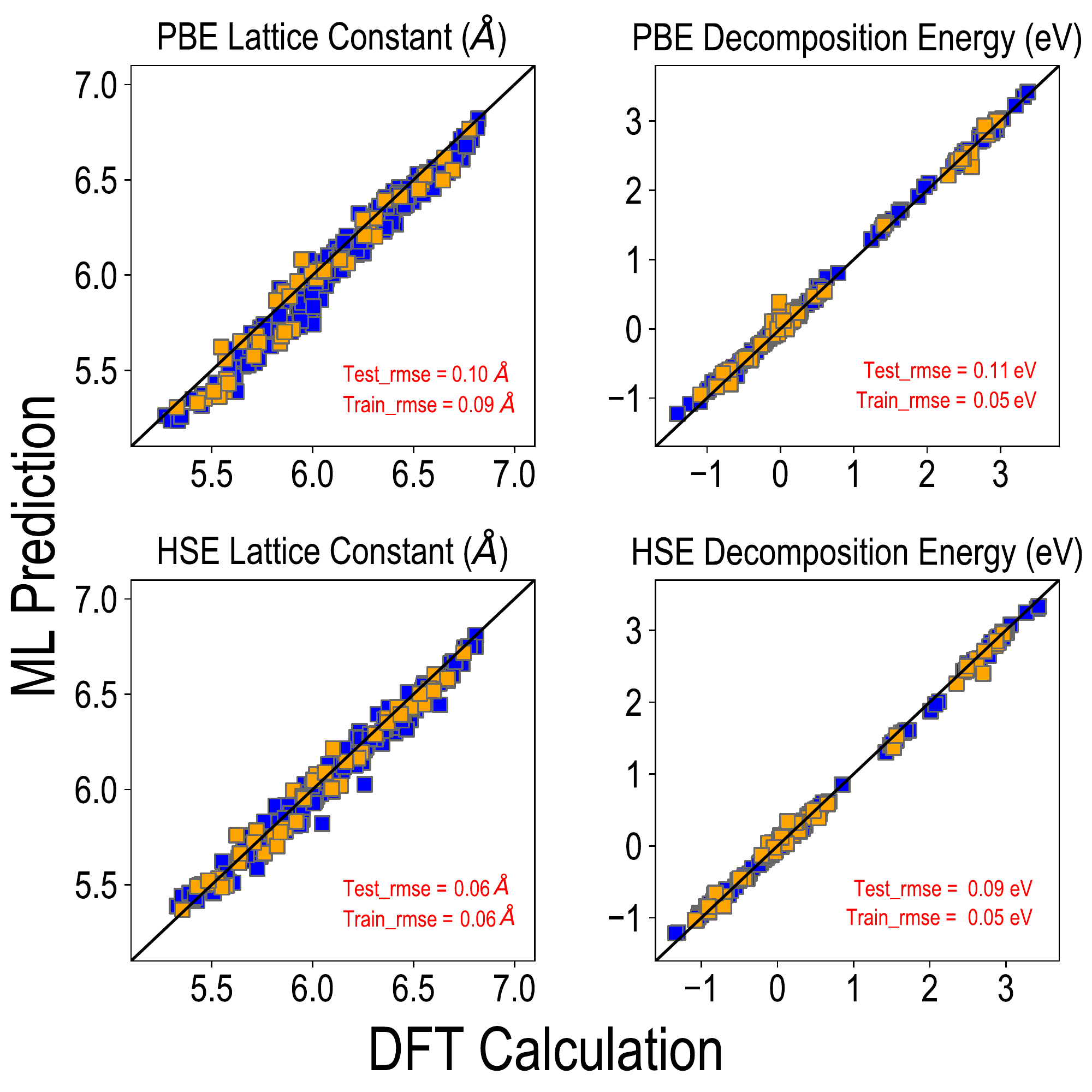}
\caption{\label{Fig:NN_struct} 
Neural network models trained for (a) PBE lattice constant, (b) PBE $\Delta$H$_{decomp}$, (c) HSE06 lattice constant, and (d) HSE06 $\Delta$H$_{decomp}$.}
\end{figure}

\subsection*{Neural Network Models}

\begin{figure}[!t]
\centering
\includegraphics[width=0.90\columnwidth]{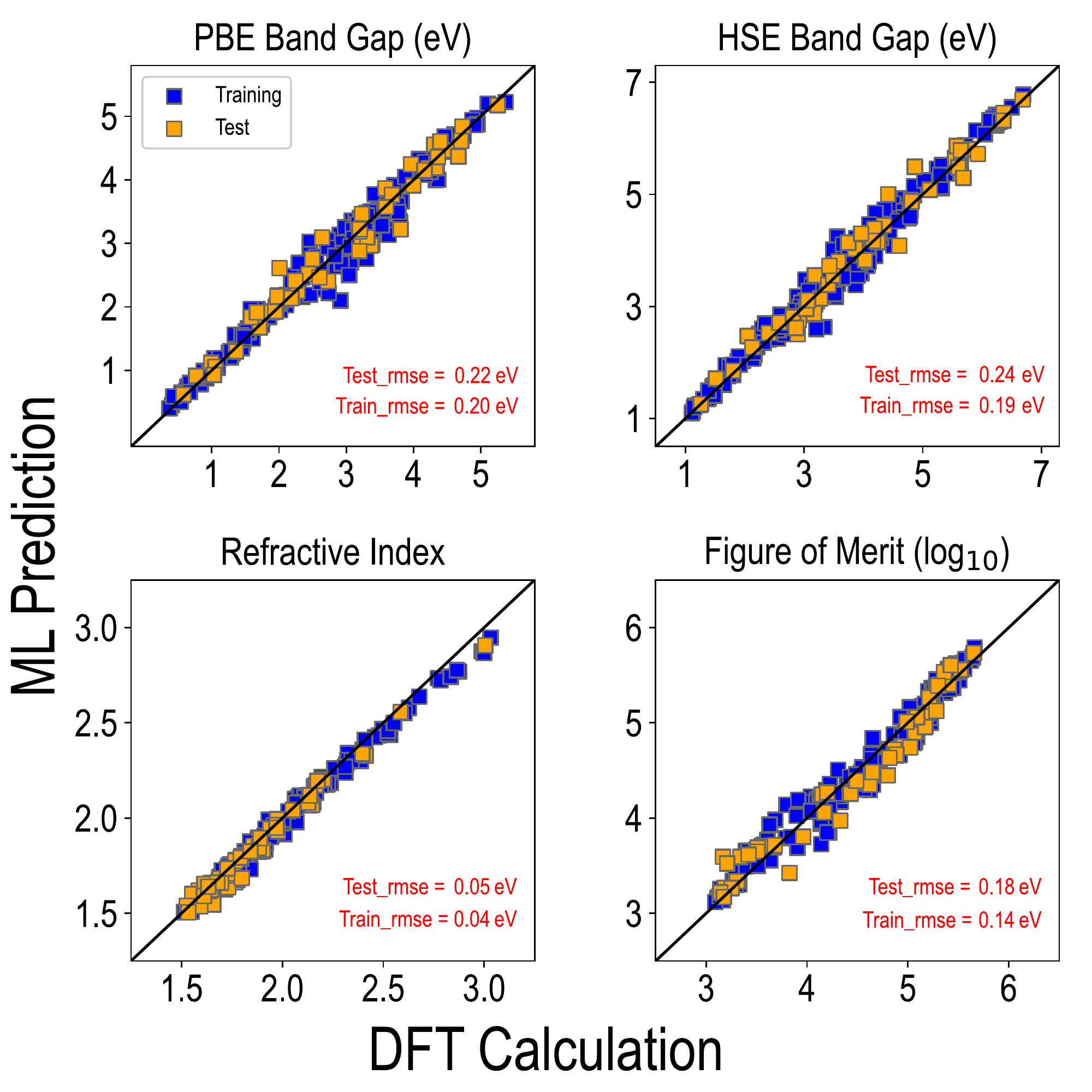}
\caption{\label{Fig:NN_elec} 
Neural network models trained for (a) PBE E$_{gap}$, (b) HSE06 E$_{gap}$, (d) refractive index, and (c) figure of merit.}
\end{figure}

\begin{figure}[!b]
\centering
\includegraphics[width=0.90\columnwidth]{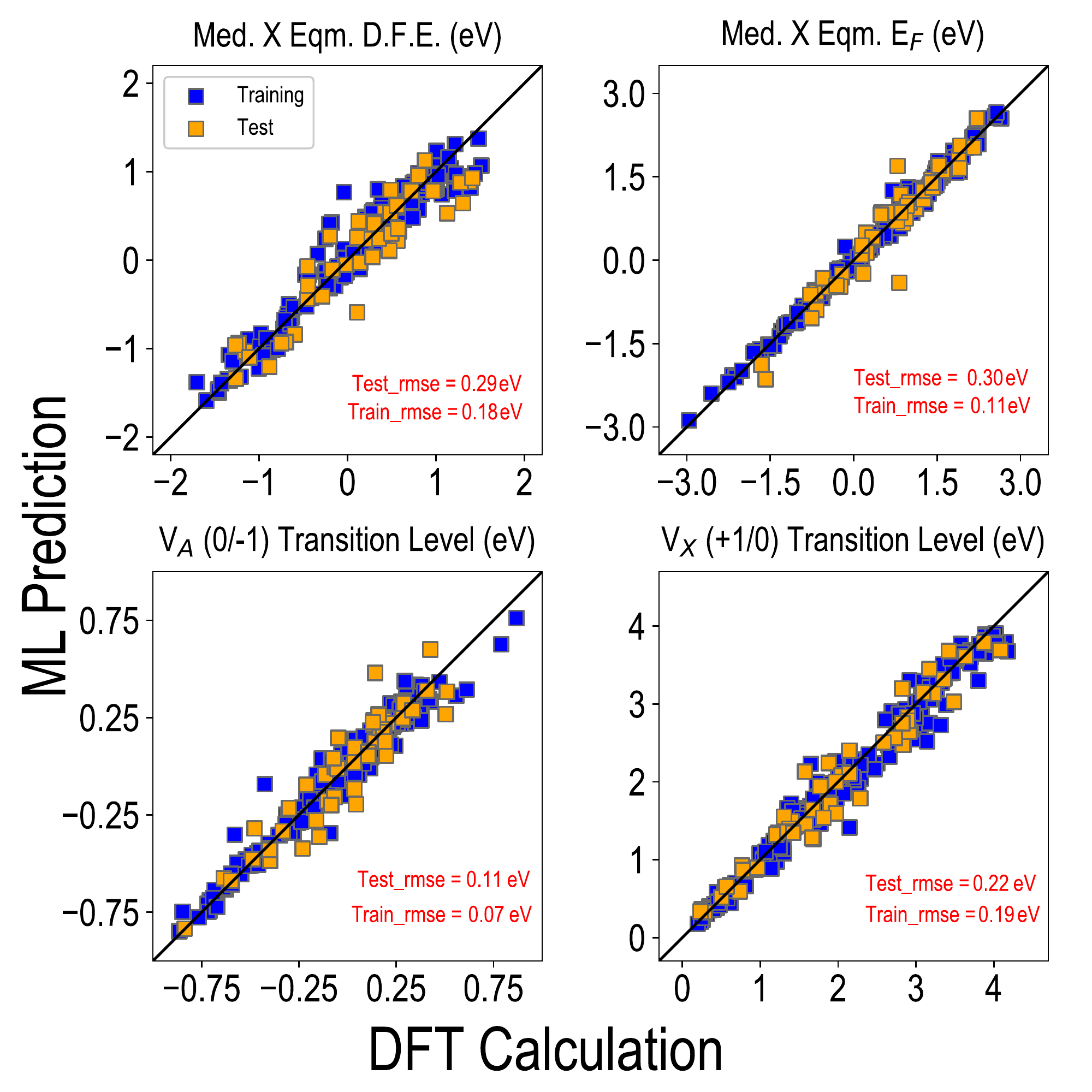}
\caption{\label{Fig:NN_defects} 
Neural network models trained for (a) equilibrium D.F.E. and (b) equilibrium E$_{F}$, both at medium halide chemical potential conditions, (c) V$_{A}$ (0/-1) charge transition level and (d) V$_{X}$ (+1/0) charge transition level, both in eV with respect to VBM.}
\end{figure}

Rigorously optimized, cross-validated, and tested NN predictive models for various properties of interest---roughly divided in terms of the type of properties---are presented in Fig. \ref{Fig:NN_struct}, Fig. \ref{Fig:NN_elec} and Fig. \ref{Fig:NN_defects}. We first report predictions for the structure and energetics of halide perovskites in terms of the lattice constant and $\Delta$H$_{decomp}$ from PBE and HSE in Fig. \ref{Fig:NN_struct}. The lattice constants are predicted with similar training and test RMSE of < 0.1 \AA, which is an error of $\sim$ 7$\%$ given the range of values adopted across the dataset. $\Delta$H$_{decomp}$ is also predicted with a high accuracy of $\sim$ 0.1 eV test RMSE for both PBE and HSE, which is a remarkably small error of $\sim$ 2.5$\%$ given the total range of values, and comparable or better than machine-learned formation energy errors in the literature \cite{ML_form1,ML_form2,ML_form3}. It should be noted from the definitions of $\Delta$H$_{decomp}$, $\Delta$H$_{form}$, and $\Delta$H$_{mix}$ that there is an interdependence between them, meaning that given $\Delta$H$_{decomp}$ and the energies of all elemental or molecular standard states and halide compounds of A, B, and X species, $\Delta$H$_{form}$ and $\Delta$H$_{mix}$ can be estimated as well. The NN models presented in Fig. \ref{Fig:NN_struct} may be applied to predict, at the PBE and HSE levels, the lattice constant, decomposition energy, formation energy, and mixing energy of every compound in the dataset of 17,955. \\

Next, models are trained for the electronic, dielectric, and optical properties, namely E$_{gap}$$^{PBE}$, E$_{gap}$$^{HSE}$, $\eta$, and FOM, and the results are presented in Fig. \ref{Fig:NN_elec}. The test RMSE for PBE and HSE band gaps are 0.22 eV and 0.24 eV respectively, which are similar to or better than most ML-based band gap predictions in the literature \cite{MLHP7,MK5,MK6,Gaps_ML1,Gaps_ML2,Gaps_ML3}. The errors in band gap predictions lie around 5$\%$ of the respective ranges of PBE and HSE computed values. $\eta$ can be predicted with a very high accuracy, showing a test RMSE of 0.05, corresponding to $\sim$ 3$\%$ of the range of values in the dataset, while the FOM test RMSE converges to 0.18 (in log$_{10}$), which is $\sim$ 5$\%$ of the range of values. Thus, the band gaps, electronic dielectric constant or refractive index, and PV figure of merit can all be predicted for the halide perovskite dataset with an error of 5$\%$ or less, and these predictions can be used to screen compounds with suitable band gaps and large PV figures of merit. Finally, we look at the defect properties, and train separate models for the equilibrium D.F.E. and E$_{F}$ under X-rich, medium-X, and B-rich chemical potential conditions, and transition levels V$_{A}$ (0/-1) and V$_{X}$ (+1/0). In Fig. \ref{Fig:NN_defects}, NN models are presented for D.F.E. and E$_{F}$ under medium-X conditions, and for V$_{A}$ (0/-1) and V$_{X}$ (+1/0). \\

Prediction performances for defect properties reveal slightly higher errors as compared to the structural, energetic, electronic, and optical properties discussed earlier. There are additional factors at play when it comes to determining the behavior of point defects, such as the defect-induced distortions in the perovskite structure, which means that the descriptors that have worked so well thus far---elemental/molecular properties of A, B, and X species---may need to be fortified with more information for better predictions. Our past work shows that ML-defect predictions are significantly improved by using cheaper unit cell defect calculated properties as descriptors in addition to elemental properties of the defect atom \cite{MK1,MK3}. Nevertheless, keeping in mind the desire for simplicity in an ML model and the convenience of common inputs for predicting multiple properties, we report the best predictions here for all defect properties using only elemental/molecular descriptors. As shown in Fig. \ref{Fig:NN_defects}, D.F.E. and E$_{F}$ under medium-X chemical potential conditions can both be predicted with a test RMSE of $\sim$ 0.3 eV, which is an error of less than 10$\%$ for both properties. However, a larger difference between the training and test RMSE values are seen here as compared to earlier properties, indicating a possible tendency for over-fitting despite rigorous optimization. NN models for V$_{A}$ (0/-1) and V$_{X}$ (+1/0) show test RMSE of 0.11 eV and 0.22 eV respectively, which are both less than 10$\%$ of the total ranges of values. Predictions for D.F.E. and E$_{F}$ under X-rich and B-rich conditions are presented in Fig. \ref{Fig:SI_NN_defects}, revealing similar accuracies as the models presented in Fig. \ref{Fig:NN_defects}. Thus, predictions can reliably be made, generally with an error of no more than 10$\%$, of the entire vacancy defect formation energy picture (considering only A-site and X-site vacancies) of all halide perovskite compounds, and this information can be used to screen for defect tolerant compounds. The training and test set prediction RMSE values for every property is listed in Table \ref{table:rmse}. \\

\begin{table}
\centering
  \caption{\ NN model training and test prediction RMSEs for every property.}
  \label{table:rmse}
  \begin{tabular}{ccc}
    \hline
   &   &   \\
\textbf{Property}  &  \textbf{Training Set RMSE}  &  \textbf{Test Set RMSE} \\
   &   &   \\
\hline
   &   &   \\
      PBE Lattice Constant  & 0.09 \AA  & 0.10 \AA  \\
      HSE Lattice Constant  & 0.06 \AA  & 0.06 \AA  \\
      $\Delta$H$_{decomp}$ (PBE) & 0.05 eV  & 0.11 eV  \\
      $\Delta$H$_{decomp}$ (HSE)  & 0.05 eV  & 0.09 eV  \\
      E$_{gap}$$^{PBE}$  & 0.20 eV  & 0.22 eV  \\
      E$_{gap}$$^{HSE}$  & 0.19 eV  & 0.24 eV  \\
      Refractive Index  & 0.04  & 0.05  \\
      PV FOM (log$_{10}$)  & 0.14  & 0.18  \\
      X-rich D.F.E.  & 0.12 eV  & 0.23 eV  \\
      X-rich E$_{F}$  & 0.06 eV  & 0.19 eV  \\
      Medium-X D.F.E.  & 0.18 eV  & 0.29 eV  \\
      Medium-X E$_{F}$  & 0.11 eV  & 0.30 eV  \\
      B-rich D.F.E.  & 0.11 eV  & 0.30 eV  \\
      B-rich E$_{F}$  & 0.11 eV  & 0.25 eV  \\
      V$_{A}$ (0/-1) & 0.07 eV  & 0.11 eV  \\
      V$_{X}$ (+1/0)  & 0.19 eV  & 0.22 eV  \\
   &   &   \\
    \hline
  \end{tabular}
\end{table}

\subsection*{High-Throughput Prediction and Screening}

\begin{figure}[!b]
\centering
\includegraphics[width=\columnwidth]{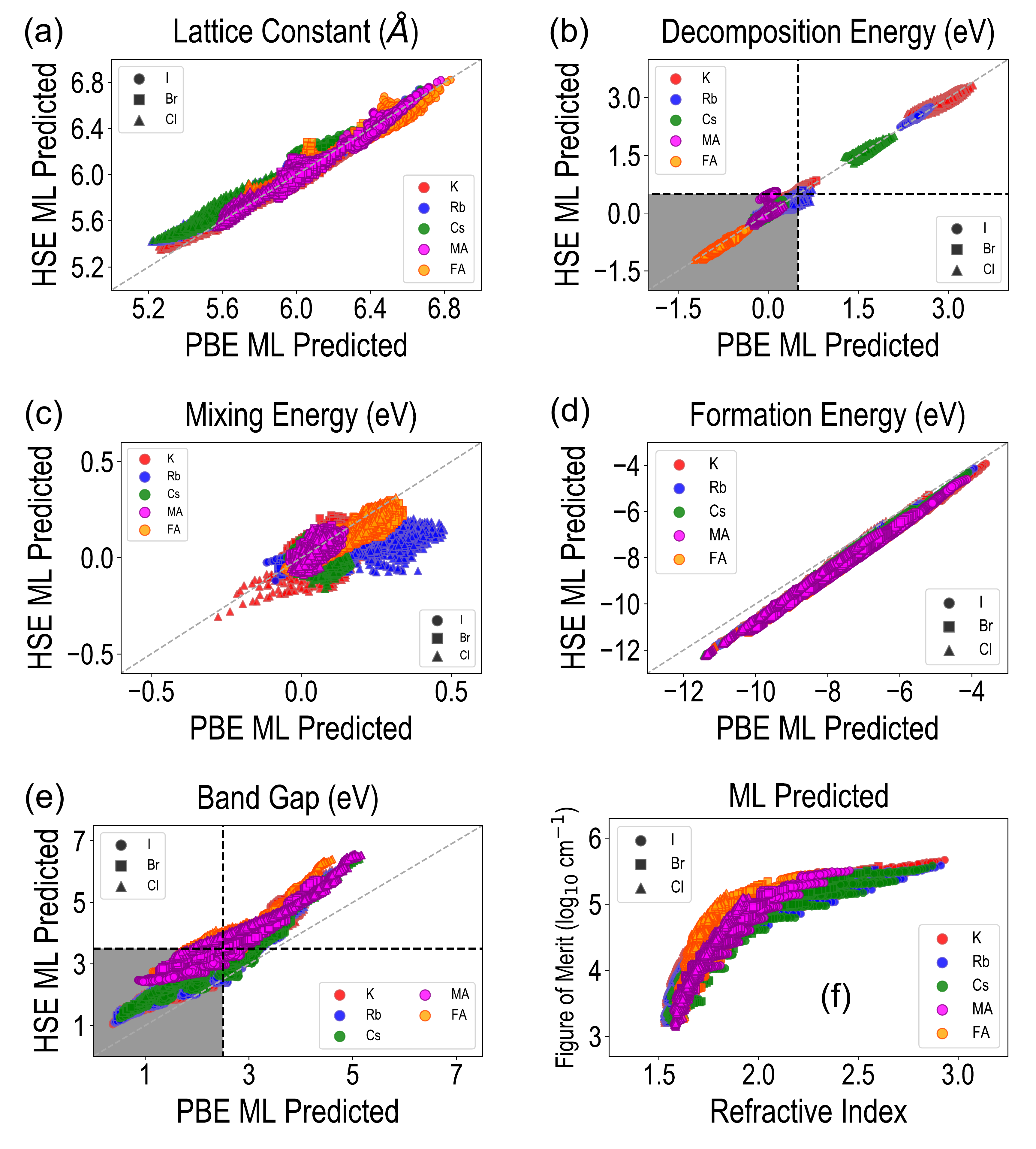}
\caption{\label{Fig:PBE_vs_HSE_ML} 
A visualization of the ML predicted PBE and HSE lattice constants (a), $\Delta$H$_{decomp}$ (b), $\Delta$H$_{mix}$ (c), $\Delta$H$_{form}$ (d), E$_{gap}$ (e), and refractive index vs figure of merit (f), for the chemical space of 17,955 perovskite compounds.}
\end{figure}

\begin{figure}[!t]
\centering
\includegraphics[width=\linewidth]{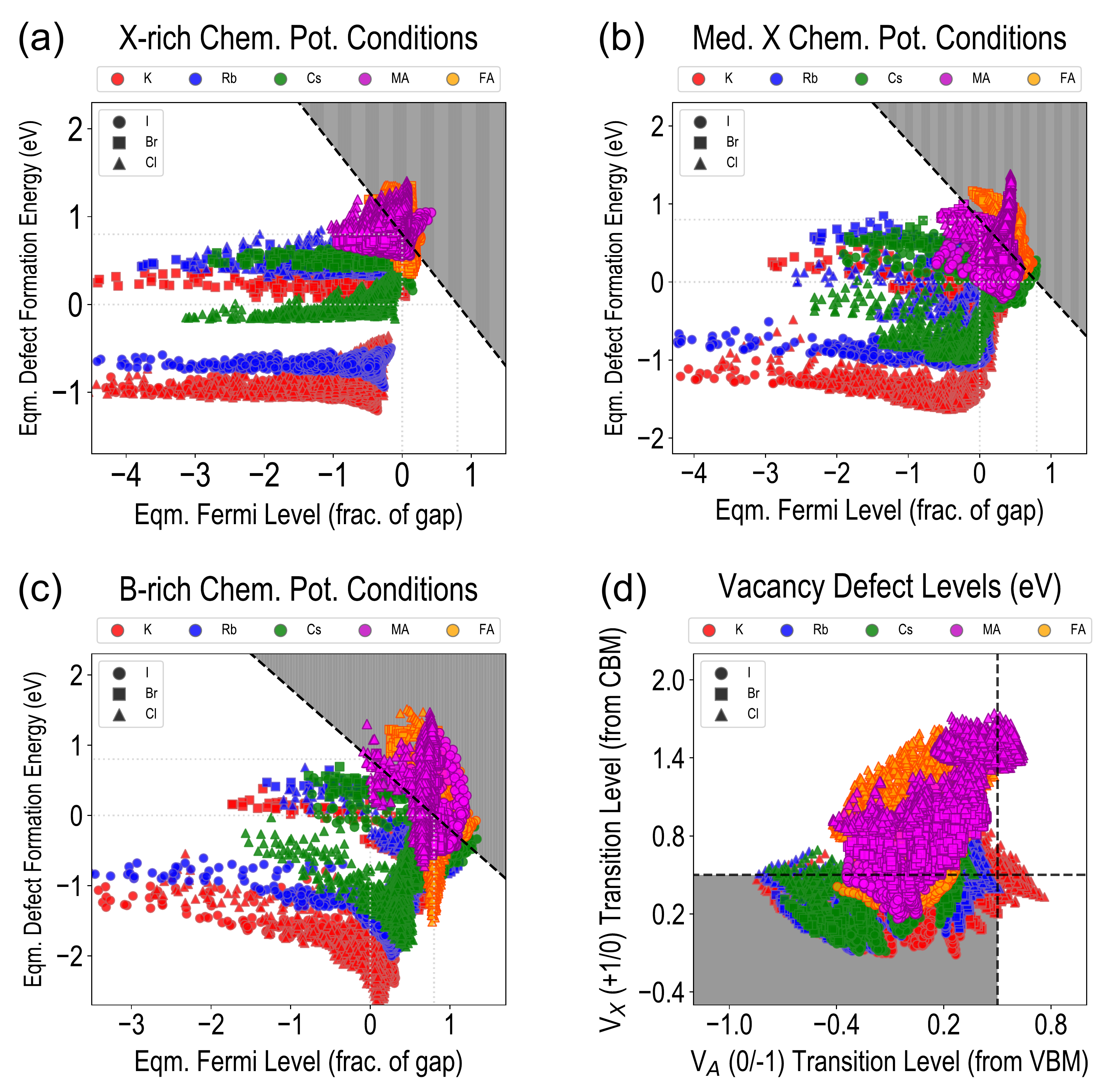}
\caption{\label{Fig:Defect_data_ML} 
ML predicted defect properties for the chemical space of 17,955 perovskite compounds: Equilibrium D.F.E. vs E$_{F}$/E$_{gap}$ plots are shown for (a) halide rich, (b) medium halide, and (c) B-rich chemical potential conditions. The shaded region represents materials that satisfy the criterion D.F.E. + E$_{F}$/E$_{gap}$ > 1.0 eV and are thus defect tolerant; for high-throughput screening, this threshold is relaxed to 0.8 eV. The V$_{A}$ (0/-1) and V$_{X}$ (+1/0) charge transition levels are plotted in (d), with respect to the VBM and CBM respectively, and the shaded region represents cases with defect levels < 0.2 eV from band edges while the remaining materials have deep vacancy defect levels.}
\end{figure}

The optimized NN models were deployed for predictions of every property of interest over the entire space of 17,955 halide perovskite compositions. The ML predicted property space is visualized in Fig. \ref{Fig:PBE_vs_HSE_ML} and Fig. \ref{Fig:Defect_data_ML}, similar to the way DFT data was visualized earlier. Since regression is an interpolative approach, the ML predicted ranges of each property are similar to the DFT dataset, especially since the chemical space of 17,955 compounds is well represented within the 229 compounds chosen for DFT computations. The trends in structural, energetic, electronic, optical and defect properties remain the same as before, with many more compounds occupying the shaded regions of favorability---immediately highlighting the advantage of a DFT+ML high-throughput screening approach. \\

As shown in Fig. \ref{Fig:PBE_vs_HSE_ML}(a), (b) and (d), the lattice constant, $\Delta$H$_{decomp}$, and $\Delta$H$_{form}$ at the PBE and HSE levels yield similar predictions, with a lot more MA and FA compounds in the low $\Delta$H$_{decomp}$ region. $\Delta$H$_{mix}$$^{PBE}$ and $\Delta$H$_{mix}$$^{HSE}$ in Fig. \ref{Fig:PBE_vs_HSE_ML}(c) remain uncorrelated for many compounds but across the dataset, $\Delta$H$_{mix}$ values lie below the desired threshold of 0.5 eV. ML predicted E$_{gap}$$^{PBE}$ and E$_{gap}$$^{HSE}$ plotted in Fig. \ref{Fig:PBE_vs_HSE_ML}(e) also show similar trends as in the DFT dataset, with several hundred more fresh compounds in the desirable region. Further, the ML predicted FOM vs $\eta$ plot captures a large number of Cs, Rb, and K-containing perovskites---many of them occupying the desired band gap region---that do quite well when it comes to absorption within the solar spectrum. The ML predicted defect properties presented in Fig. \ref{Fig:Defect_data_ML}, again, capture the same trends as in the DFT dataset. A large majority of Cs, Rb and K perovskites fail to satisfy the D.F.E. / E$_{F}$ criterion and will thus not be defect-tolerant, while the number of FA and MA compounds with large enough defect energies increase from X-rich (a) to medium-X (b) to B-rich (c) conditions. The V$_{A}$ (0/-1) vs V$_{X}$ (+1/0) plot in Fig. \ref{Fig:Defect_data_ML}(d) once again shows a predominance of Cs, Rb, and K compounds in the desired shallow defect level region, with an enormous number of FA and MA bromides and chlorides containing deeper level vacancy defects. \\

\begin{figure*}[!t]
\centering
\includegraphics[width=0.90\linewidth]{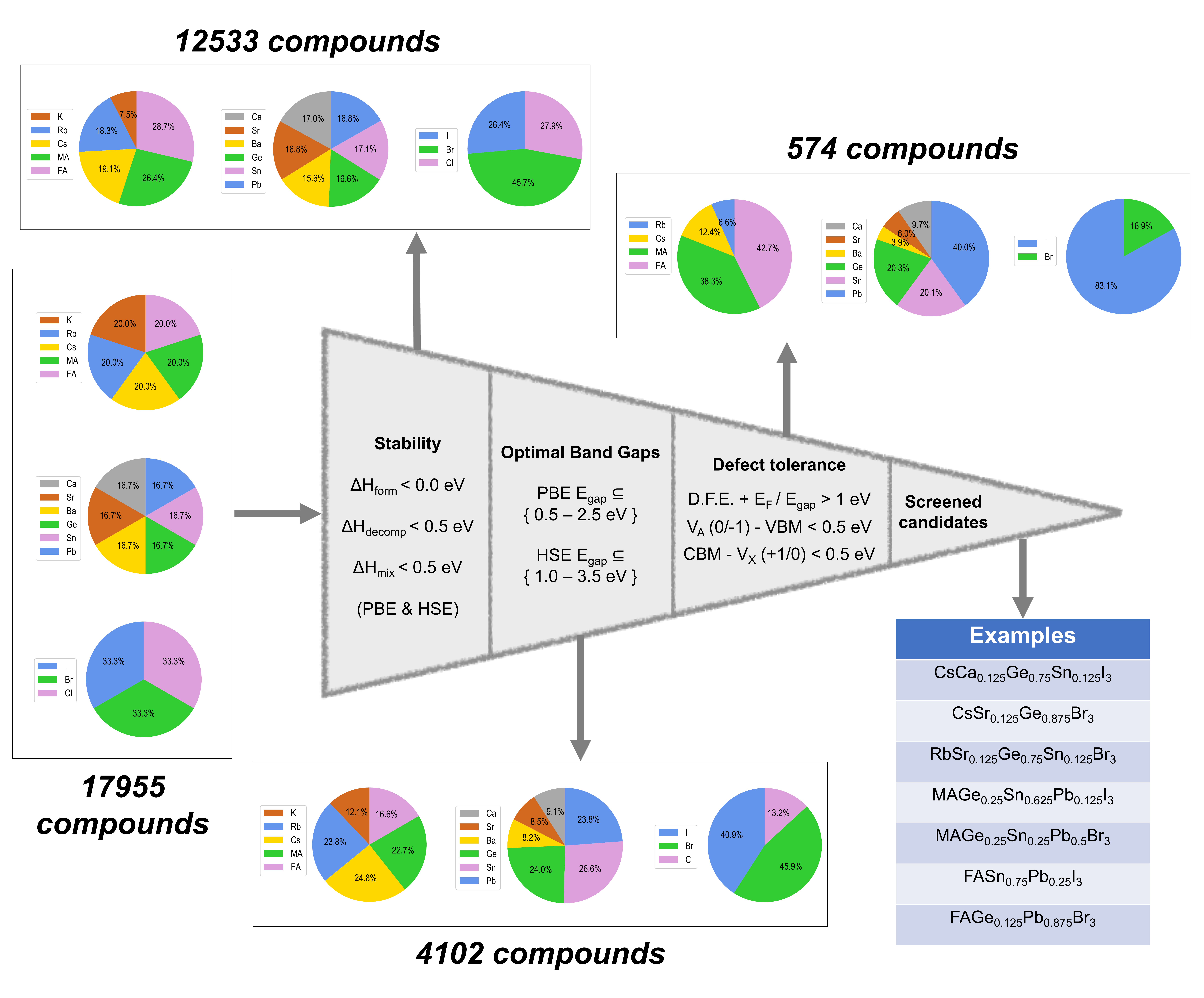}
\caption{\label{Fig:Screening_process} 
A visualization of the perovskite compositional space through the screening process, in terms of the fractions of each type of A, B and X atoms remaining in the set of compounds. Also shown are a few examples of screened compounds.}
\end{figure*}

A closer look at the DFT/ML data reveals reasons for the popularity of (FA,MA)(Pb,Sn)(I,Br)$_3$ perovskites, which have high FOM values of $>$ 5.2, suitable band gaps between 1 eV and 2 eV, low decomposition energies, and high defect formation energies. A majority of purely inorganic Cs, Rb and K-based perovskites show high FOM values and fewer deep defect levels, but are plagued by instability to decomposition and may need to be further stabilized, e.g., by entropic contributions with more B-site cations. Based on the screening funnel identified in Fig. \ref{Fig:outline}(c), we use the ML predicted properties across the dataset of 17,955 perovskites to identify compounds with sufficiently low formation, decomposition, and mixing energies (from PBE and HSE), band gaps in suitable ranges (from PBE and HSE), and defect tolerance based on equilibrium defect formation energies and transition levels. This screening process is presented in Fig. \ref{Fig:Screening_process}, using pie charts to display statistics and the frequency of occurrence of each A, B, and X species, and listing some examples of screened compounds. It is seen that 12,533 compounds out of 17,955 are identified as stable, 4102 of which have suitable E$_{gap}$$^{PBE}$ and E$_{gap}$$^{HSE}$ values, and finally, 574 compounds are left which display robust vacancy defect-tolerance as well. \\

It is interesting to note the prevalence of certain species and the absence of certain other species through the screening process. A majority of the stable compounds are FA and MA-based, while a large number of K-based compounds are eliminated during the stability screening. All B-site atoms are equally represented at this stage while Br is much more prevalent than I or Cl, indicating that the bromides of FA and MA with any combinations of B atoms are more likely to be stable than other combinations. At the band gap screening stage, there is a rearrangement of the frequencies of occurrences: more FA and MA based compounds are eliminated, so that there is more of a balance between the 5 types of A species. Pb, Sn, and Ge are now the majority occurring B-site atoms, and fading numbers of Ca, Sr, and Ba are seen, while the number of chlorides further reduces, presumably because of their larger than desired band gaps. Iodides and bromides are both well represented at this point. After the defect tolerance screening stage, it can be seen that all the K-based compounds are eliminated, there are small numbers of Cs and Rb compounds left, and a majority of the perovskites have MA or FA at the A-site. Pb is the most common B-site element followed by Sn and Ge, with group II elements Ca, Sr, and Ba playing the role of low fraction substituents in certain compositions. All the chlorides are also eliminated and > 83$\%$ of the compounds are now iodides, indicating that MA and FA-based iodide perovskites are much more likely to be defect-tolerant than the corresponding bromides. A few selected compounds from the screened list of candidates are also identified as examples in Fig. \ref{Fig:Screening_process}. \\

\begin{figure}[!t]
\centering
\includegraphics[width=\linewidth]{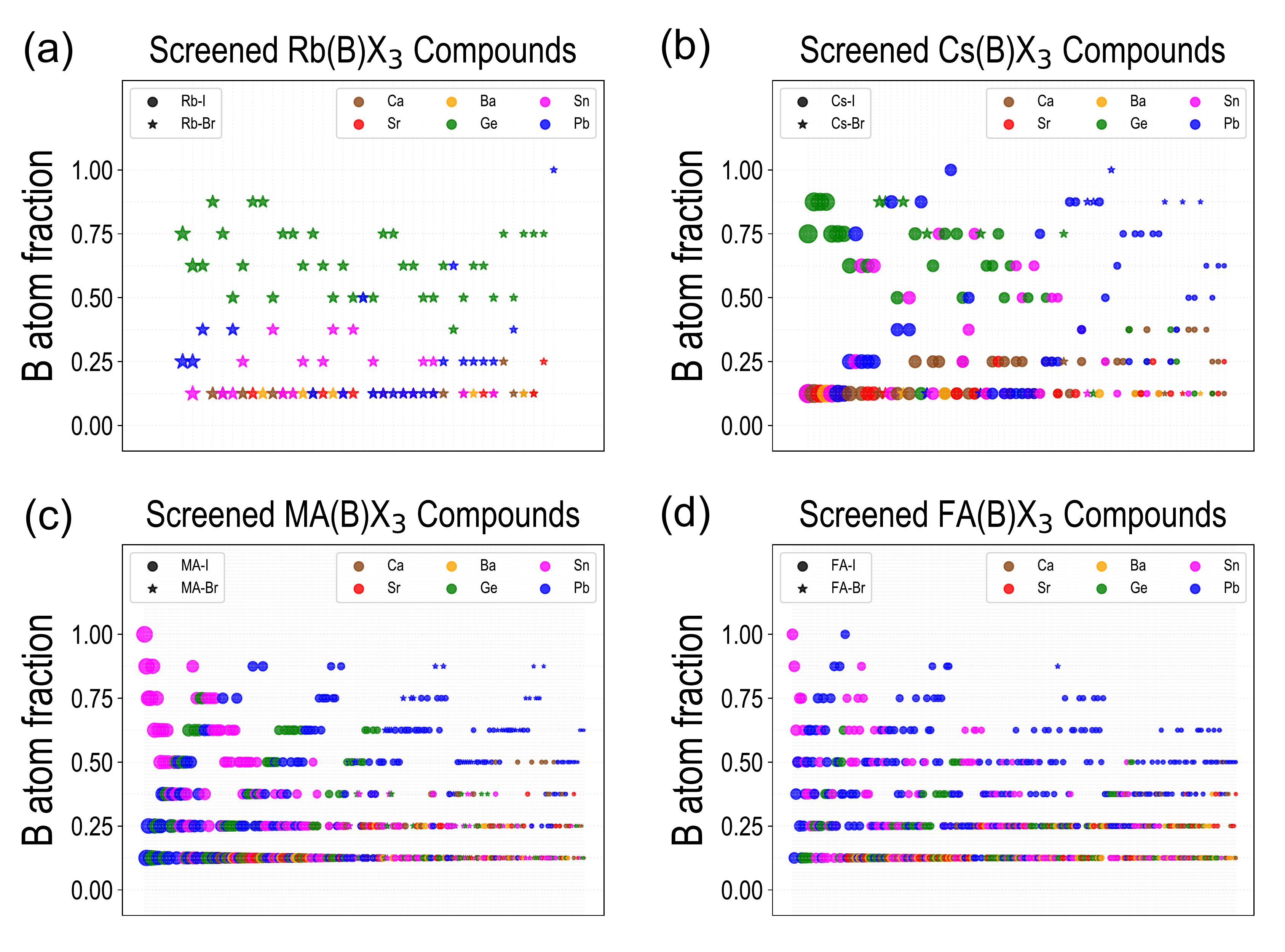}
\caption{\label{Fig:Screened_space} 
Visualizing the compositional space of the screened (a) Rb, (b) Cs, (c) MA and (d) FA iodide and bromide perovskites. The size of each scatter point is proportional to the calculated figure of merit.}
\end{figure}

\begin{table*}[h]
\centering
  \caption{\ Selected examples of screened perovskites.}
  \label{table:screened}
  \begin{tabular}{ccccccc}
    \hline
      &   &   &   &   &   &   \\
\textbf{Perovskite Formula}  &  \textbf{HSE $\Delta$H$_{decomp}$}  &  \textbf{HSE $\Delta$H$_{mix}$}  &  \textbf{PBE E$_{gap}$}  &  \textbf{HSE E$_{gap}$}  &  \textbf{PV FOM}  &  \textbf{Eqm. Cond.} \\
   &  \textbf{(eV p.f.u.)}  &  \textbf{(eV p.f.u.)}  &  \textbf{(eV)}  &  \textbf{(eV)}  &  \textbf{(log$_{10}$ cm$^{-1}$)}  &  \textbf{(Med. X)} \\
      &   &   &   &   &   &   \\
\hline
      &   &   &   &   &   &   \\
CsCa$_{0.125}$Ge$_{0.75}$Sn$_{0.125}$I$_{3}$  &  0.09  &  0.06  &  0.95  &  1.60  &  5.35  &  mod. p-type \\
CsCa$_{0.125}$Ge$_{0.875}$I$_{3}$  &  0.12  &  0.08  &  1.04  &  1.66  &  5.35  &  intrinsic \\
CsSr$_{0.125}$Ge$_{0.875}$I$_{3}$  &  0.12  &  0.07  &  1.05  &  1.65  &  5.34  &  intrinsic \\
CsBa$_{0.125}$Ge$_{0.875}$I$_{3}$  &  0.15  &  0.06  &  1.05  &  1.63  &  5.35  &  mod. p-type \\
CsSr$_{0.125}$Ge$_{0.75}$Sn$_{0.125}$I$_{3}$  &  0.10  &  0.06  &  1.00  &  1.62  &  5.34  &  mod. p-type \\
RbGe$_{0.75}$Pb$_{0.25}$Br$_{3}$  &  -0.04  &  -0.02  &  0.56  &  1.51  &  5.41  &  very p-type \\
CsCa$_{0.125}$Ge$_{0.75}$Pb$_{0.125}$I$_{3}$  &  0.12  &  0.07  &  0.96  &  1.63  &  5.33  &  mod. p-type \\
RbGe$_{0.625}$Sn$_{0.125}$Pb$_{0.25}$Br$_{3}$  &  -0.03  &  -0.03  &  0.57  &  1.48  &  5.40  &  very p-type \\
MASnI$_{3}$  &  0.35  &  0.00  &  0.87  &  2.47  &  5.51  &  very p-type \\
MASn$_{0.875}$Pb$_{0.125}$I$_{3}$  &  0.37  &  0.00  &  0.90  &  2.46  &  5.48  &  very p-type \\
CsSr$_{0.125}$Ge$_{0.75}$Pb$_{0.125}$I$_{3}$  &  0.13  &  0.08  &  1.01  &  1.65  &  5.33  &  mod. p-type \\
CsCa$_{0.125}$Ge$_{0.625}$Pb$_{0.25}$I$_{3}$  &  0.12  &  0.07  &  0.96  &  1.64  &  5.30  &  mod. p-type \\
MASn$_{0.75}$Pb$_{0.25}$I$_{3}$  &  0.39  &  0.00  &  0.94  &  2.46  &  5.47  &  very p-type \\
MAGe$_{0.125}$Sn$_{0.75}$Pb$_{0.125}$I$_{3}$  &  0.41  &  0.04  &  0.88  &  2.46  &  5.48  &  very p-type \\
MAGe$_{0.125}$Sn$_{0.875}$I$_{3}$  &  0.40  &  0.05  &  0.86  &  2.47  &  5.49  &  very p-type \\
FACa$_{0.375}$Ba$_{0.125}$Pb$_{0.5}$I$_{3}$  &  -0.43  &  -0.05  &  2.29  &  3.23  &  4.76  &  mod. n-type \\
MACa$_{0.25}$Ge$_{0.25}$Pb$_{0.5}$Br$_{3}$  &  -0.04  &  0.13  &  2.07  &  3.35  &  4.74  &  mod. p-type \\
FACa$_{0.25}$Sr$_{0.125}$Ba$_{0.125}$Pb$_{0.5}$I$_{3}$  &  -0.44  &  -0.03  &  2.32  &  3.24  &  4.76  &  mod. n-type \\
FACa$_{0.25}$Sr$_{0.25}$Pb$_{0.5}$I$_{3}$  &  -0.44  &  -0.02  &  2.31  &  3.24  &  4.75  &  mod. n-type \\
MACa$_{0.125}$Sr$_{0.125}$Ge$_{0.25}$Pb$_{0.5}$Br$_{3}$  &  -0.04  &  0.13  &  2.13  &  3.38  &  4.73  &  mod. p-type \\
FACa$_{0.125}$Sr$_{0.25}$Ba$_{0.125}$Pb$_{0.5}$I$_{3}$  &  -0.45  &  -0.02  &  2.37  &  3.26  &  4.76  &  mod. n-type \\
FACa$_{0.125}$Sr$_{0.375}$Pb$_{0.5}$I$_{3}$  &  -0.44  &  -0.01  &  2.36  &  3.26  &  4.75  &  mod. n-type \\
MACa$_{0.25}$Sn$_{0.125}$Pb$_{0.625}$Br$_{3}$  &  -0.07  &  0.06  &  2.23  &  3.47  &  4.74  &  mod. p-type \\
MACa$_{0.25}$Ge$_{0.125}$Pb$_{0.625}$Br$_{3}$  &  -0.05  &  0.10  &  2.18  &  3.44  &  4.73  &  mod. p-type \\
MACa$_{0.125}$Sr$_{0.125}$Ge$_{0.125}$Pb$_{0.625}$Br$_{3}$  &  -0.05  &  0.09  &  2.25  &  3.48  &  4.72  &  mod. p-type \\
      &   &   &   &   &   &   \\
    \hline
  \end{tabular}
\end{table*}

\begin{figure*}[h]
\centering
\includegraphics[width=\linewidth]{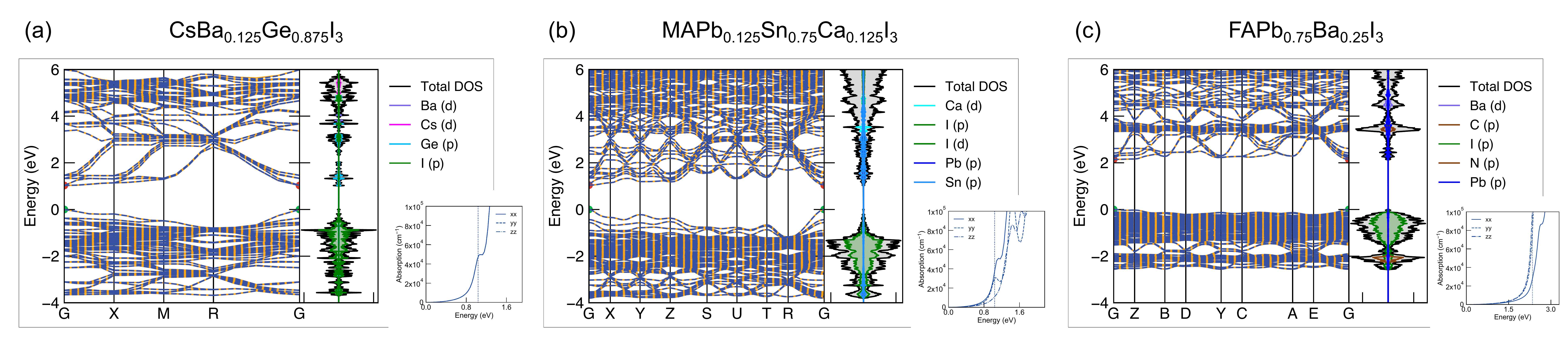}
\caption{\label{Fig:Band_struct} 
PBE-computed electronic band structure, density of states, and anisotropic optical absorption spectra (also showing band gaps as vertical dotted lines) of three compounds selected from the screened list: (a) CsBa$_{0.125}$Ge$_{0.875}$I$_{3}$, (b) MAPb$_{0.125}$Sn$_{0.75}$Ca$_{0.125}$I$_{3}$, and (c) FAPb$_{0.75}$Ba$_{0.25}$I$_{3}$.}
\end{figure*}

To further scrutinize the screened space of compounds, the fractions (corresponding to the amount of mixing at the B-site) of different types of B-site cations are plotted for all the screened iodides and bromides of FA, MA, Cs, and Rb in Fig. \ref{Fig:Screened_space}. It is noted from Fig. \ref{Fig:Screened_space}(a) that in Rb-based compounds, Pb and Sn occur in smaller fractions of 0.125--0.375, whereas Ge occurs in fractions of 0.500--0.875, and Ca/Sr/Ba only ever occur in fractions of 0.125. In Cs-based compounds captured in Fig. \ref{Fig:Screened_space}(b), Pb and Sn occur in both lower and higher fractions (0.125--1.0) although Ge continues to adopt higher fractions as well, while Ca/Sr/Ba occur in fractions of 0.125 or 0.25. Trends in MA- and FA-based compounds in Fig. \ref{Fig:Screened_space}(c) and (d) are similar to each other, with a clear prevalence of both Pb and Sn atoms from low to high fractions (0.125--1.0), relatively fewer occurrences of Ge in fractions of 0.25--0.75, and some occurrences of Ca/Sr/Ba in fractions of 0.125--0.5. An increase of the radius of A-site species from Rb to Cs to MA to FA predictably leads to a higher occurrence of larger B-site cations like Pb and Sn that also show suitable band gaps and defect tolerance. It is seen that group II cations usually prefer smaller substitutions of more common B-site atoms like Pb and Sn. The symbol sizes in Fig. \ref{Fig:Screened_space} are proportional to the PV FOM, revealing that a number of Cs-based perovskites with small fractions of Sn/Ca/Sr/Ba/Pb and large fractions of Ge, as well as a number of MA-based compounds with small fractions of Pb/Ge and larger fractions of Sn, show some of the largest absorption within the solar spectrum range. \\

A selected list of 25 screened compounds are presented in Table \ref{table:screened}, along with their ML predicted decomposition and mixing energies (at the HSE level), PBE and HSE band gaps, PV FOM (log$_{10}$), and nature of equilibrium conductivity under medium-halide chemical potential conditions as determined by A-site and X-site vacancies. Some of the compounds listed or their close derivatives are part of the experimental literature and have been suggested as candidates for solar cell absorption. Additional computations were performed on some of the screened compounds. The electronic band structures of three selected novel mixed cation perovskites, namely CsBa$_{0.125}$Ge$_{0.875}$I$_{3}$, MAPb$_{0.125}$Sn$_{0.75}$Ca$_{0.125}$I$_{3}$, and FAPb$_{0.75}$Ba$_{0.25}$I$_{3}$, were computed from PBE and are presented in Fig. \ref{Fig:Band_struct}, along with the electronic density of states and the optical absorption spectrum, which shows the absorption coefficients (in cm$^{-1}$) plotted against incident photon energy (in eV), and also shows E$_{gap}$$^{PBE}$ as a vertical line on the x-axis. The three compounds have direct band gaps of 1.03 eV, 1.04 eV and 2.14 eV, respectively. \\

\begin{figure*}[btp]
\centering
\includegraphics[width=0.80\linewidth]{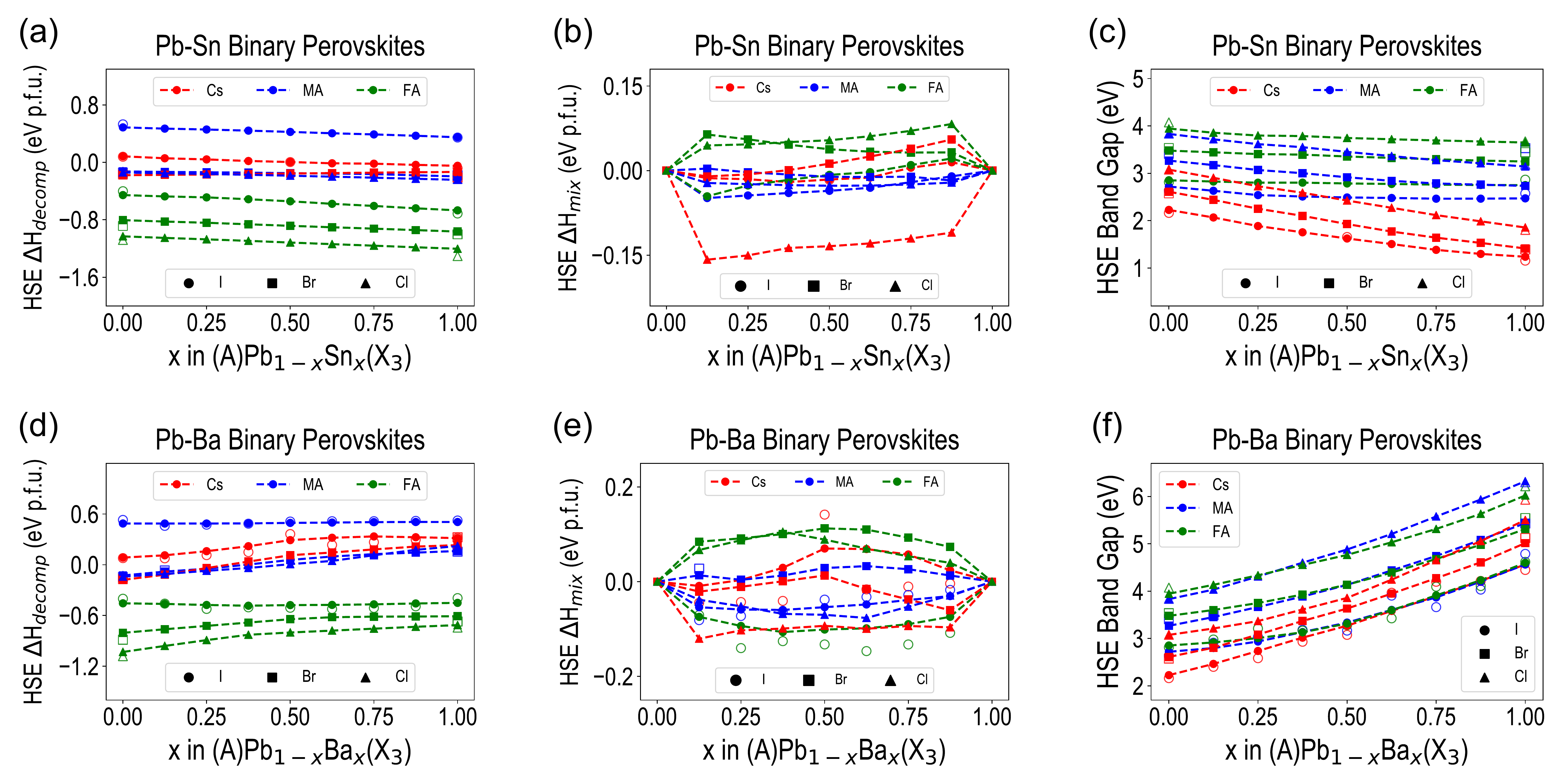}
\caption{\label{Fig:ML_binaries} 
ML (HSE) predicted decomposition energies, mixing energies (with added entropic contribution) and band gaps of Pb-Sn (a, b, c) and Pb-Ba (d, e, f) binary ABX$_{3}$ perovskites with A = Cs, MA or FA and X = I, Br or Cl, plotted as a function of Sn or Ba content. DFT calculated data points, where available, have been shown using open scatter points.}
\end{figure*}

\begin{figure}[h]
\centering
\includegraphics[width=\linewidth]{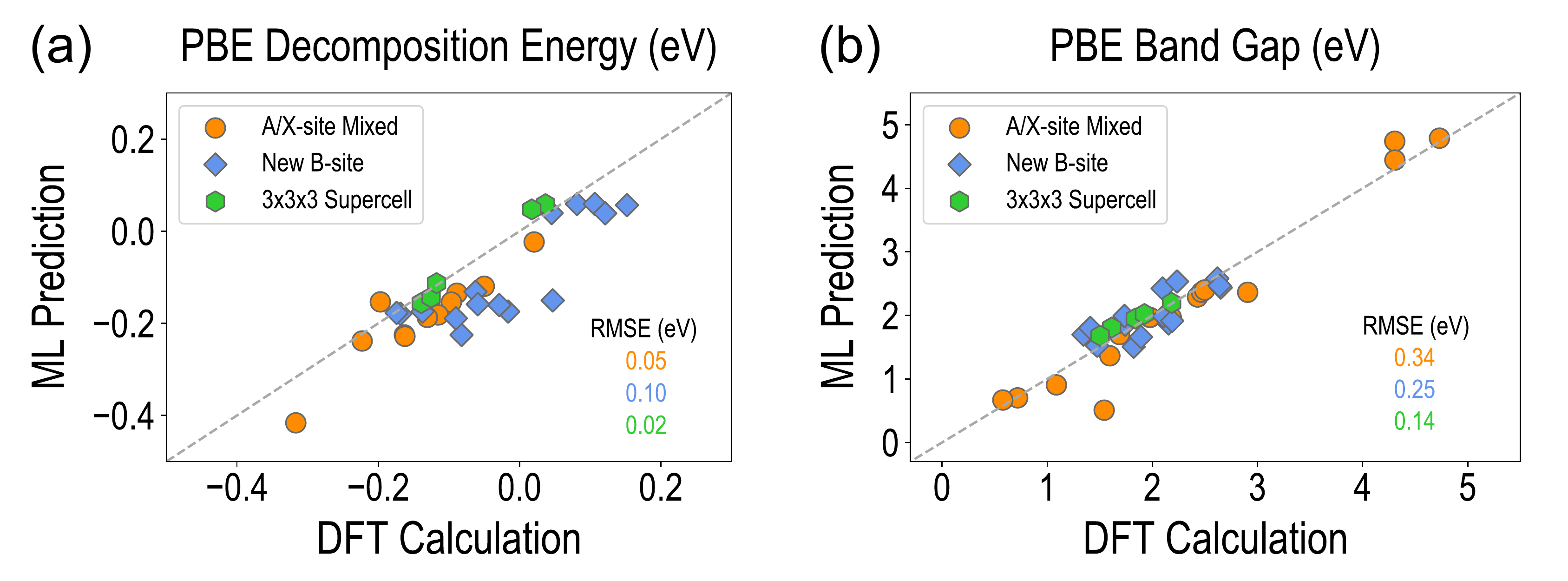}
\caption{\label{Fig:ML_outside_pred} 
ML (PBE) predicted decomposition energies and band gaps compared with DFT computed values for out-of-sample data points, namely new A-site and X-site mixed compounds (orange), B-site mixing with new elements (blue), and alloys simulated in a 3$\times$3$\times$3 MAPbX$_{3}$ supercell (green). RMSE values for the three sub-datasets are also shown.}
\end{figure}

Next, we looked at two binaries of interest, Pb-Sn and Pb-Ba, of the iodide, bromide and chloride perovskites of Cs, MA, and FA. Fig. \ref{Fig:ML_binaries} shows the ML predicted decomposition energies, mixing energies and band gaps of entire series of Pb-Sn and Pb-Ba compositions, at the HSE level. DFT ground truth values are also shown where available. As seen from Fig. \ref{Fig:ML_binaries}(a), $\Delta$H$_{decomp}$$^{HSE}$ shows a decreasing trend for Pb-Sn compounds for all A-X combinations with increasing Sn content, while $\Delta$H$_{mix}$$^{HSE}$ in Fig. \ref{Fig:ML_binaries}(b) shows a more irregular trend with both downward and upward facing curves. CsPb$_{1-x}$Sn$_{x}$Cl$_{3}$ and MAPb$_{1-x}$Sn$_{x}$I$_{3}$ show increased stability for intermediate compositions compared to the end points, whereas FAPb$_{1-x}$Sn$_{x}$Br$_{3}$ and FAPb$_{1-x}$Sn$_{x}$Cl$_{3}$ lead to intermediate compositions with reduced stability. Band gaps of all Pb-Sn binaries, plotted in Fig. \ref{Fig:ML_binaries}(c), show a decreasing trend with increasing Sn, with the change much more drastic in the all inorganic CsPb$_{1-x}$Sn$_{x}$X$_{3}$ compounds compared to their organic-inorganic counterparts. From Fig. \ref{Fig:ML_binaries}(f), it can be seen that the band gaps of Pb-Ba binaries show the opposite trend, increasing all the way from Pb to Ba with increasing Ba content, for all A-X combinations. $\Delta$H$_{decomp}$$^{HSE}$ of Pb-Ba binaries in Fig. \ref{Fig:ML_binaries}(d) show irregular trends while $\Delta$H$_{mix}$$^{HSE}$ in Fig. \ref{Fig:ML_binaries}(e) shows both downward and upward facing trends. CsPb$_{1-x}$Ba$_{x}$Cl$_{3}$, FAPb$_{1-x}$Ba$_{x}$I$_{3}$, MAPb$_{1-x}$Ba$_{x}$I$_{3}$, and MAPb$_{1-x}$Ba$_{x}$Cl$_{3}$ show more stable intermediate compositions than the respective end points while FAPb$_{1-x}$Ba$_{x}$Br$_{3}$ and FAPb$_{1-x}$Ba$_{x}$Cl$_{3}$ compositions show the opposite trend. ML predicted properties can be used to quickly produce such plots for any desired binaries, ternaries, and more. \\

\subsection*{Predictive Power and Future Work}

The data-driven design framework demonstrated in this work, powered by high-throughput DFT computations and neural networks, works quite well for a constrained problem where only certain atoms, compositions, phases, and levels of theory are at play. Since ML approaches being applied here are interpolative in nature, a natural question arises over how far the applicability of these predictive models could be stretched. Do they only work well for the set of $\sim$ 18,000 compounds selected here, or could they be used for other related compositions which may be considered \textit{out-of-sample} by some definitions? A few examples of such out-of-sample data points may include:

\begin{enumerate}

    \item Extensions to more A-site species such ethylammonium (EA) and dimethylammonium (DMA), and B-site species such as lanthanides (La, Eu), Cd, Zn, etc.
    
    \item Mixing fractions other than n/8 (where n is an integer between 0 and 8, necessitated by the consideration of 2$\times$2$\times$2 supercells for all computations).
    
    \item Mixing at A-site or X-site, as well as simultaneous mixing at two or three sites.
    
    \item Consideration of non-cubic perovskite phases, such as orthorhombic, tetragonal and hexagonal.

\end{enumerate}

In theory, using the suitable weighted averages of elemental/molecular properties as input, our predictive models can be applied to compositions such as MACa$_{0.06}$Ba$_{0.47}$Sn$_{0.22}$Pb$_{0.25}$I$_{3}$ and CsPb$_{0.33}$Sn$_{0.33}$Ge$_{0.33}$Br$_{3}$ which involve arbitrary cation-site mixing, as well as compounds with new types of atoms at different sites or mixing at A and X sites. To test the true predictive power of current models on out-of-sample points, we performed some additional calculations and collected some data from past work. Three types of new data points were included: (a) 15 compounds with random A-site and X-site mixing (considering the same set of A, B and X atoms as before), (b) 15 compounds with new B-site cations (Mg, Si, Cd, Zn, etc.) mixed in fractions of 1/8 with Pb at the B-site in MA(B)(X$_{3}$) compounds, collected from our past works \cite{MK1,MK2,MK8}, and (c) 5 new MAPbX$_{3}$ 3$\times$3$\times$3 supercell calculations with Sn/Ba/Sr mixing at the Pb-site. Fig. \ref{Fig:ML_outside_pred} shows the ML predicted decomposition energies and band gaps (at the PBE level) plotted against DFT results---showing excellent correspondence. The 35 compounds considered for testing out-of-sample predictive power are listed in Table \ref{table:SI_outside}. \\

The 3$\times$3$\times$3 supercell calculations, which represent new types of B-site mixing fractions, are very accurately predicted, with RMSE values of 0.02 eV and 0.14 eV for $\Delta$H$_{decomp}$$^{PBE}$ and E$_{gap}$$^{PBE}$, respectively. For both A/X-site mixing and new B-site systems, $\Delta$H$_{decomp}$$^{PBE}$ RMSE values are quite low at 0.05 eV and 0.10 eV respectively. Band gaps are slightly worse for A/X-site mixing points with an RMSE of 0.34 eV, while predictions for new B-site points lie at 0.25 eV. These results are remarkable, and already prove that for pseudo-cubic halide perovskites, ML models based on tabulated elemental or molecular properties are quite sufficient to universally predict the stability and band gaps of any possible composition, even if significantly different from the training dataset. We expect that inclusion of these out-of-sample points in the ML training step would further improve these predictions. Presently, we are extending the dataset to include many more A-site mixed, X-site mixed, and multi-site mixed points, compositions from mixing in larger supercells, as well as alternative perovskite phases. \\

Another limitation in this work is the lack of consideration of entropic contributions to the decomposition energies, which may stabilize previously unstable or metastable mixed compositions. While the detailed calculation of vibrational, electronic and configurational entropies from first principles thermodynamics is beyond the scope of this work, we can obtain rough estimates for the mixing entropy contributions in alloys \cite{Sampson} as T$\Delta$S = k$_{B}$T*$\sum$$_{i}$x$_{i}$*ln(x$_{i}$), where x$_{i}$ is the mixing fraction (between 0 and 1) of the i$^{th}$ element at the B-site, k$_{B}$ is the Boltzmann constant, and T is the growth temperature. Fig. \ref{Fig:SI_mix_entropy} shows this value in eV, calculated at 100$\degree$C for the entire dataset of 17,955 perovskites, plotted against their ML-predicted $\Delta$H$_{decomp}$$^{PBE}$. It can be seen that the mixing entropy contribution ranges between -0.06 eV and 0 eV, meaning there would only be a sight shift in the stability threshold (shown using dashed vertical lines in Fig. \ref{Fig:SI_mix_entropy}) if it is added to $\Delta$H$_{decomp}$. and very few compounds would be added that were eliminated during the screening test earlier. Future work will involve better estimates of entropy terms for best materials as well as for better screening. \\

A note should also be made about the inherent uncertainties of using density functional theory. Universal benchmarking of DFT functionals for different properties of interest in the halide perovskite chemical space is not a solved problem. Although the HSE06 functional has been used here to improve the estimates, as compared to PBE, of the structure, energetics, and certainly the electronic properties, further improvements may be necessary in terms of tuning the mixing parameter in HSE06 (fixed at 0.25 in this work), consideration of spin-orbit coupling for band gaps and defect formation energies, and the use of GW approximations for both electronic and optical properties \cite{GW}. Since most of these extensions involve fairly expensive computations, they can initially be performed on a restricted set of compounds---such as the screened set of 574 in this work---prior to being applied on a much larger dataset. This also opens the door to the use of multi-fidelity learning approaches, wherein large amounts of lower-accuracy data (e.g., PBE calculations) combined with more modest higher-accuracy estimates (e.g., HSE06 or GW calculations, or even experimental measurements) can lead to very accurate ML models for high-fidelity predictions \cite{MK3,MK4}.
    
Despite the many promising aspects of a DFT+ML screening process studied here, there are still significant hurdles towards experimental realization. There is a heavy imbalance between compounds studied from DFT and related approaches, and compounds that are actually synthesized, characterized and tested for their electrical, electronic, optical, and dielectric properties. Additional variables in the set of descriptors and output properties would be added in future work that account for the likelihood of synthesis of screened compounds and possible reaction pathways to make them happen. Current work also ignores the presence of metastable structures, or perovskite phases that constitute unexplored local minima in the energy landscape. The NN framework only takes the atoms and composition into account, and makes predictions for what is assumed to be the global minimum energy structure of the given halide perovskite composition. Future extensions will include computations on metastable phases and retraining of NN models to reflect that, explicitly studying how the free energy is influenced by polymorphism, configurational entropy, surface energies, octahedral rotation and distortion, and non-stoichiometric compounds.

\section*{Conclusions}

In summary, a data-driven framework was developed for the on-demand prediction and design of novel pseudo-cubic mixed-cation halide perovskites, using high-throughput DFT computations and rigorously optimized neural network models. DFT data was generated for multiple properties, including the lattice constant, formation/decomposition/mixing energy, band gap, refractive index, photovoltaic figure of merit based on optical absorption spectrum, and the formation energies and charge transition levels of vacancy defects, using two different levels of theory, for $\sim$ 1.3$\%$ of the entire possible set of compounds. NN regression models were trained on the DFT data, using the elemental or molecular properties of A, B and X species constituents as descriptors, and following standard machine learning practices, leading to predictions of all properties over the entire dataset. Screening of compounds with suitable stability, band gaps and defect tolerance reveals the predominance of MA and FA-based iodides, with B-site mixing involving larger fractions of Pb and Sn and smaller fractions of Ge, Ba, Sr, and Ca. Selected compounds were subjected to additional computations and certain perovskite binaries of interest were examined in terms of their stability and band gaps. The predictive models also showed impressive accuracy for out-of-sample points, which included A-site and X-site mixing, new types of B-site cations, and larger supercell calculations, indicating the robustness and wide applicability of the design framework. We highlighted the limitations of this work and the extensions being pursued for continuous improvement. The next few years will undoubtedly witness innovative applications of methods rooted in AI, ML, and data science on halide perovskite chemical spaces, and the accelerated design of novel structures, compositions, and synthesis pathways for next generation optoelectronics, power devices, and related applications. \\

\section*{Conflicts of interest}
There are no conflicts to declare.

\section*{Acknowledgements}
Extensive discussions with and scientific feedback from UC San Diego researchers David Fenning and Rishi Kumar are acknowledged. This work was performed, partly at the Center for Nanoscale Materials, a U.S. Department of Energy Office of Science User Facility, and supported by the U.S. Department of Energy, Office of Science, under Contract No. DE-AC02-06CH11357, and partly at Purdue University, under startup account F.10023800.05.002. This research used resources of the National Energy Research Scientific Computing Center, a DOE Office of Science User Facility supported by the Office of Science of the U.S. Department of Energy under Contract No. DE-AC02-05CH11231. We gratefully acknowledge the computing resources provided on Bebop, a high-performance computing cluster operated by the Laboratory Computing Resource Center at Argonne National Laboratory.

\balance

\bibliography{rsc} 

\providecommand*{\mcitethebibliography}{\thebibliography}
\csname @ifundefined\endcsname{endmcitethebibliography}
{\let\endmcitethebibliography\endthebibliography}{}
\begin{mcitethebibliography}{117}
\providecommand*{\natexlab}[1]{#1}
\providecommand*{\mciteSetBstSublistMode}[1]{}
\providecommand*{\mciteSetBstMaxWidthForm}[2]{}
\providecommand*{\mciteBstWouldAddEndPuncttrue}
  {\def\EndOfBibitem{\unskip.}}
\providecommand*{\mciteBstWouldAddEndPunctfalse}
  {\let\EndOfBibitem\relax}
\providecommand*{\mciteSetBstMidEndSepPunct}[3]{}
\providecommand*{\mciteSetBstSublistLabelBeginEnd}[3]{}
\providecommand*{\EndOfBibitem}{}
\mciteSetBstSublistMode{f}
\mciteSetBstMaxWidthForm{subitem}
{(\emph{\alph{mcitesubitemcount}})}
\mciteSetBstSublistLabelBeginEnd{\mcitemaxwidthsubitemform\space}
{\relax}{\relax}

\bibitem[Saeki and Kranthiraja(2019)]{ML2}
A.~Saeki and K.~Kranthiraja, \emph{Japanese Journal of Applied Physics}, 2019,
  \textbf{59}, SD0801\relax
\mciteBstWouldAddEndPuncttrue
\mciteSetBstMidEndSepPunct{\mcitedefaultmidpunct}
{\mcitedefaultendpunct}{\mcitedefaultseppunct}\relax
\EndOfBibitem
\bibitem[Takahashi \emph{et~al.}(2020)Takahashi, Kumagai, Miyamoto, Mochizuki,
  and Oba]{ML3}
A.~Takahashi, Y.~Kumagai, J.~Miyamoto, Y.~Mochizuki and F.~Oba, \emph{Phys.
  Rev. Materials}, 2020, \textbf{4}, 103801\relax
\mciteBstWouldAddEndPuncttrue
\mciteSetBstMidEndSepPunct{\mcitedefaultmidpunct}
{\mcitedefaultendpunct}{\mcitedefaultseppunct}\relax
\EndOfBibitem
\bibitem[Mannodi-Kanakkithodi \emph{et~al.}(2018)Mannodi-Kanakkithodi,
  Chandrasekaran, Kim, Huan, Pilania, Botu, and Ramprasad]{MK6}
A.~Mannodi-Kanakkithodi, A.~Chandrasekaran, C.~Kim, T.~D. Huan, G.~Pilania,
  V.~Botu and R.~Ramprasad, \emph{Materials Today}, 2018, \textbf{21},
  785--796\relax
\mciteBstWouldAddEndPuncttrue
\mciteSetBstMidEndSepPunct{\mcitedefaultmidpunct}
{\mcitedefaultendpunct}{\mcitedefaultseppunct}\relax
\EndOfBibitem
\bibitem[Schlexer-Lamoureux \emph{et~al.}(2019)Schlexer-Lamoureux, Winther,
  Garrido-Torres, Streibel, Zhao, Bajdich, Abild-Pedersen, and Bligaard]{ML1}
P.~Schlexer-Lamoureux, K.~T. Winther, J.~A. Garrido-Torres, V.~Streibel,
  M.~Zhao, M.~Bajdich, F.~Abild-Pedersen and T.~Bligaard, \emph{ChemCatChem},
  2019, \textbf{11}, 3581--3601\relax
\mciteBstWouldAddEndPuncttrue
\mciteSetBstMidEndSepPunct{\mcitedefaultmidpunct}
{\mcitedefaultendpunct}{\mcitedefaultseppunct}\relax
\EndOfBibitem
\bibitem[Munshi \emph{et~al.}(2021)Munshi, Chen, Chien, and
  Balasubramanian]{ML5}
J.~Munshi, W.~Chen, T.~Chien and G.~Balasubramanian, \emph{Journal of Chemical
  Information and Modeling}, 2021, \textbf{61}, 134--142\relax
\mciteBstWouldAddEndPuncttrue
\mciteSetBstMidEndSepPunct{\mcitedefaultmidpunct}
{\mcitedefaultendpunct}{\mcitedefaultseppunct}\relax
\EndOfBibitem
\bibitem[Feng \emph{et~al.}(2020)Feng, Wu, and Deng]{ML6}
H.-J. Feng, K.~Wu and Z.-Y. Deng, \emph{Cell Reports Physical Science}, 2020,
  \textbf{1}, 100179\relax
\mciteBstWouldAddEndPuncttrue
\mciteSetBstMidEndSepPunct{\mcitedefaultmidpunct}
{\mcitedefaultendpunct}{\mcitedefaultseppunct}\relax
\EndOfBibitem
\bibitem[Schmidt \emph{et~al.}(2019)Schmidt, Marques, Botti, and Marques]{ML4}
J.~Schmidt, M.~R.~G. Marques, S.~Botti and M.~A.~L. Marques, \emph{npj
  Computational Materials}, 2019, \textbf{5}, 83\relax
\mciteBstWouldAddEndPuncttrue
\mciteSetBstMidEndSepPunct{\mcitedefaultmidpunct}
{\mcitedefaultendpunct}{\mcitedefaultseppunct}\relax
\EndOfBibitem
\bibitem[Ramprasad \emph{et~al.}(2017)Ramprasad, Batra, Pilania,
  Mannodi-Kanakkithodi, and Kim]{MK5}
R.~Ramprasad, R.~Batra, G.~Pilania, A.~Mannodi-Kanakkithodi and C.~Kim,
  \emph{npj Computational Materials}, 2017, \textbf{3}, 54\relax
\mciteBstWouldAddEndPuncttrue
\mciteSetBstMidEndSepPunct{\mcitedefaultmidpunct}
{\mcitedefaultendpunct}{\mcitedefaultseppunct}\relax
\EndOfBibitem
\bibitem[Choubisa \emph{et~al.}(2020)Choubisa, Askerka, Ryczko, Voznyy, Mills,
  Tamblyn, and Sargent]{MLHP1}
H.~Choubisa, M.~Askerka, K.~Ryczko, O.~Voznyy, K.~Mills, I.~Tamblyn and E.~H.
  Sargent, \emph{Matter}, 2020, \textbf{3}, 433--448\relax
\mciteBstWouldAddEndPuncttrue
\mciteSetBstMidEndSepPunct{\mcitedefaultmidpunct}
{\mcitedefaultendpunct}{\mcitedefaultseppunct}\relax
\EndOfBibitem
\bibitem[Wu and Wang(2020)]{MLHP2}
T.~Wu and J.~Wang, \emph{ACS Applied Materials and Interfaces}, 2020,
  \textbf{12}, 57821--57831\relax
\mciteBstWouldAddEndPuncttrue
\mciteSetBstMidEndSepPunct{\mcitedefaultmidpunct}
{\mcitedefaultendpunct}{\mcitedefaultseppunct}\relax
\EndOfBibitem
\bibitem[Herbol \emph{et~al.}(2018)Herbol, Hu, Frazier, Clancy, and
  Poloczek]{MLHP3}
H.~C. Herbol, W.~Hu, P.~Frazier, P.~Clancy and M.~Poloczek, \emph{npj
  Computational Materials}, 2018, \textbf{4}, 51\relax
\mciteBstWouldAddEndPuncttrue
\mciteSetBstMidEndSepPunct{\mcitedefaultmidpunct}
{\mcitedefaultendpunct}{\mcitedefaultseppunct}\relax
\EndOfBibitem
\bibitem[Lu \emph{et~al.}(2018)Lu, Zhou, Ouyang, Guo, Li, and Wang]{MLHP4}
S.~Lu, Q.~Zhou, Y.~Ouyang, Y.~Guo, Q.~Li and J.~Wang, \emph{Nature
  Communications}, 2018, \textbf{9}, 3405\relax
\mciteBstWouldAddEndPuncttrue
\mciteSetBstMidEndSepPunct{\mcitedefaultmidpunct}
{\mcitedefaultendpunct}{\mcitedefaultseppunct}\relax
\EndOfBibitem
\bibitem[Pu \emph{et~al.}(2021)Pu, Xiao, Wang, Li, and Wang]{MLHP5}
W.~Pu, W.~Xiao, J.~Wang, X.~Li and L.~Wang, \emph{Materials and Design}, 2021,
  \textbf{198}, 109387\relax
\mciteBstWouldAddEndPuncttrue
\mciteSetBstMidEndSepPunct{\mcitedefaultmidpunct}
{\mcitedefaultendpunct}{\mcitedefaultseppunct}\relax
\EndOfBibitem
\bibitem[Nakajima and Sawada(2017)]{MLHP6}
T.~Nakajima and K.~Sawada, \emph{The Journal of Physical Chemistry Letters},
  2017, \textbf{8}, 4826--4831\relax
\mciteBstWouldAddEndPuncttrue
\mciteSetBstMidEndSepPunct{\mcitedefaultmidpunct}
{\mcitedefaultendpunct}{\mcitedefaultseppunct}\relax
\EndOfBibitem
\bibitem[Mannodi-Kanakkithodi \emph{et~al.}(2019)Mannodi-Kanakkithodi, Park,
  Jeon, Cao, Gosztola, Martinson, and Chan]{MK1}
A.~Mannodi-Kanakkithodi, J.-S. Park, N.~Jeon, D.~H. Cao, D.~J. Gosztola,
  A.~B.~F. Martinson and M.~K.~Y. Chan, \emph{Chemistry of Materials}, 2019,
  \textbf{31}, 3599--3612\relax
\mciteBstWouldAddEndPuncttrue
\mciteSetBstMidEndSepPunct{\mcitedefaultmidpunct}
{\mcitedefaultendpunct}{\mcitedefaultseppunct}\relax
\EndOfBibitem
\bibitem[Mannodi-Kanakkithodi \emph{et~al.}(2020)Mannodi-Kanakkithodi, Park,
  Martinson, and Chan]{MK2}
A.~Mannodi-Kanakkithodi, J.-S. Park, A.~B.~F. Martinson and M.~K.~Y. Chan,
  \emph{The Journal of Physical Chemistry C}, 2020, \textbf{124},
  16729--16738\relax
\mciteBstWouldAddEndPuncttrue
\mciteSetBstMidEndSepPunct{\mcitedefaultmidpunct}
{\mcitedefaultendpunct}{\mcitedefaultseppunct}\relax
\EndOfBibitem
\bibitem[Gladkikh \emph{et~al.}(2020)Gladkikh, Kim, Hajibabaei, Jana, Myung,
  and Kim]{MLHP7}
V.~Gladkikh, D.~Y. Kim, A.~Hajibabaei, A.~Jana, C.~W. Myung and K.~S. Kim,
  \emph{The Journal of Physical Chemistry C}, 2020, \textbf{124},
  8905--8918\relax
\mciteBstWouldAddEndPuncttrue
\mciteSetBstMidEndSepPunct{\mcitedefaultmidpunct}
{\mcitedefaultendpunct}{\mcitedefaultseppunct}\relax
\EndOfBibitem
\bibitem[Saidi \emph{et~al.}(2020)Saidi, Shadid, and Castelli]{MLHP8}
W.~A. Saidi, W.~Shadid and I.~E. Castelli, \emph{npj Computational Materials},
  2020, \textbf{6}, 36\relax
\mciteBstWouldAddEndPuncttrue
\mciteSetBstMidEndSepPunct{\mcitedefaultmidpunct}
{\mcitedefaultendpunct}{\mcitedefaultseppunct}\relax
\EndOfBibitem
\bibitem[Sun \emph{et~al.}(2019)Sun, Hartono, Ren, Oviedo, Buscemi, Layurova,
  Chen, Ogunfunmi, Thapa, Ramasamy, Settens, DeCost, Kusne, Liu, Tian, Peters,
  Correa-Baena, and Buonassisi]{MLHP9}
S.~Sun, N.~T. Hartono, Z.~D. Ren, F.~Oviedo, A.~M. Buscemi, M.~Layurova, D.~X.
  Chen, T.~Ogunfunmi, J.~Thapa, S.~Ramasamy, C.~Settens, B.~L. DeCost, A.~G.
  Kusne, Z.~Liu, S.~I. Tian, I.~M. Peters, J.-P. Correa-Baena and
  T.~Buonassisi, \emph{Joule}, 2019, \textbf{3}, 1437--1451\relax
\mciteBstWouldAddEndPuncttrue
\mciteSetBstMidEndSepPunct{\mcitedefaultmidpunct}
{\mcitedefaultendpunct}{\mcitedefaultseppunct}\relax
\EndOfBibitem
\bibitem[Chen \emph{et~al.}(2020)Chen, Wang, Li, Hou, and Yin]{MLHP10}
X.~Chen, C.~Wang, Z.~Li, Z.~Hou and W.-J. Yin, \emph{Science China Materials},
  2020, \textbf{63}, 1024--1035\relax
\mciteBstWouldAddEndPuncttrue
\mciteSetBstMidEndSepPunct{\mcitedefaultmidpunct}
{\mcitedefaultendpunct}{\mcitedefaultseppunct}\relax
\EndOfBibitem
\bibitem[Oba and Kumagai(2018)]{Semi}
F.~Oba and Y.~Kumagai, \emph{Applied Physics Express}, 2018, \textbf{11},
  060101\relax
\mciteBstWouldAddEndPuncttrue
\mciteSetBstMidEndSepPunct{\mcitedefaultmidpunct}
{\mcitedefaultendpunct}{\mcitedefaultseppunct}\relax
\EndOfBibitem
\bibitem[Muhammad \emph{et~al.}(2020)Muhammad, Liu, Ahmad, Jalali~Asadabadi,
  Franchini, and Ahmad]{C1}
Z.~Muhammad, P.~Liu, R.~Ahmad, S.~Jalali~Asadabadi, C.~Franchini and I.~Ahmad,
  \emph{Phys. Chem. Chem. Phys.}, 2020, \textbf{22}, 11943--11955\relax
\mciteBstWouldAddEndPuncttrue
\mciteSetBstMidEndSepPunct{\mcitedefaultmidpunct}
{\mcitedefaultendpunct}{\mcitedefaultseppunct}\relax
\EndOfBibitem
\bibitem[Kim \emph{et~al.}(2020)Kim, Eom, Ha, Hong, and Kim]{C2}
S.~Kim, T.~Eom, Y.-S. Ha, K.-H. Hong and H.~Kim, \emph{Chemistry of Materials},
  2020, \textbf{32}, 4265--4272\relax
\mciteBstWouldAddEndPuncttrue
\mciteSetBstMidEndSepPunct{\mcitedefaultmidpunct}
{\mcitedefaultendpunct}{\mcitedefaultseppunct}\relax
\EndOfBibitem
\bibitem[Boukhvalov \emph{et~al.}(2020)Boukhvalov, Zhidkov, Akbulatov,
  Kukharenko, Cholakh, Stevenson, Troshin, and Kurmaev]{C3}
D.~W. Boukhvalov, I.~S. Zhidkov, A.~F. Akbulatov, A.~I. Kukharenko, S.~O.
  Cholakh, K.~J. Stevenson, P.~A. Troshin and E.~Z. Kurmaev, \emph{The Journal
  of Physical Chemistry A}, 2020, \textbf{124}, 135--140\relax
\mciteBstWouldAddEndPuncttrue
\mciteSetBstMidEndSepPunct{\mcitedefaultmidpunct}
{\mcitedefaultendpunct}{\mcitedefaultseppunct}\relax
\EndOfBibitem
\bibitem[Schelhas \emph{et~al.}(2019)Schelhas, Li, Christians, Goyal, Kairys,
  Harvey, Kim, Stone, Luther, Zhu, Stevanovic, and Berry]{C4}
L.~T. Schelhas, Z.~Li, J.~A. Christians, A.~Goyal, P.~Kairys, S.~P. Harvey,
  D.~H. Kim, K.~H. Stone, J.~M. Luther, K.~Zhu, V.~Stevanovic and J.~J. Berry,
  \emph{Energy Environ. Sci.}, 2019, \textbf{12}, 1341--1348\relax
\mciteBstWouldAddEndPuncttrue
\mciteSetBstMidEndSepPunct{\mcitedefaultmidpunct}
{\mcitedefaultendpunct}{\mcitedefaultseppunct}\relax
\EndOfBibitem
\bibitem[Zhu \emph{et~al.}(2019)Zhu, Ye, Zhao, and Qiu]{C5}
S.~Zhu, J.~Ye, Y.~Zhao and Y.~Qiu, \emph{The Journal of Physical Chemistry C},
  2019, \textbf{123}, 20476--20487\relax
\mciteBstWouldAddEndPuncttrue
\mciteSetBstMidEndSepPunct{\mcitedefaultmidpunct}
{\mcitedefaultendpunct}{\mcitedefaultseppunct}\relax
\EndOfBibitem
\bibitem[C6(2020)]{C6}
\emph{Spectrochimica Acta Part A: Molecular and Biomolecular Spectroscopy},
  2020, \textbf{239}, 118493\relax
\mciteBstWouldAddEndPuncttrue
\mciteSetBstMidEndSepPunct{\mcitedefaultmidpunct}
{\mcitedefaultendpunct}{\mcitedefaultseppunct}\relax
\EndOfBibitem
\bibitem[Basera \emph{et~al.}(2020)Basera, Kumar, Saini, and Bhattacharya]{C7}
P.~Basera, M.~Kumar, S.~Saini and S.~Bhattacharya, \emph{Phys. Rev. B}, 2020,
  \textbf{101}, 054108\relax
\mciteBstWouldAddEndPuncttrue
\mciteSetBstMidEndSepPunct{\mcitedefaultmidpunct}
{\mcitedefaultendpunct}{\mcitedefaultseppunct}\relax
\EndOfBibitem
\bibitem[Banerjee \emph{et~al.}(2019)Banerjee, Chakraborty, and Ahuja]{C8}
A.~Banerjee, S.~Chakraborty and R.~Ahuja, \emph{ACS Applied Energy Materials},
  2019, \textbf{2}, 6990--6997\relax
\mciteBstWouldAddEndPuncttrue
\mciteSetBstMidEndSepPunct{\mcitedefaultmidpunct}
{\mcitedefaultendpunct}{\mcitedefaultseppunct}\relax
\EndOfBibitem
\bibitem[Hossain \emph{et~al.}(2020)Hossain, Khanom, Israt, Hossain, Hossain,
  and Ahmed]{C9}
K.~Hossain, S.~Khanom, F.~Israt, M.~Hossain, M.~Hossain and F.~Ahmed,
  \emph{Solid State Communications}, 2020, \textbf{320}, 114024\relax
\mciteBstWouldAddEndPuncttrue
\mciteSetBstMidEndSepPunct{\mcitedefaultmidpunct}
{\mcitedefaultendpunct}{\mcitedefaultseppunct}\relax
\EndOfBibitem
\bibitem[Bechtel and Van~der Ven(2018)]{C10}
J.~S. Bechtel and A.~Van~der Ven, \emph{Phys. Rev. Materials}, 2018,
  \textbf{2}, 045401\relax
\mciteBstWouldAddEndPuncttrue
\mciteSetBstMidEndSepPunct{\mcitedefaultmidpunct}
{\mcitedefaultendpunct}{\mcitedefaultseppunct}\relax
\EndOfBibitem
\bibitem[Rajeswarapalanichamy \emph{et~al.}(2020)Rajeswarapalanichamy,
  Amudhavalli, Padmavathy, and Iyakutti]{C11}
R.~Rajeswarapalanichamy, A.~Amudhavalli, R.~Padmavathy and K.~Iyakutti,
  \emph{Materials Science and Engineering: B}, 2020, \textbf{258}, 114560\relax
\mciteBstWouldAddEndPuncttrue
\mciteSetBstMidEndSepPunct{\mcitedefaultmidpunct}
{\mcitedefaultendpunct}{\mcitedefaultseppunct}\relax
\EndOfBibitem
\bibitem[Ornelas-Cruz \emph{et~al.}(2020)Ornelas-Cruz, Trejo, Oviedo-Roa,
  Salazar, Carvajal, Miranda, and Cruz-Irisson]{C12}
I.~Ornelas-Cruz, A.~Trejo, R.~Oviedo-Roa, F.~Salazar, E.~Carvajal, A.~Miranda
  and M.~Cruz-Irisson, \emph{Computational Materials Science}, 2020,
  \textbf{178}, 109619\relax
\mciteBstWouldAddEndPuncttrue
\mciteSetBstMidEndSepPunct{\mcitedefaultmidpunct}
{\mcitedefaultendpunct}{\mcitedefaultseppunct}\relax
\EndOfBibitem
\bibitem[Boziki \emph{et~al.}(2020)Boziki, Kubicki, Mishra, Meloni, Emsley,
  Grätzel, and Rothlisberger]{C13}
A.~Boziki, D.~J. Kubicki, A.~Mishra, S.~Meloni, L.~Emsley, M.~Grätzel and
  U.~Rothlisberger, \emph{Chemistry of Materials}, 2020, \textbf{32},
  2605--2614\relax
\mciteBstWouldAddEndPuncttrue
\mciteSetBstMidEndSepPunct{\mcitedefaultmidpunct}
{\mcitedefaultendpunct}{\mcitedefaultseppunct}\relax
\EndOfBibitem
\bibitem[Guan \emph{et~al.}(2020)Guan, Xu, Liang, Han, Guo, Wang, and Li]{C14}
L.~Guan, X.~Xu, Y.~Liang, S.~Han, J.~Guo, J.~Wang and X.~Li, \emph{Physics
  Letters A}, 2020, \textbf{384}, 126173\relax
\mciteBstWouldAddEndPuncttrue
\mciteSetBstMidEndSepPunct{\mcitedefaultmidpunct}
{\mcitedefaultendpunct}{\mcitedefaultseppunct}\relax
\EndOfBibitem
\bibitem[Shakil \emph{et~al.}(2020)Shakil, Akram, Zeba, Ahmad, Gillani, and
  Gadhi]{C15}
M.~Shakil, A.~Akram, I.~Zeba, R.~Ahmad, S.~S.~A. Gillani and M.~A. Gadhi,
  \emph{Materials Research Express}, 2020, \textbf{7}, 025513\relax
\mciteBstWouldAddEndPuncttrue
\mciteSetBstMidEndSepPunct{\mcitedefaultmidpunct}
{\mcitedefaultendpunct}{\mcitedefaultseppunct}\relax
\EndOfBibitem
\bibitem[Liu \emph{et~al.}(2020)Liu, Li, Zeng, and Meng]{C16}
H.~Liu, X.~Li, Y.~Zeng and L.~Meng, \emph{Computational Materials Science},
  2020, \textbf{177}, 109576\relax
\mciteBstWouldAddEndPuncttrue
\mciteSetBstMidEndSepPunct{\mcitedefaultmidpunct}
{\mcitedefaultendpunct}{\mcitedefaultseppunct}\relax
\EndOfBibitem
\bibitem[Chang \emph{et~al.}(2019)Chang, Chen, Wang, Wang, Chen, and Yuan]{C17}
J.~Chang, H.~Chen, G.~Wang, B.~Wang, X.~Chen and H.~Yuan, \emph{RSC Adv.},
  2019, \textbf{9}, 7015--7024\relax
\mciteBstWouldAddEndPuncttrue
\mciteSetBstMidEndSepPunct{\mcitedefaultmidpunct}
{\mcitedefaultendpunct}{\mcitedefaultseppunct}\relax
\EndOfBibitem
\bibitem[Tang \emph{et~al.}(2017)Tang, Xu, Zhang, Hu, Lau, and Liu]{C18}
Z.-K. Tang, Z.-F. Xu, D.-Y. Zhang, S.-X. Hu, W.-M. Lau and L.-M. Liu,
  \emph{Scientific Reports}, 2017, \textbf{7}, 7843\relax
\mciteBstWouldAddEndPuncttrue
\mciteSetBstMidEndSepPunct{\mcitedefaultmidpunct}
{\mcitedefaultendpunct}{\mcitedefaultseppunct}\relax
\EndOfBibitem
\bibitem[Dalpian \emph{et~al.}(2019)Dalpian, Zhao, Kazmerski, and Zunger]{C19}
G.~M. Dalpian, X.-G. Zhao, L.~Kazmerski and A.~Zunger, \emph{Chemistry of
  Materials}, 2019, \textbf{31}, 2497--2506\relax
\mciteBstWouldAddEndPuncttrue
\mciteSetBstMidEndSepPunct{\mcitedefaultmidpunct}
{\mcitedefaultendpunct}{\mcitedefaultseppunct}\relax
\EndOfBibitem
\bibitem[Hao \emph{et~al.}(2014)Hao, Stoumpos, Chang, and Kanatzidis]{E1}
F.~Hao, C.~C. Stoumpos, R.~P.~H. Chang and M.~G. Kanatzidis, \emph{Journal of
  the American Chemical Society}, 2014, \textbf{136}, 8094--8099\relax
\mciteBstWouldAddEndPuncttrue
\mciteSetBstMidEndSepPunct{\mcitedefaultmidpunct}
{\mcitedefaultendpunct}{\mcitedefaultseppunct}\relax
\EndOfBibitem
\bibitem[Wu \emph{et~al.}(2019)Wu, Kuo, Jhuang, Chen, Lai, and Chen]{E2}
M.-J. Wu, C.-C. Kuo, L.-S. Jhuang, P.-H. Chen, Y.-F. Lai and F.-C. Chen,
  \emph{Advanced Energy Materials}, 2019, \textbf{9}, 1901863\relax
\mciteBstWouldAddEndPuncttrue
\mciteSetBstMidEndSepPunct{\mcitedefaultmidpunct}
{\mcitedefaultendpunct}{\mcitedefaultseppunct}\relax
\EndOfBibitem
\bibitem[Khatun \emph{et~al.}(2020)Khatun, Maiti, and Pal]{E3}
S.~Khatun, A.~Maiti and A.~J. Pal, \emph{Applied Physics Letters}, 2020,
  \textbf{116}, 012104\relax
\mciteBstWouldAddEndPuncttrue
\mciteSetBstMidEndSepPunct{\mcitedefaultmidpunct}
{\mcitedefaultendpunct}{\mcitedefaultseppunct}\relax
\EndOfBibitem
\bibitem[Ding \emph{et~al.}(2019)Ding, Du, Zhou, Yuan, Cheng, Jing, Yao, Zhang,
  He, Cui, Zhan, and Sun]{E4}
J.~Ding, S.~Du, T.~Zhou, Y.~Yuan, X.~Cheng, L.~Jing, Q.~Yao, J.~Zhang, Q.~He,
  H.~Cui, X.~Zhan and H.~Sun, \emph{The Journal of Physical Chemistry C}, 2019,
  \textbf{123}, 14969--14975\relax
\mciteBstWouldAddEndPuncttrue
\mciteSetBstMidEndSepPunct{\mcitedefaultmidpunct}
{\mcitedefaultendpunct}{\mcitedefaultseppunct}\relax
\EndOfBibitem
\bibitem[Mozur \emph{et~al.}(2020)Mozur, Hope, Trowbridge, Halat, Daemen,
  Maughan, Prisk, Grey, and Neilson]{E5}
E.~M. Mozur, M.~A. Hope, J.~C. Trowbridge, D.~M. Halat, L.~L. Daemen, A.~E.
  Maughan, T.~R. Prisk, C.~P. Grey and J.~R. Neilson, \emph{Chemistry of
  Materials}, 2020, \textbf{32}, 6266--6277\relax
\mciteBstWouldAddEndPuncttrue
\mciteSetBstMidEndSepPunct{\mcitedefaultmidpunct}
{\mcitedefaultendpunct}{\mcitedefaultseppunct}\relax
\EndOfBibitem
\bibitem[Ogomi \emph{et~al.}(2014)Ogomi, Morita, Tsukamoto, Saitho, Fujikawa,
  Shen, Toyoda, Yoshino, Pandey, Ma, and Hayase]{E6}
Y.~Ogomi, A.~Morita, S.~Tsukamoto, T.~Saitho, N.~Fujikawa, Q.~Shen, T.~Toyoda,
  K.~Yoshino, S.~S. Pandey, T.~Ma and S.~Hayase, \emph{The Journal of Physical
  Chemistry Letters}, 2014, \textbf{5}, 1004--1011\relax
\mciteBstWouldAddEndPuncttrue
\mciteSetBstMidEndSepPunct{\mcitedefaultmidpunct}
{\mcitedefaultendpunct}{\mcitedefaultseppunct}\relax
\EndOfBibitem
\bibitem[Wang \emph{et~al.}(2019)Wang, Zhang, Wang, Li, Meng, You, Yin, and
  Wu]{E7}
Y.~Wang, X.~Zhang, D.~Wang, X.~Li, J.~Meng, J.~You, Z.~Yin and J.~Wu, \emph{ACS
  Applied Materials and Interfaces}, 2019, \textbf{11}, 28005--28012\relax
\mciteBstWouldAddEndPuncttrue
\mciteSetBstMidEndSepPunct{\mcitedefaultmidpunct}
{\mcitedefaultendpunct}{\mcitedefaultseppunct}\relax
\EndOfBibitem
\bibitem[Greenland \emph{et~al.}(2020)Greenland, Shnier, Rajendran, Smith,
  Game, Wamwangi, Turnbull, Samuel, Billing, and Lidzey]{E8}
C.~Greenland, A.~Shnier, S.~K. Rajendran, J.~A. Smith, O.~S. Game, D.~Wamwangi,
  G.~A. Turnbull, I.~D.~W. Samuel, D.~G. Billing and D.~G. Lidzey,
  \emph{Advanced Energy Materials}, 2020, \textbf{10}, 1901350\relax
\mciteBstWouldAddEndPuncttrue
\mciteSetBstMidEndSepPunct{\mcitedefaultmidpunct}
{\mcitedefaultendpunct}{\mcitedefaultseppunct}\relax
\EndOfBibitem
\bibitem[Jia \emph{et~al.}(2019)Jia, Zuo, Tao, Sun, Zhao, Yang, Cheng, Wang,
  Yuan, Yang, Gao, Xing, Wei, Zhang, Yip, Liu, Shen, Yin, Han, Liu, Wang, Luo,
  Tan, Jin, and Ding]{E9}
X.~Jia, C.~Zuo, S.~Tao, K.~Sun, Y.~Zhao, S.~Yang, M.~Cheng, M.~Wang, Y.~Yuan,
  J.~Yang, F.~Gao, G.~Xing, Z.~Wei, L.~Zhang, H.-L. Yip, M.~Liu, Q.~Shen,
  L.~Yin, L.~Han, S.~Liu, L.~Wang, J.~Luo, H.~Tan, Z.~Jin and L.~Ding,
  \emph{Science Bulletin}, 2019, \textbf{64}, 1532 -- 1539\relax
\mciteBstWouldAddEndPuncttrue
\mciteSetBstMidEndSepPunct{\mcitedefaultmidpunct}
{\mcitedefaultendpunct}{\mcitedefaultseppunct}\relax
\EndOfBibitem
\bibitem[Prasanna \emph{et~al.}(2019)Prasanna, Leijtens, Dunfield, Raiford,
  Wolf, Swifter, Werner, Eperon, de~Paula, Palmstrom, Boyd, van Hest, Bent,
  Teeter, Berry, and McGehee]{E10}
R.~Prasanna, T.~Leijtens, S.~P. Dunfield, J.~A. Raiford, E.~J. Wolf, S.~A.
  Swifter, J.~Werner, G.~E. Eperon, C.~de~Paula, A.~F. Palmstrom, C.~C. Boyd,
  M.~F. A.~M. van Hest, S.~F. Bent, G.~Teeter, J.~J. Berry and M.~D. McGehee,
  \emph{Nature Energy}, 2019, \textbf{4}, 939 -- 947\relax
\mciteBstWouldAddEndPuncttrue
\mciteSetBstMidEndSepPunct{\mcitedefaultmidpunct}
{\mcitedefaultendpunct}{\mcitedefaultseppunct}\relax
\EndOfBibitem
\bibitem[Serrano-Sánchez \emph{et~al.}(2020)Serrano-Sánchez, Conesa,
  Rodrigues, Marini, Martínez, and Alonso]{E11}
F.~Serrano-Sánchez, J.~Conesa, J.~Rodrigues, C.~Marini, J.~Martínez and
  J.~Alonso, \emph{Journal of Alloys and Compounds}, 2020, \textbf{821},
  153414\relax
\mciteBstWouldAddEndPuncttrue
\mciteSetBstMidEndSepPunct{\mcitedefaultmidpunct}
{\mcitedefaultendpunct}{\mcitedefaultseppunct}\relax
\EndOfBibitem
\bibitem[Yin \emph{et~al.}(2020)Yin, Fu, Zhou, Song, Li, Zhang, Wang,
  Mariyappan, Alshehri, Ahamad, and Yamauchi]{E12}
Y.~Yin, S.~Fu, S.~Zhou, Y.~Song, L.~Li, M.~Zhang, J.~Wang, P.~Mariyappan, S.~M.
  Alshehri, T.~Ahamad and Y.~Yamauchi, \emph{Electronic Materials Letters},
  2020, \textbf{16}, 224--230\relax
\mciteBstWouldAddEndPuncttrue
\mciteSetBstMidEndSepPunct{\mcitedefaultmidpunct}
{\mcitedefaultendpunct}{\mcitedefaultseppunct}\relax
\EndOfBibitem
\bibitem[Zhang \emph{et~al.}(2020)Zhang, Li, Zhang, and Guo]{E13}
H.~Zhang, R.~Li, M.~Zhang and M.~Guo, \emph{Ceramics International}, 2020,
  \textbf{46}, 14038 -- 14047\relax
\mciteBstWouldAddEndPuncttrue
\mciteSetBstMidEndSepPunct{\mcitedefaultmidpunct}
{\mcitedefaultendpunct}{\mcitedefaultseppunct}\relax
\EndOfBibitem
\bibitem[Rybin \emph{et~al.}(2020)Rybin, Ghosh, Tisdale, Shrestha, Yoho, Vo,
  Even, Katan, Nie, Neukirch, and Tretiak]{E14}
N.~Rybin, D.~Ghosh, J.~Tisdale, S.~Shrestha, M.~Yoho, D.~Vo, J.~Even, C.~Katan,
  W.~Nie, A.~J. Neukirch and S.~Tretiak, \emph{Chemistry of Materials}, 2020,
  \textbf{32}, 1854--1863\relax
\mciteBstWouldAddEndPuncttrue
\mciteSetBstMidEndSepPunct{\mcitedefaultmidpunct}
{\mcitedefaultendpunct}{\mcitedefaultseppunct}\relax
\EndOfBibitem
\bibitem[Subedi \emph{et~al.}(2020)Subedi, Li, Junda, Song, Yan, and
  Podraza]{E15}
B.~Subedi, C.~Li, M.~M. Junda, Z.~Song, Y.~Yan and N.~J. Podraza, \emph{The
  Journal of Chemical Physics}, 2020, \textbf{152}, 064705\relax
\mciteBstWouldAddEndPuncttrue
\mciteSetBstMidEndSepPunct{\mcitedefaultmidpunct}
{\mcitedefaultendpunct}{\mcitedefaultseppunct}\relax
\EndOfBibitem
\bibitem[Xiao \emph{et~al.}(2020)Xiao, Wang, Wu, Wu, Ren, Xiong, and Yang]{E16}
Z.~Xiao, Q.~Wang, X.~Wu, Y.~Wu, J.~Ren, Z.~Xiong and X.~Yang, \emph{Organic
  Electronics}, 2020, \textbf{77}, 105546\relax
\mciteBstWouldAddEndPuncttrue
\mciteSetBstMidEndSepPunct{\mcitedefaultmidpunct}
{\mcitedefaultendpunct}{\mcitedefaultseppunct}\relax
\EndOfBibitem
\bibitem[Liashenko \emph{et~al.}(2019)Liashenko, Cherotchenko, Pushkarev,
  Pakštas, Naujokaitis, Khubezhov, Polozkov, Agapev, Zakhidov, Shelykh, and
  Makarov]{E17}
T.~G. Liashenko, E.~D. Cherotchenko, A.~P. Pushkarev, V.~Pakštas,
  A.~Naujokaitis, S.~A. Khubezhov, R.~G. Polozkov, K.~B. Agapev, A.~A.
  Zakhidov, I.~A. Shelykh and S.~V. Makarov, \emph{Phys. Chem. Chem. Phys.},
  2019, \textbf{21}, 18930--18938\relax
\mciteBstWouldAddEndPuncttrue
\mciteSetBstMidEndSepPunct{\mcitedefaultmidpunct}
{\mcitedefaultendpunct}{\mcitedefaultseppunct}\relax
\EndOfBibitem
\bibitem[Alam \emph{et~al.}(2019)Alam, Wegner, Pouget, Amidani, Kvashnina,
  Aldakov, and Reiss]{E18}
F.~Alam, K.~D. Wegner, S.~Pouget, L.~Amidani, K.~Kvashnina, D.~Aldakov and
  P.~Reiss, \emph{The Journal of Chemical Physics}, 2019, \textbf{151},
  231101\relax
\mciteBstWouldAddEndPuncttrue
\mciteSetBstMidEndSepPunct{\mcitedefaultmidpunct}
{\mcitedefaultendpunct}{\mcitedefaultseppunct}\relax
\EndOfBibitem
\bibitem[Nagane \emph{et~al.}(2018)Nagane, Ghosh, Hoye, Zhao, Ahmad, Walker,
  Islam, Ogale, and Sadhanala]{E19}
S.~Nagane, D.~Ghosh, R.~L.~Z. Hoye, B.~Zhao, S.~Ahmad, A.~B. Walker, M.~S.
  Islam, S.~Ogale and A.~Sadhanala, \emph{The Journal of Physical Chemistry C},
  2018, \textbf{122}, 5940--5947\relax
\mciteBstWouldAddEndPuncttrue
\mciteSetBstMidEndSepPunct{\mcitedefaultmidpunct}
{\mcitedefaultendpunct}{\mcitedefaultseppunct}\relax
\EndOfBibitem
\bibitem[Szostak \emph{et~al.}(2019)Szostak, Silva, Turren-Cruz, Soares,
  Freitas, Hagfeldt, Tolentino, and Nogueira]{E20}
R.~Szostak, J.~C. Silva, S.-H. Turren-Cruz, M.~M. Soares, R.~O. Freitas,
  A.~Hagfeldt, H.~C.~N. Tolentino and A.~F. Nogueira, \emph{Science Advances},
  2019, \textbf{5}, eaaw6619\relax
\mciteBstWouldAddEndPuncttrue
\mciteSetBstMidEndSepPunct{\mcitedefaultmidpunct}
{\mcitedefaultendpunct}{\mcitedefaultseppunct}\relax
\EndOfBibitem
\bibitem[Charles \emph{et~al.}(2020)Charles, Weller, Rieger, Hatcher, Henry,
  Feldmann, Wolverson, and Wilson]{E21}
B.~Charles, M.~T. Weller, S.~Rieger, L.~E. Hatcher, P.~F. Henry, J.~Feldmann,
  D.~Wolverson and C.~C. Wilson, \emph{Chemistry of Materials}, 2020,
  \textbf{32}, 2282--2291\relax
\mciteBstWouldAddEndPuncttrue
\mciteSetBstMidEndSepPunct{\mcitedefaultmidpunct}
{\mcitedefaultendpunct}{\mcitedefaultseppunct}\relax
\EndOfBibitem
\bibitem[Handa \emph{et~al.}(2019)Handa, Wakamiya, and Kanemitsu]{E22}
T.~Handa, A.~Wakamiya and Y.~Kanemitsu, \emph{APL Materials}, 2019, \textbf{7},
  080903\relax
\mciteBstWouldAddEndPuncttrue
\mciteSetBstMidEndSepPunct{\mcitedefaultmidpunct}
{\mcitedefaultendpunct}{\mcitedefaultseppunct}\relax
\EndOfBibitem
\bibitem[Franssen \emph{et~al.}(2020)Franssen, van Heumen, and Kentgens]{E23}
W.~M.~J. Franssen, C.~M.~M. van Heumen and A.~P.~M. Kentgens, \emph{Inorganic
  Chemistry}, 2020, \textbf{59}, 3730--3739\relax
\mciteBstWouldAddEndPuncttrue
\mciteSetBstMidEndSepPunct{\mcitedefaultmidpunct}
{\mcitedefaultendpunct}{\mcitedefaultseppunct}\relax
\EndOfBibitem
\bibitem[Beal \emph{et~al.}(2020)Beal, Hagström, Barrier, Gold-Parker,
  Prasanna, Bush, Passarello, Schelhas, Brüning, Tassone, Steinrück, McGehee,
  Toney, and Nogueira]{E24}
R.~E. Beal, N.~Z. Hagström, J.~Barrier, A.~Gold-Parker, R.~Prasanna, K.~A.
  Bush, D.~Passarello, L.~T. Schelhas, K.~Brüning, C.~J. Tassone, H.-G.
  Steinrück, M.~D. McGehee, M.~F. Toney and A.~F. Nogueira, \emph{Matter},
  2020, \textbf{2}, 207 -- 219\relax
\mciteBstWouldAddEndPuncttrue
\mciteSetBstMidEndSepPunct{\mcitedefaultmidpunct}
{\mcitedefaultendpunct}{\mcitedefaultseppunct}\relax
\EndOfBibitem
\bibitem[Chang \emph{et~al.}(2020)Chang, Tseng, Chiang, Wu, Chen, Chen, Yuan,
  and Chen]{E25}
S.~H. Chang, P.-C. Tseng, S.-E. Chiang, J.-R. Wu, Y.-T. Chen, C.-J. Chen, C.-T.
  Yuan and S.-H. Chen, \emph{Solar Energy Materials and Solar Cells}, 2020,
  \textbf{210}, 110478\relax
\mciteBstWouldAddEndPuncttrue
\mciteSetBstMidEndSepPunct{\mcitedefaultmidpunct}
{\mcitedefaultendpunct}{\mcitedefaultseppunct}\relax
\EndOfBibitem
\bibitem[Kim \emph{et~al.}(2019)Kim, Lee, Nam, Yun, Lee, Kim, Noh, Lee, Kim,
  Lee, and Heo]{E26}
S.-Y. Kim, H.-C. Lee, Y.~Nam, Y.~Yun, S.-H. Lee, D.~H. Kim, J.~H. Noh, J.-H.
  Lee, D.-H. Kim, S.~Lee and Y.-W. Heo, \emph{Acta Materialia}, 2019,
  \textbf{181}, 460 -- 469\relax
\mciteBstWouldAddEndPuncttrue
\mciteSetBstMidEndSepPunct{\mcitedefaultmidpunct}
{\mcitedefaultendpunct}{\mcitedefaultseppunct}\relax
\EndOfBibitem
\bibitem[Yang \emph{et~al.}(2020)Yang, Zhang, Yang, Eperon, and Ginger]{E27}
Z.~Yang, X.~Zhang, W.~Yang, G.~E. Eperon and D.~S. Ginger, \emph{Chemistry of
  Materials}, 2020, \textbf{32}, 2782--2794\relax
\mciteBstWouldAddEndPuncttrue
\mciteSetBstMidEndSepPunct{\mcitedefaultmidpunct}
{\mcitedefaultendpunct}{\mcitedefaultseppunct}\relax
\EndOfBibitem
\bibitem[Pham \emph{et~al.}(2019)Pham, Duong, Rickard, Kremer, Weber, and
  Wong-Leung]{E28}
H.~T. Pham, T.~Duong, W.~D.~A. Rickard, F.~Kremer, K.~J. Weber and
  J.~Wong-Leung, \emph{The Journal of Physical Chemistry C}, 2019,
  \textbf{123}, 26718--26726\relax
\mciteBstWouldAddEndPuncttrue
\mciteSetBstMidEndSepPunct{\mcitedefaultmidpunct}
{\mcitedefaultendpunct}{\mcitedefaultseppunct}\relax
\EndOfBibitem
\bibitem[Zhang \emph{et~al.}(2014)Zhang, Yu, Lyu, Wang, Yun, and Wang]{E29}
M.~Zhang, H.~Yu, M.~Lyu, Q.~Wang, J.-H. Yun and L.~Wang, \emph{Chem. Commun.},
  2014, \textbf{50}, 11727--11730\relax
\mciteBstWouldAddEndPuncttrue
\mciteSetBstMidEndSepPunct{\mcitedefaultmidpunct}
{\mcitedefaultendpunct}{\mcitedefaultseppunct}\relax
\EndOfBibitem
\bibitem[McMeekin \emph{et~al.}(2016)McMeekin, Sadoughi, Rehman, Eperon,
  Saliba, H{\"o}rantner, Haghighirad, Sakai, Korte, Rech, Johnston, Herz, and
  Snaith]{E30}
D.~P. McMeekin, G.~Sadoughi, W.~Rehman, G.~E. Eperon, M.~Saliba, M.~T.
  H{\"o}rantner, A.~Haghighirad, N.~Sakai, L.~Korte, B.~Rech, M.~B. Johnston,
  L.~M. Herz and H.~J. Snaith, \emph{Science}, 2016, \textbf{351},
  151--155\relax
\mciteBstWouldAddEndPuncttrue
\mciteSetBstMidEndSepPunct{\mcitedefaultmidpunct}
{\mcitedefaultendpunct}{\mcitedefaultseppunct}\relax
\EndOfBibitem
\bibitem[Yang \emph{et~al.}(2016)Yang, Rajagopal, Jo, Chueh, Williams, Huang,
  Katahara, Hillhouse, and Jen]{E31}
Z.~Yang, A.~Rajagopal, S.~B. Jo, C.-C. Chueh, S.~Williams, C.-C. Huang, J.~K.
  Katahara, H.~W. Hillhouse and A.~K.-Y. Jen, \emph{Nano Letters}, 2016,
  \textbf{16}, 7739--7747\relax
\mciteBstWouldAddEndPuncttrue
\mciteSetBstMidEndSepPunct{\mcitedefaultmidpunct}
{\mcitedefaultendpunct}{\mcitedefaultseppunct}\relax
\EndOfBibitem
\bibitem[Basera \emph{et~al.}(2020)Basera, Kumar, Saini, and
  Bhattacharya]{Comput_mixed_1}
P.~Basera, M.~Kumar, S.~Saini and S.~Bhattacharya, \emph{Phys. Rev. B}, 2020,
  \textbf{101}, 054108\relax
\mciteBstWouldAddEndPuncttrue
\mciteSetBstMidEndSepPunct{\mcitedefaultmidpunct}
{\mcitedefaultendpunct}{\mcitedefaultseppunct}\relax
\EndOfBibitem
\bibitem[Takahashi \emph{et~al.}(2011)Takahashi, Obara, Lin, Takahashi, Naito,
  Inabe, Ishibashi, and Terakura]{Sn_Pb_perovs_1}
Y.~Takahashi, R.~Obara, Z.-Z. Lin, Y.~Takahashi, T.~Naito, T.~Inabe,
  S.~Ishibashi and K.~Terakura, \emph{Dalton Trans.}, 2011, \textbf{40},
  5563--5568\relax
\mciteBstWouldAddEndPuncttrue
\mciteSetBstMidEndSepPunct{\mcitedefaultmidpunct}
{\mcitedefaultendpunct}{\mcitedefaultseppunct}\relax
\EndOfBibitem
\bibitem[Stranks and Snaith(2015)]{Sn_Pb_perovs_2}
S.~D. Stranks and H.~J. Snaith, \emph{Nature Nanotechnology}, 2015,
  \textbf{10}, 391--402\relax
\mciteBstWouldAddEndPuncttrue
\mciteSetBstMidEndSepPunct{\mcitedefaultmidpunct}
{\mcitedefaultendpunct}{\mcitedefaultseppunct}\relax
\EndOfBibitem
\bibitem[Prasanna \emph{et~al.}(2019)Prasanna, Leijtens, Dunfield, Raiford,
  Wolf, Swifter, Werner, Eperon, de~Paula, Palmstrom, Boyd, van Hest, Bent,
  Teeter, Berry, and McGehee]{Sn_Pb_perovs_3}
R.~Prasanna, T.~Leijtens, S.~P. Dunfield, J.~A. Raiford, E.~J. Wolf, S.~A.
  Swifter, J.~Werner, G.~E. Eperon, C.~de~Paula, A.~F. Palmstrom, C.~C. Boyd,
  M.~F. A.~M. van Hest, S.~F. Bent, G.~Teeter, J.~J. Berry and M.~D. McGehee,
  \emph{Nature Energy}, 2019, \textbf{4}, 939--947\relax
\mciteBstWouldAddEndPuncttrue
\mciteSetBstMidEndSepPunct{\mcitedefaultmidpunct}
{\mcitedefaultendpunct}{\mcitedefaultseppunct}\relax
\EndOfBibitem
\bibitem[Hao \emph{et~al.}(2014)Hao, Stoumpos, Chang, and
  Kanatzidis]{Sn_Pb_perovs_4}
F.~Hao, C.~C. Stoumpos, R.~P.~H. Chang and M.~G. Kanatzidis, \emph{Journal of
  the American Chemical Society}, 2014, \textbf{136}, 8094--8099\relax
\mciteBstWouldAddEndPuncttrue
\mciteSetBstMidEndSepPunct{\mcitedefaultmidpunct}
{\mcitedefaultendpunct}{\mcitedefaultseppunct}\relax
\EndOfBibitem
\bibitem[Yin \emph{et~al.}(2020)Yin, Fu, Zhou, Song, Li, Zhang, Wang,
  Mariyappan, Alshehri, Ahamad, and Yamauchi]{Sn_Pb_perovs_5}
Y.~Yin, S.~Fu, S.~Zhou, Y.~Song, L.~Li, M.~Zhang, J.~Wang, P.~Mariyappan, S.~M.
  Alshehri, T.~Ahamad and Y.~Yamauchi, \emph{Electronic Materials Letters},
  2020, \textbf{16}, 224--230\relax
\mciteBstWouldAddEndPuncttrue
\mciteSetBstMidEndSepPunct{\mcitedefaultmidpunct}
{\mcitedefaultendpunct}{\mcitedefaultseppunct}\relax
\EndOfBibitem
\bibitem[Handa \emph{et~al.}(2019)Handa, Wakamiya, and
  Kanemitsu]{Sn_Pb_perovs_6}
T.~Handa, A.~Wakamiya and Y.~Kanemitsu, \emph{APL Materials}, 2019, \textbf{7},
  080903\relax
\mciteBstWouldAddEndPuncttrue
\mciteSetBstMidEndSepPunct{\mcitedefaultmidpunct}
{\mcitedefaultendpunct}{\mcitedefaultseppunct}\relax
\EndOfBibitem
\bibitem[Yang \emph{et~al.}(2020)Yang, Zhang, Yang, Eperon, and
  Ginger]{Sn_Pb_perovs_7}
Z.~Yang, X.~Zhang, W.~Yang, G.~E. Eperon and D.~S. Ginger, \emph{Chemistry of
  Materials}, 2020, \textbf{32}, 2782--2794\relax
\mciteBstWouldAddEndPuncttrue
\mciteSetBstMidEndSepPunct{\mcitedefaultmidpunct}
{\mcitedefaultendpunct}{\mcitedefaultseppunct}\relax
\EndOfBibitem
\bibitem[Mannodi-Kanakkithodi and Chan(2021)]{MK4}
A.~Mannodi-Kanakkithodi and M.~K. Chan, \emph{Trends in Chemistry}, 2021,
  \textbf{3}, 79--82\relax
\mciteBstWouldAddEndPuncttrue
\mciteSetBstMidEndSepPunct{\mcitedefaultmidpunct}
{\mcitedefaultendpunct}{\mcitedefaultseppunct}\relax
\EndOfBibitem
\bibitem[Chan and Ceder(2010)]{ChanBandGap}
M.~Chan and G.~Ceder, \emph{Physical review letters}, 2010, \textbf{105},
  196403\relax
\mciteBstWouldAddEndPuncttrue
\mciteSetBstMidEndSepPunct{\mcitedefaultmidpunct}
{\mcitedefaultendpunct}{\mcitedefaultseppunct}\relax
\EndOfBibitem
\bibitem[Burke and Wagner(2013)]{GGA}
K.~Burke and L.~O. Wagner, \emph{International Journal of Quantum Chemistry},
  2013, \textbf{113}, 96--101\relax
\mciteBstWouldAddEndPuncttrue
\mciteSetBstMidEndSepPunct{\mcitedefaultmidpunct}
{\mcitedefaultendpunct}{\mcitedefaultseppunct}\relax
\EndOfBibitem
\bibitem[Perdew \emph{et~al.}(1996)Perdew, Burke, and Ernzerhof]{PBE}
J.~P. Perdew, K.~Burke and M.~Ernzerhof, \emph{Phys. Rev. Lett.}, 1996,
  \textbf{77}, 3865--3868\relax
\mciteBstWouldAddEndPuncttrue
\mciteSetBstMidEndSepPunct{\mcitedefaultmidpunct}
{\mcitedefaultendpunct}{\mcitedefaultseppunct}\relax
\EndOfBibitem
\bibitem[Park \emph{et~al.}(2018)Park, Kim, Xie, and Walsh]{SOC1}
J.~S. Park, S.~Kim, Z.~Xie and A.~Walsh, \emph{Nature Reviews Materials}, 2018,
  \textbf{3}, 194--210\relax
\mciteBstWouldAddEndPuncttrue
\mciteSetBstMidEndSepPunct{\mcitedefaultmidpunct}
{\mcitedefaultendpunct}{\mcitedefaultseppunct}\relax
\EndOfBibitem
\bibitem[Ganose \emph{et~al.}(2018)Ganose, Matsumoto, Buckeridge, and
  Scanlon]{SOC2}
A.~M. Ganose, S.~Matsumoto, J.~Buckeridge and D.~O. Scanlon, \emph{Chemistry of
  Materials}, 2018, \textbf{30}, 3827--3835\relax
\mciteBstWouldAddEndPuncttrue
\mciteSetBstMidEndSepPunct{\mcitedefaultmidpunct}
{\mcitedefaultendpunct}{\mcitedefaultseppunct}\relax
\EndOfBibitem
\bibitem[Amat \emph{et~al.}(2014)Amat, Mosconi, Ronca, Quarti, Umari,
  Nazeeruddin, Grätzel, and De~Angelis]{SOC3}
A.~Amat, E.~Mosconi, E.~Ronca, C.~Quarti, P.~Umari, M.~K. Nazeeruddin,
  M.~Grätzel and F.~De~Angelis, \emph{Nano Letters}, 2014, \textbf{14},
  3608--3616\relax
\mciteBstWouldAddEndPuncttrue
\mciteSetBstMidEndSepPunct{\mcitedefaultmidpunct}
{\mcitedefaultendpunct}{\mcitedefaultseppunct}\relax
\EndOfBibitem
\bibitem[Shi \emph{et~al.}(2015)Shi, Yin, Hong, Zhu, and Yan]{Yanfa}
T.~Shi, W.-J. Yin, F.~Hong, K.~Zhu and Y.~Yan, \emph{Applied Physics Letters},
  2015, \textbf{106}, 103902\relax
\mciteBstWouldAddEndPuncttrue
\mciteSetBstMidEndSepPunct{\mcitedefaultmidpunct}
{\mcitedefaultendpunct}{\mcitedefaultseppunct}\relax
\EndOfBibitem
\bibitem[Wei \emph{et~al.}(1990)Wei, Ferreira, Bernard, and Zunger]{SQS1}
S.-H. Wei, L.~G. Ferreira, J.~E. Bernard and A.~Zunger, \emph{Phys. Rev. B},
  1990, \textbf{42}, 9622--9649\relax
\mciteBstWouldAddEndPuncttrue
\mciteSetBstMidEndSepPunct{\mcitedefaultmidpunct}
{\mcitedefaultendpunct}{\mcitedefaultseppunct}\relax
\EndOfBibitem
\bibitem[Jiang \emph{et~al.}(2016)Jiang, Nahas, Xu, Prosandeev, Wang, and
  Bellaiche]{SQS2}
Z.~Jiang, Y.~Nahas, B.~Xu, S.~Prosandeev, D.~Wang and L.~Bellaiche,
  \emph{Journal of Physics: Condensed Matter}, 2016, \textbf{28}, 475901\relax
\mciteBstWouldAddEndPuncttrue
\mciteSetBstMidEndSepPunct{\mcitedefaultmidpunct}
{\mcitedefaultendpunct}{\mcitedefaultseppunct}\relax
\EndOfBibitem
\bibitem[Kresse and Hafner(1994)]{vasp1}
G.~Kresse and J.~Hafner, \emph{Phys. Rev. B}, 1994, \textbf{49},
  14251--14269\relax
\mciteBstWouldAddEndPuncttrue
\mciteSetBstMidEndSepPunct{\mcitedefaultmidpunct}
{\mcitedefaultendpunct}{\mcitedefaultseppunct}\relax
\EndOfBibitem
\bibitem[Kresse and Furthm\"uller(1996)]{vasp2}
G.~Kresse and J.~Furthm\"uller, \emph{Phys. Rev. B}, 1996, \textbf{54},
  11169--11186\relax
\mciteBstWouldAddEndPuncttrue
\mciteSetBstMidEndSepPunct{\mcitedefaultmidpunct}
{\mcitedefaultendpunct}{\mcitedefaultseppunct}\relax
\EndOfBibitem
\bibitem[Bl\"ochl(1994)]{PAW}
P.~E. Bl\"ochl, \emph{Phys. Rev. B}, 1994, \textbf{50}, 17953--17979\relax
\mciteBstWouldAddEndPuncttrue
\mciteSetBstMidEndSepPunct{\mcitedefaultmidpunct}
{\mcitedefaultendpunct}{\mcitedefaultseppunct}\relax
\EndOfBibitem
\bibitem[Gueymard(2004)]{AM1.5}
C.~A. Gueymard, \emph{Solar Energy}, 2004, \textbf{76}, 423--453\relax
\mciteBstWouldAddEndPuncttrue
\mciteSetBstMidEndSepPunct{\mcitedefaultmidpunct}
{\mcitedefaultendpunct}{\mcitedefaultseppunct}\relax
\EndOfBibitem
\bibitem[Jain \emph{et~al.}(2013)Jain, Ong, Hautier, Chen, Richards, Dacek,
  Cholia, Gunter, Skinner, Ceder, and Persson]{MP}
A.~Jain, S.~P. Ong, G.~Hautier, W.~Chen, W.~D. Richards, S.~Dacek, S.~Cholia,
  D.~Gunter, D.~Skinner, G.~Ceder and K.~A. Persson, \emph{APL Materials},
  2013, \textbf{1}, 011002\relax
\mciteBstWouldAddEndPuncttrue
\mciteSetBstMidEndSepPunct{\mcitedefaultmidpunct}
{\mcitedefaultendpunct}{\mcitedefaultseppunct}\relax
\EndOfBibitem
\bibitem[Wen \emph{et~al.}(2021)Wen, Blau, Spotte-Smith, Dwaraknath, and
  Persson]{NN1}
M.~Wen, S.~M. Blau, E.~W.~C. Spotte-Smith, S.~Dwaraknath and K.~A. Persson,
  \emph{Chem. Sci.}, 2021, \textbf{12}, 1858--1868\relax
\mciteBstWouldAddEndPuncttrue
\mciteSetBstMidEndSepPunct{\mcitedefaultmidpunct}
{\mcitedefaultendpunct}{\mcitedefaultseppunct}\relax
\EndOfBibitem
\bibitem[Noh \emph{et~al.}(2019)Noh, Kim, Stein, Sanchez-Lengeling, Gregoire,
  Aspuru-Guzik, and Jung]{NN2}
J.~Noh, J.~Kim, H.~S. Stein, B.~Sanchez-Lengeling, J.~M. Gregoire,
  A.~Aspuru-Guzik and Y.~Jung, \emph{Matter}, 2019, \textbf{1},
  1370--1384\relax
\mciteBstWouldAddEndPuncttrue
\mciteSetBstMidEndSepPunct{\mcitedefaultmidpunct}
{\mcitedefaultendpunct}{\mcitedefaultseppunct}\relax
\EndOfBibitem
\bibitem[Xie and Grossman(2018)]{NN3}
T.~Xie and J.~C. Grossman, \emph{Phys. Rev. Lett.}, 2018, \textbf{120},
  145301\relax
\mciteBstWouldAddEndPuncttrue
\mciteSetBstMidEndSepPunct{\mcitedefaultmidpunct}
{\mcitedefaultendpunct}{\mcitedefaultseppunct}\relax
\EndOfBibitem
\bibitem[Scott \emph{et~al.}(2007)Scott, Coveney, Kilner, Rossiny, and
  Alford]{NN4}
D.~Scott, P.~Coveney, J.~Kilner, J.~Rossiny and N.~N. Alford, \emph{Journal of
  the European Ceramic Society}, 2007, \textbf{27}, 4425--4435\relax
\mciteBstWouldAddEndPuncttrue
\mciteSetBstMidEndSepPunct{\mcitedefaultmidpunct}
{\mcitedefaultendpunct}{\mcitedefaultseppunct}\relax
\EndOfBibitem
\bibitem[Rajan(2015)]{NN5}
K.~Rajan, \emph{Annual Review of Materials Research}, 2015, \textbf{45},
  153--169\relax
\mciteBstWouldAddEndPuncttrue
\mciteSetBstMidEndSepPunct{\mcitedefaultmidpunct}
{\mcitedefaultendpunct}{\mcitedefaultseppunct}\relax
\EndOfBibitem
\bibitem[Tsai \emph{et~al.}(2020)Tsai, Hsia, Yang, Liu, and Fang]{NN_hp1}
C.-W. Tsai, C.-H. Hsia, S.-J. Yang, S.-J. Liu and Z.-Y. Fang, \emph{Applied
  Soft Computing}, 2020, \textbf{88}, 106068\relax
\mciteBstWouldAddEndPuncttrue
\mciteSetBstMidEndSepPunct{\mcitedefaultmidpunct}
{\mcitedefaultendpunct}{\mcitedefaultseppunct}\relax
\EndOfBibitem
\bibitem[{Cho} \emph{et~al.}(2020){Cho}, {Kim}, {Lee}, {Choi}, {Lee}, and
  {Rhee}]{NN_hp2}
H.~{Cho}, Y.~{Kim}, E.~{Lee}, D.~{Choi}, Y.~{Lee} and W.~{Rhee}, \emph{IEEE
  Access}, 2020, \textbf{8}, 52588--52608\relax
\mciteBstWouldAddEndPuncttrue
\mciteSetBstMidEndSepPunct{\mcitedefaultmidpunct}
{\mcitedefaultendpunct}{\mcitedefaultseppunct}\relax
\EndOfBibitem
\bibitem[Freysoldt \emph{et~al.}(2009)Freysoldt, Neugebauer, and Van~de
  Walle]{Corr1}
C.~Freysoldt, J.~Neugebauer and C.~G. Van~de Walle, \emph{Phys. Rev. Lett.},
  2009, \textbf{102}, 016402\relax
\mciteBstWouldAddEndPuncttrue
\mciteSetBstMidEndSepPunct{\mcitedefaultmidpunct}
{\mcitedefaultendpunct}{\mcitedefaultseppunct}\relax
\EndOfBibitem
\bibitem[Freysoldt \emph{et~al.}(2014)Freysoldt, Grabowski, Hickel, Neugebauer,
  Kresse, Janotti, and Van~de Walle]{Corr2}
C.~Freysoldt, B.~Grabowski, T.~Hickel, J.~Neugebauer, G.~Kresse, A.~Janotti and
  C.~G. Van~de Walle, \emph{Rev. Mod. Phys.}, 2014, \textbf{86}, 253--305\relax
\mciteBstWouldAddEndPuncttrue
\mciteSetBstMidEndSepPunct{\mcitedefaultmidpunct}
{\mcitedefaultendpunct}{\mcitedefaultseppunct}\relax
\EndOfBibitem
\bibitem[Mannodi-Kanakkithodi \emph{et~al.}(2020)Mannodi-Kanakkithodi,
  Toriyama, Sen, Davis, Klie, and Chan]{MK3}
A.~Mannodi-Kanakkithodi, M.~Y. Toriyama, F.~G. Sen, M.~J. Davis, R.~F. Klie and
  M.~K. Chan, \emph{npj Computational Materials}, 2020, \textbf{6}, 39\relax
\mciteBstWouldAddEndPuncttrue
\mciteSetBstMidEndSepPunct{\mcitedefaultmidpunct}
{\mcitedefaultendpunct}{\mcitedefaultseppunct}\relax
\EndOfBibitem
\bibitem[Mannodi-Kanakkithodi \emph{et~al.}(2021)Mannodi-Kanakkithodi, Xiang,
  Jacoby, Biegaj, Dunham, Gamelin, and Chan]{MK7}
A.~Mannodi-Kanakkithodi, X.~Xiang, L.~Jacoby, R.~Biegaj, S.~Dunham, D.~Gamelin
  and M.~Chan, \emph{Research Square}, 2021,  preprint doi:
  https://doi.org/10.21203/rs.3.rs--723035/v1\relax
\mciteBstWouldAddEndPuncttrue
\mciteSetBstMidEndSepPunct{\mcitedefaultmidpunct}
{\mcitedefaultendpunct}{\mcitedefaultseppunct}\relax
\EndOfBibitem
\bibitem[Sampson \emph{et~al.}(2017)Sampson, Park, Schaller, Chan, and
  Martinson]{Sampson}
M.~D. Sampson, J.~S. Park, R.~D. Schaller, M.~K.~Y. Chan and A.~B.~F.
  Martinson, \emph{J. Mater. Chem. A}, 2017, \textbf{5}, 3578--3588\relax
\mciteBstWouldAddEndPuncttrue
\mciteSetBstMidEndSepPunct{\mcitedefaultmidpunct}
{\mcitedefaultendpunct}{\mcitedefaultseppunct}\relax
\EndOfBibitem
\bibitem[Bartel \emph{et~al.}(2019)Bartel, Sutton, Goldsmith, Ouyang, Musgrave,
  Ghiringhelli, and Scheffler]{Tol1}
C.~J. Bartel, C.~Sutton, B.~R. Goldsmith, R.~Ouyang, C.~B. Musgrave, L.~M.
  Ghiringhelli and M.~Scheffler, \emph{Science Advances}, 2019, \textbf{5},
  eaav0693\relax
\mciteBstWouldAddEndPuncttrue
\mciteSetBstMidEndSepPunct{\mcitedefaultmidpunct}
{\mcitedefaultendpunct}{\mcitedefaultseppunct}\relax
\EndOfBibitem
\bibitem[Travis \emph{et~al.}(2016)Travis, Glover, Bronstein, Scanlon, and
  Palgrave]{Tol2}
W.~Travis, E.~N.~K. Glover, H.~Bronstein, D.~O. Scanlon and R.~G. Palgrave,
  \emph{Chem. Sci.}, 2016, \textbf{7}, 4548--4556\relax
\mciteBstWouldAddEndPuncttrue
\mciteSetBstMidEndSepPunct{\mcitedefaultmidpunct}
{\mcitedefaultendpunct}{\mcitedefaultseppunct}\relax
\EndOfBibitem
\bibitem[Li \emph{et~al.}(2016)Li, Yang, Park, Wei, Berry, and Zhu]{Tol3}
Z.~Li, M.~Yang, J.-S. Park, S.-H. Wei, J.~J. Berry and K.~Zhu, \emph{Chemistry
  of Materials}, 2016, \textbf{28}, 284--292\relax
\mciteBstWouldAddEndPuncttrue
\mciteSetBstMidEndSepPunct{\mcitedefaultmidpunct}
{\mcitedefaultendpunct}{\mcitedefaultseppunct}\relax
\EndOfBibitem
\bibitem[Peterson and Brgoch(2021)]{ML_form1}
G.~G.~C. Peterson and J.~Brgoch, \emph{Journal of Physics: Energy}, 2021,
  \textbf{3}, 022002\relax
\mciteBstWouldAddEndPuncttrue
\mciteSetBstMidEndSepPunct{\mcitedefaultmidpunct}
{\mcitedefaultendpunct}{\mcitedefaultseppunct}\relax
\EndOfBibitem
\bibitem[Mao \emph{et~al.}(2021)Mao, Yang, Sheng, Wang, Ouyang, Ye, Yang, and
  Zhang]{ML_form2}
Y.~Mao, H.~Yang, Y.~Sheng, J.~Wang, R.~Ouyang, C.~Ye, J.~Yang and W.~Zhang,
  \emph{ACS Omega}, 2021, \textbf{6}, 14533--14541\relax
\mciteBstWouldAddEndPuncttrue
\mciteSetBstMidEndSepPunct{\mcitedefaultmidpunct}
{\mcitedefaultendpunct}{\mcitedefaultseppunct}\relax
\EndOfBibitem
\bibitem[Chen \emph{et~al.}(2019)Chen, Ye, Zuo, Zheng, and Ong]{ML_form3}
C.~Chen, W.~Ye, Y.~Zuo, C.~Zheng and S.~P. Ong, \emph{Chemistry of Materials},
  2019, \textbf{31}, 3564--3572\relax
\mciteBstWouldAddEndPuncttrue
\mciteSetBstMidEndSepPunct{\mcitedefaultmidpunct}
{\mcitedefaultendpunct}{\mcitedefaultseppunct}\relax
\EndOfBibitem
\bibitem[Pilania \emph{et~al.}(2016)Pilania, Mannodi-Kanakkithodi, Uberuaga,
  Ramprasad, Gubernatis, and Lookman]{Gaps_ML1}
G.~Pilania, A.~Mannodi-Kanakkithodi, B.~P. Uberuaga, R.~Ramprasad, J.~E.
  Gubernatis and T.~Lookman, \emph{Scientific Reports}, 2016, \textbf{6},
  19375\relax
\mciteBstWouldAddEndPuncttrue
\mciteSetBstMidEndSepPunct{\mcitedefaultmidpunct}
{\mcitedefaultendpunct}{\mcitedefaultseppunct}\relax
\EndOfBibitem
\bibitem[Zhuo \emph{et~al.}(2018)Zhuo, Mansouri~Tehrani, and Brgoch]{Gaps_ML2}
Y.~Zhuo, A.~Mansouri~Tehrani and J.~Brgoch, \emph{The Journal of Physical
  Chemistry Letters}, 2018, \textbf{9}, 1668--1673\relax
\mciteBstWouldAddEndPuncttrue
\mciteSetBstMidEndSepPunct{\mcitedefaultmidpunct}
{\mcitedefaultendpunct}{\mcitedefaultseppunct}\relax
\EndOfBibitem
\bibitem[Gladkikh \emph{et~al.}(2020)Gladkikh, Kim, Hajibabaei, Jana, Myung,
  and Kim]{Gaps_ML3}
V.~Gladkikh, D.~Y. Kim, A.~Hajibabaei, A.~Jana, C.~W. Myung and K.~S. Kim,
  \emph{The Journal of Physical Chemistry C}, 2020, \textbf{124},
  8905--8918\relax
\mciteBstWouldAddEndPuncttrue
\mciteSetBstMidEndSepPunct{\mcitedefaultmidpunct}
{\mcitedefaultendpunct}{\mcitedefaultseppunct}\relax
\EndOfBibitem
\bibitem[Mannodi-Kanakkithodi and Chan(2021)]{MK8}
A.~Mannodi-Kanakkithodi and M.~Chan, \emph{under review}, 2021\relax
\mciteBstWouldAddEndPuncttrue
\mciteSetBstMidEndSepPunct{\mcitedefaultmidpunct}
{\mcitedefaultendpunct}{\mcitedefaultseppunct}\relax
\EndOfBibitem
\bibitem[Aryasetiawan and Gunnarsson(1998)]{GW}
F.~Aryasetiawan and O.~Gunnarsson, \emph{Reports on Progress in Physics}, 1998,
  \textbf{61}, 237--312\relax
\mciteBstWouldAddEndPuncttrue
\mciteSetBstMidEndSepPunct{\mcitedefaultmidpunct}
{\mcitedefaultendpunct}{\mcitedefaultseppunct}\relax
\EndOfBibitem
\end{mcitethebibliography}
\bibliographystyle{rsc} 

\clearpage
\newpage
\setcounter{page}{1}


\setcounter{figure}{0}   
\setcounter{table}{0} 
\renewcommand{\thetable}{S\Roman{table}} 
\renewcommand\thefigure{S\arabic{figure}}

\begin{center}
\vspace*{0.5cm}
\Large
\textbf{Supplemental material to "Data-Driven Design of Novel Halide Perovskite Alloys"\\}
\vspace{0.5cm}
\large
Arun Mannodi-Kanakkithodi,\textsuperscript{1, 2, a)} and Maria K. Y Chan \textsuperscript{2, b)}\\
\vspace{0.3cm}

\normalsize
\textsuperscript{1}\textit{School of Materials Engineering, Purdue University, West Lafayette, Indiana 47907, USA}\\
\textsuperscript{2}\textit{Center for Nanoscale Materials, Argonne National Laboratory, Argonne, Illinois 60439, USA}\\
\end{center}

\footnote{
\textsuperscript{a}amannodi@purdue.edu\hspace{0.3cm}\textsuperscript{b}mchan@anl.gov}

\vspace{1cm}

\begin{figure*}[h]
\centering
\includegraphics[width=\linewidth]{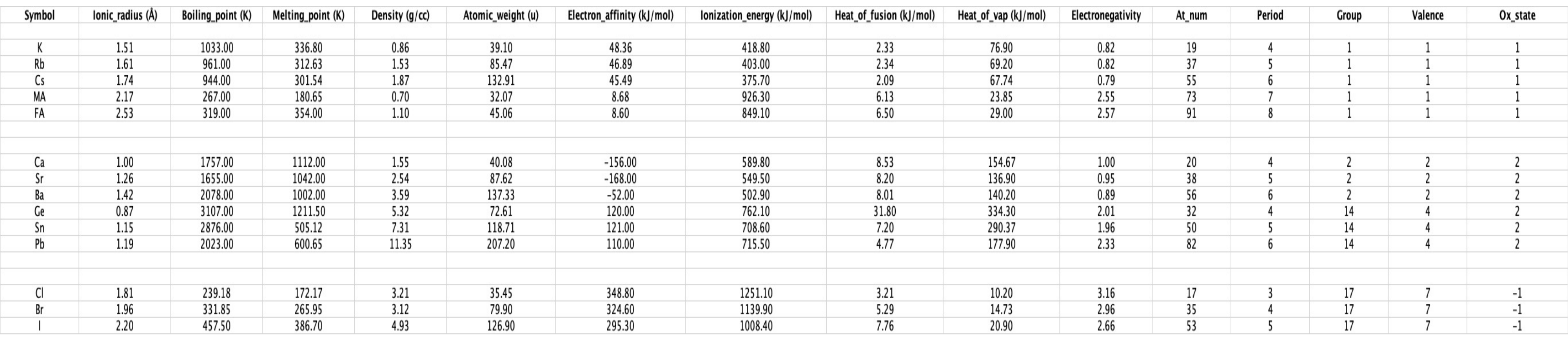}
\caption{\label{Fig:SI_elem_prop} 
List of 15 elemental/molecular properties and tabulated values used for each A, B and X-site constituent in the halide perovskite chemical space.}
\end{figure*}

\begin{figure*}
\centering
\includegraphics[width=\columnwidth]{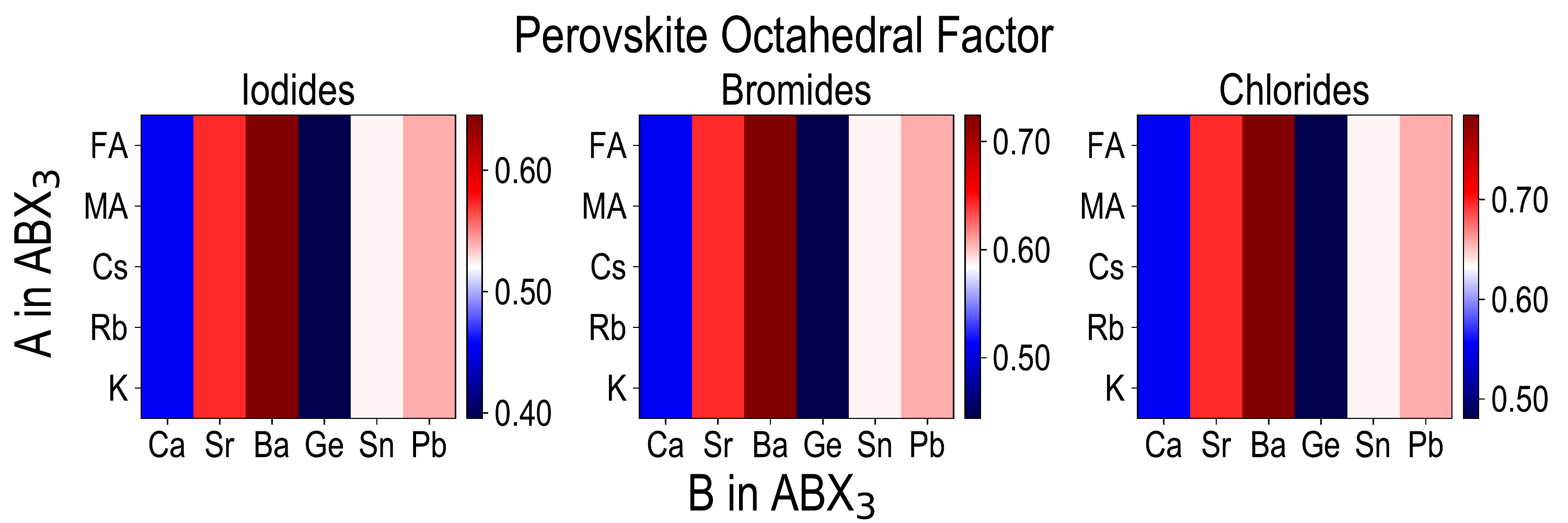}
\caption{\label{Fig:SI_oct_fact} 
Perovskite octahedral factor $\mu$ calculated for every A-B-X combination.}
\end{figure*}

\begin{figure*}
\centering
\includegraphics[width=\columnwidth]{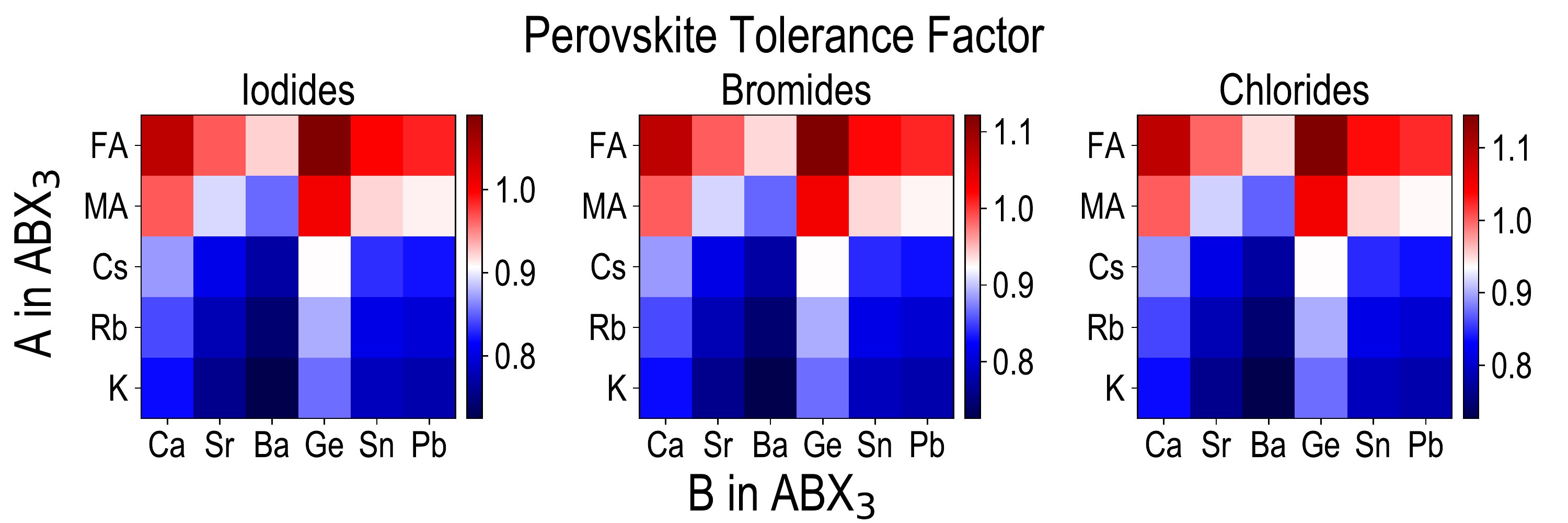}
\caption{\label{Fig:SI_tol_fact} 
Perovskite tetrahedral factor $t$ calculated for every A-B-X combination.}
\end{figure*}

\begin{figure*}[h]
\centering
\includegraphics[width=0.80\columnwidth]{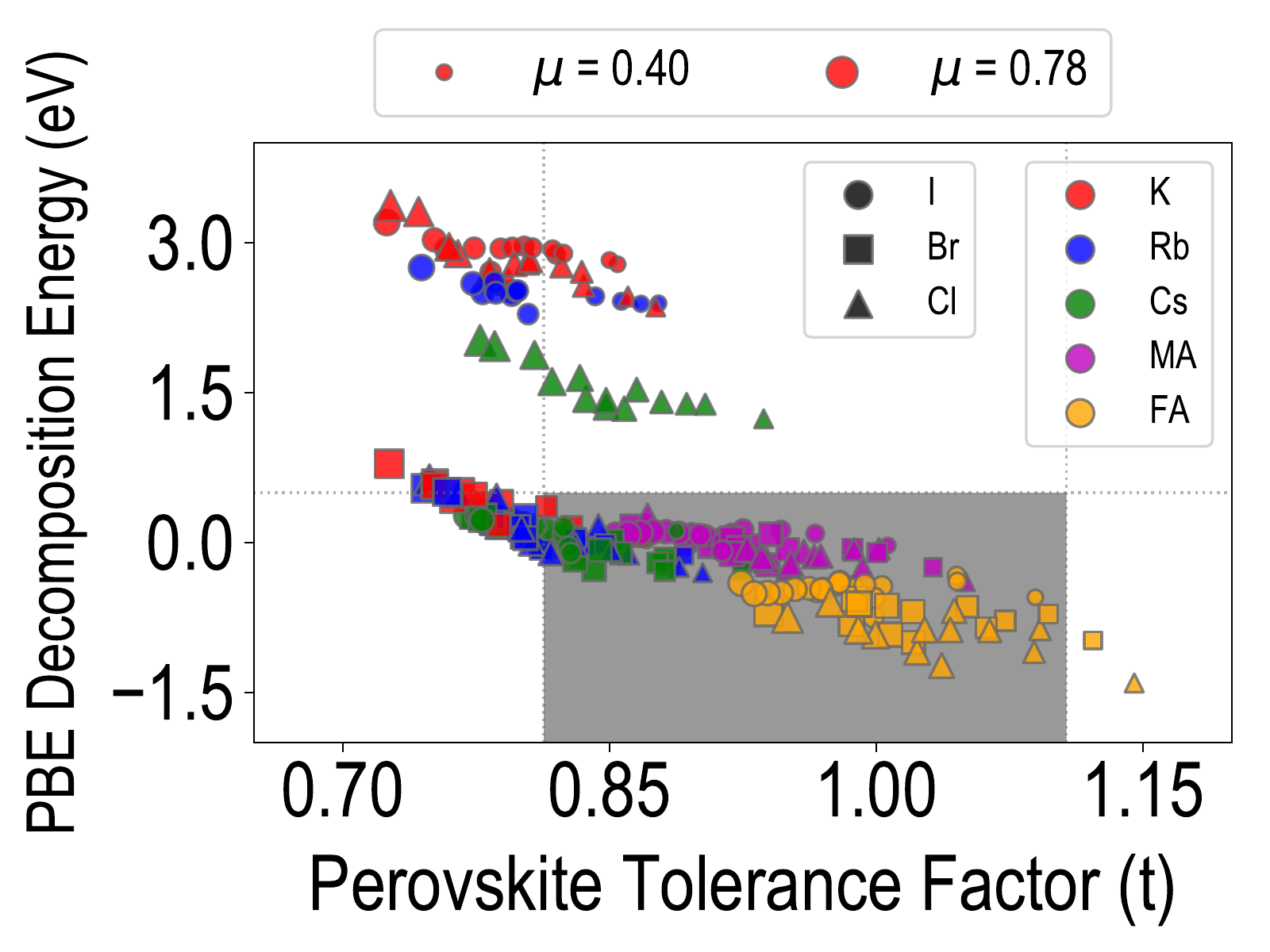}
\caption{\label{Fig:SI_pbe_tol} 
PBE computed decomposition energy ($\Delta$H$_{decomp}$) plotted against the perovskite tetrahedral factor $t$ for the entire DFT dataset. Symbol sizes are directly proportional to perovskite octahedral factor $o$. The shaded region represents compounds with low $\Delta$H$_{decomp}$ and suitable range of $t$ values.}
\end{figure*}

\begin{figure*}[h]
\centering
\includegraphics[width=0.80\columnwidth]{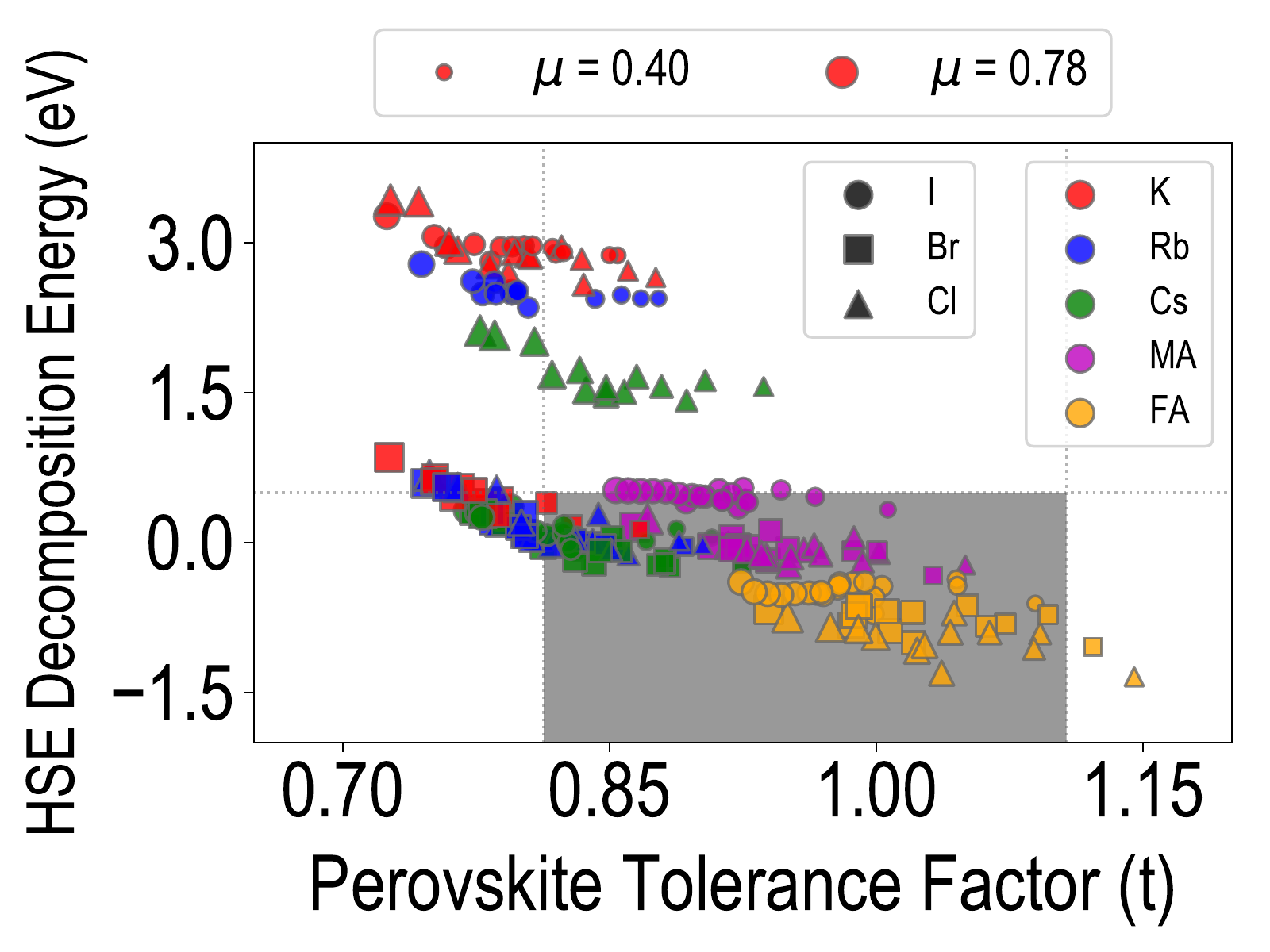}
\caption{\label{Fig:SI_hse_tol} 
HSE computed decomposition energy ($\Delta$H$_{decomp}$) plotted against the perovskite tetrahedral factor $t$ for the entire DFT dataset. Symbol sizes are directly proportional to perovskite octahedral factor $o$. The shaded region represents compounds with low $\Delta$H$_{decomp}$ and suitable range of $t$ values.}
\end{figure*}

\begin{figure*}[h]
\centering
\includegraphics[width=0.80\columnwidth]{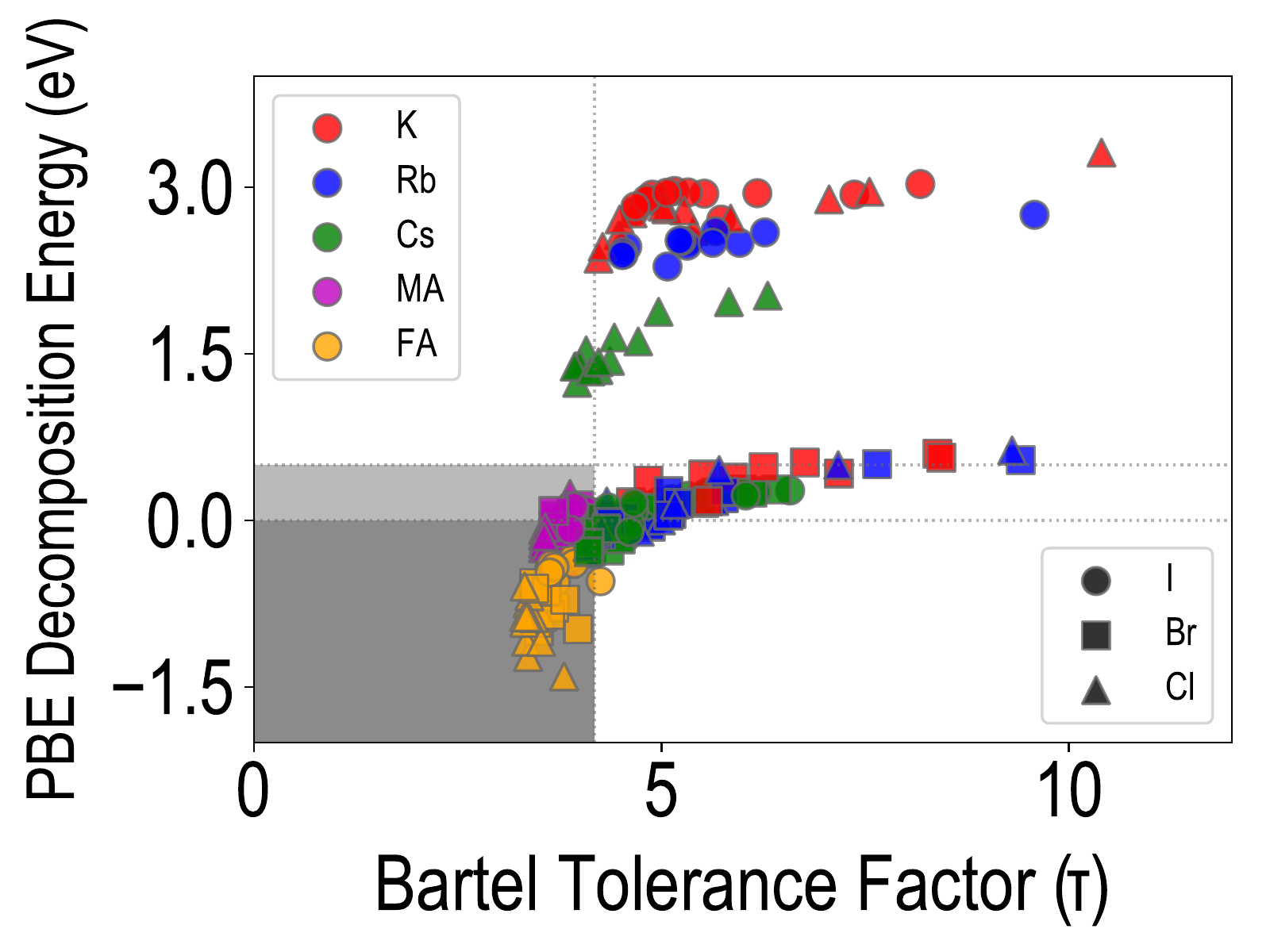}
\caption{\label{Fig:SI_pbe_tau} 
PBE computed decomposition energy ($\Delta$H$_{decomp}$) plotted against the Bartel perovskite tetrahedral factor \cite{Tol1} $\tau$ for the entire DFT dataset. The shaded region represents compounds with low $\Delta$H$_{decomp}$ and suitable $\tau$ values..}
\end{figure*}

\begin{figure*}[h]
\centering
\includegraphics[width=0.80\columnwidth]{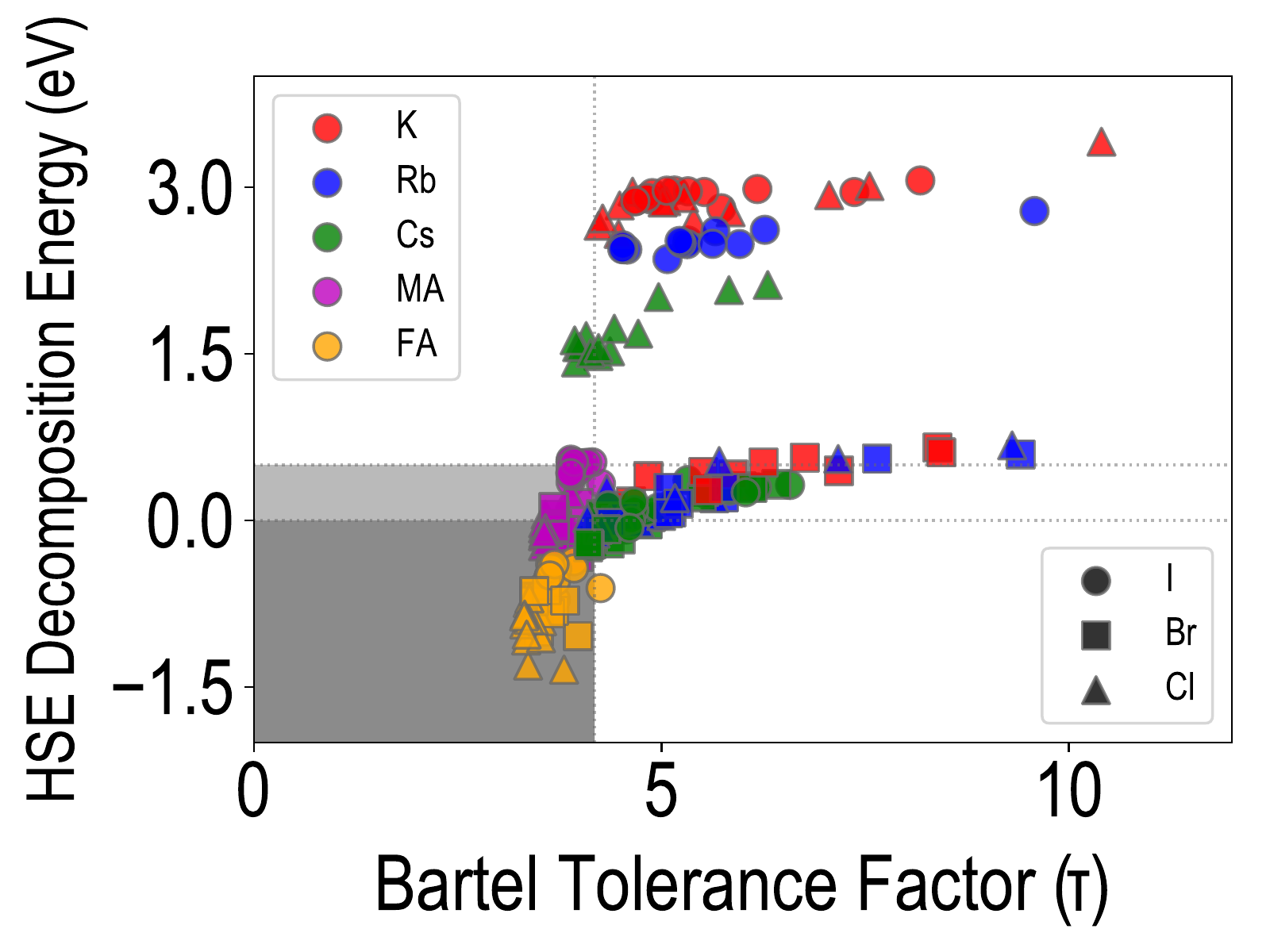}
\caption{\label{Fig:SI_hse_tau} 
HSE computed decomposition energy ($\Delta$H$_{decomp}$) plotted against the Bartel perovskite tetrahedral factor \cite{Tol1} $\tau$ for the entire DFT dataset. The shaded region represents compounds with low $\Delta$H$_{decomp}$ and suitable $\tau$ values..}
\end{figure*}

\begin{figure*}[h]
\centering
\includegraphics[width=0.75\linewidth]{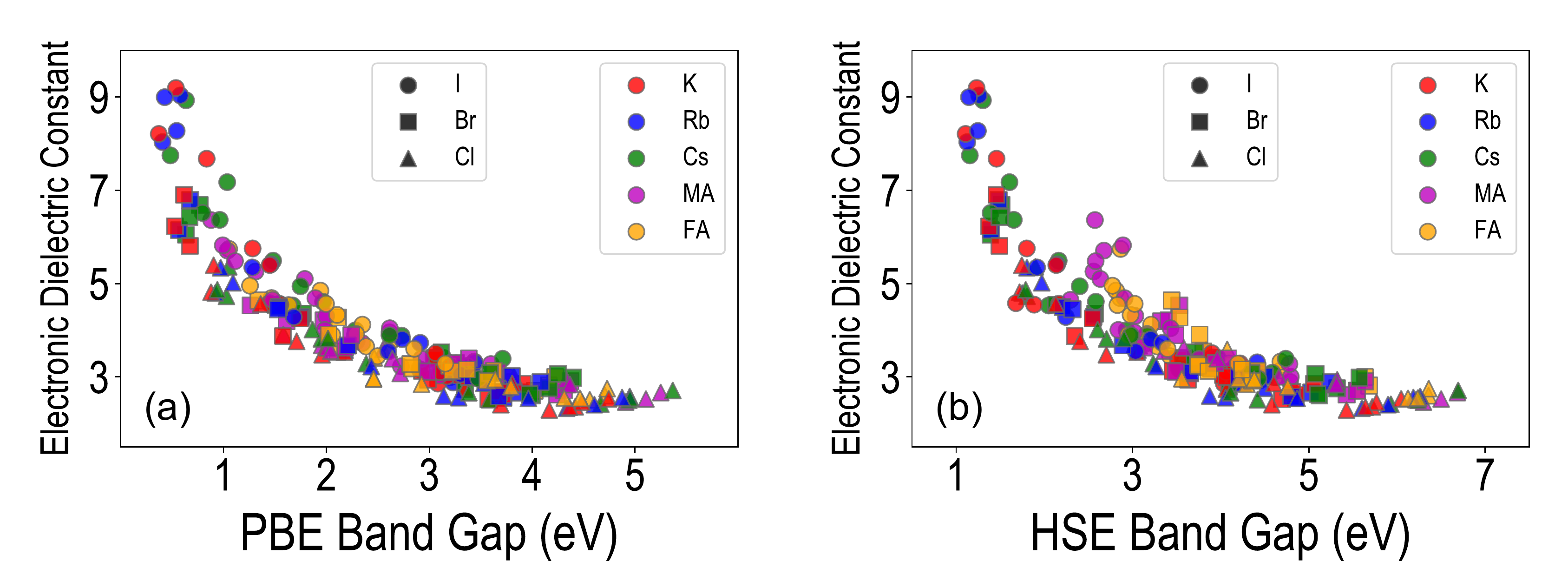}
\caption{\label{Fig:SI_gap_eps_dft} 
Electronic component of the dielectric constant plotted against the PBE and HSE band gaps for the DFT dataset of 229 compounds.}
\end{figure*}

\begin{figure*}[h]
\centering
\includegraphics[width=0.95\linewidth]{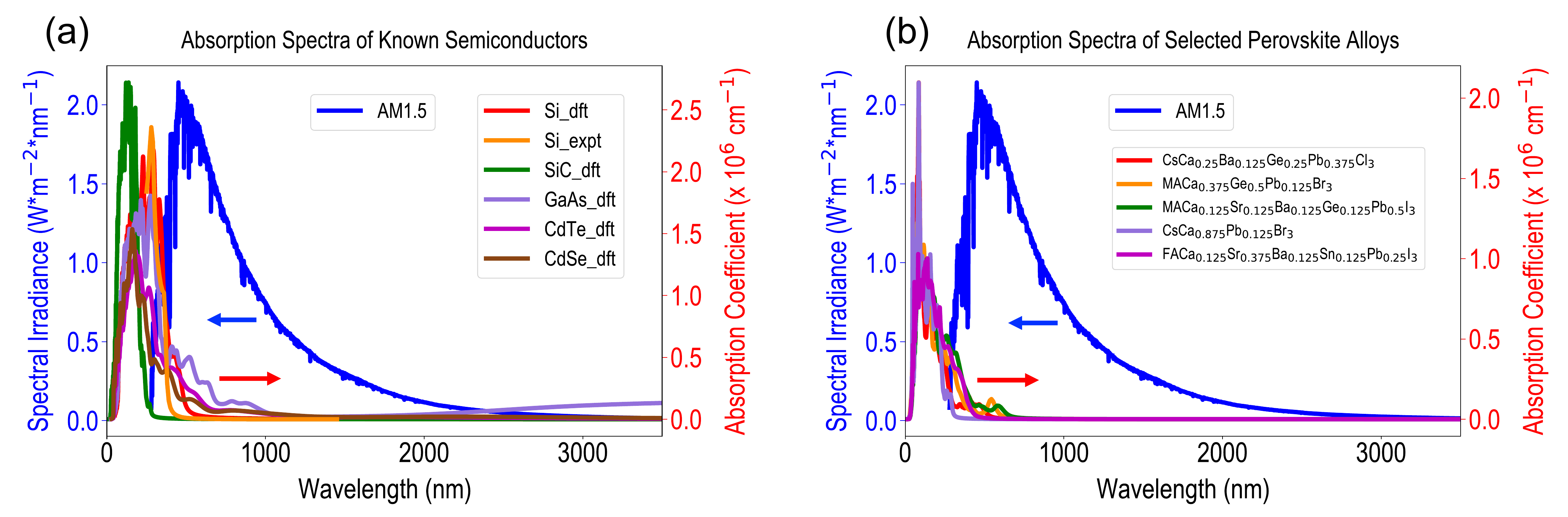}
\caption{\label{Fig:SI_abs_spectra} 
Computed optical absorption spectra (right y-axis) compared to solar spectral irradiance (left y-axis), for (a) known semiconductors Si, SiC, GaAs, CdTe and CdSe (all in zincblende structures), and (b) selected halide perovskite alloys from the current study. Experimentally measured spectrum is also reported for Si.}
\end{figure*}

\begin{table*}[h]
\centering
  \caption{\ Calculated PV figures of merit (in log$_{10}$) for known semiconductors and some selected compounds from the current study.}
  \label{table:SI_FOM}
  \begin{tabular}{cc}
    \hline
   &   \\
\textbf{Compound}  &  \textbf{log$_{10}$ (PV FOM)} \\
   &   \\
\hline
   &   \\
Si    &   5.64   \\
SiC   &   3.76   \\
GaAs   &   5.66   \\
CdTe   &   5.46   \\
CdSe   &   5.31   \\
	   &   \\
CsCa$_{0.25}$Ba$_{0.125}$Ge$_{0.25}$Pb$_{0.375}$Cl$_{3}$   &   4.72   \\
MACa$_{0.375}$Ge$_{0.5}$Pb$_{0.125}$Br$_{3}$   &   4.88   \\
MACa$_{0.125}$Sr$_{0.125}$Ba$_{0.125}$Ge$_{0.125}$Pb$_{0.5}$I$_{3}$   &   5.07   \\
CsCa$_{0.875}$Pb$_{0.125}$Br$_{3}$   &   3.84   \\
FACa$_{0.125}$Sr$_{0.375}$Ba$_{0.125}$Sn$_{0.125}$Pb$_{0.25}$I$_{3}$   &   4.88   \\
   &   \\
    \hline
  \end{tabular}
\end{table*}

\begin{figure*}[h]
\centering
\includegraphics[width=0.95\columnwidth]{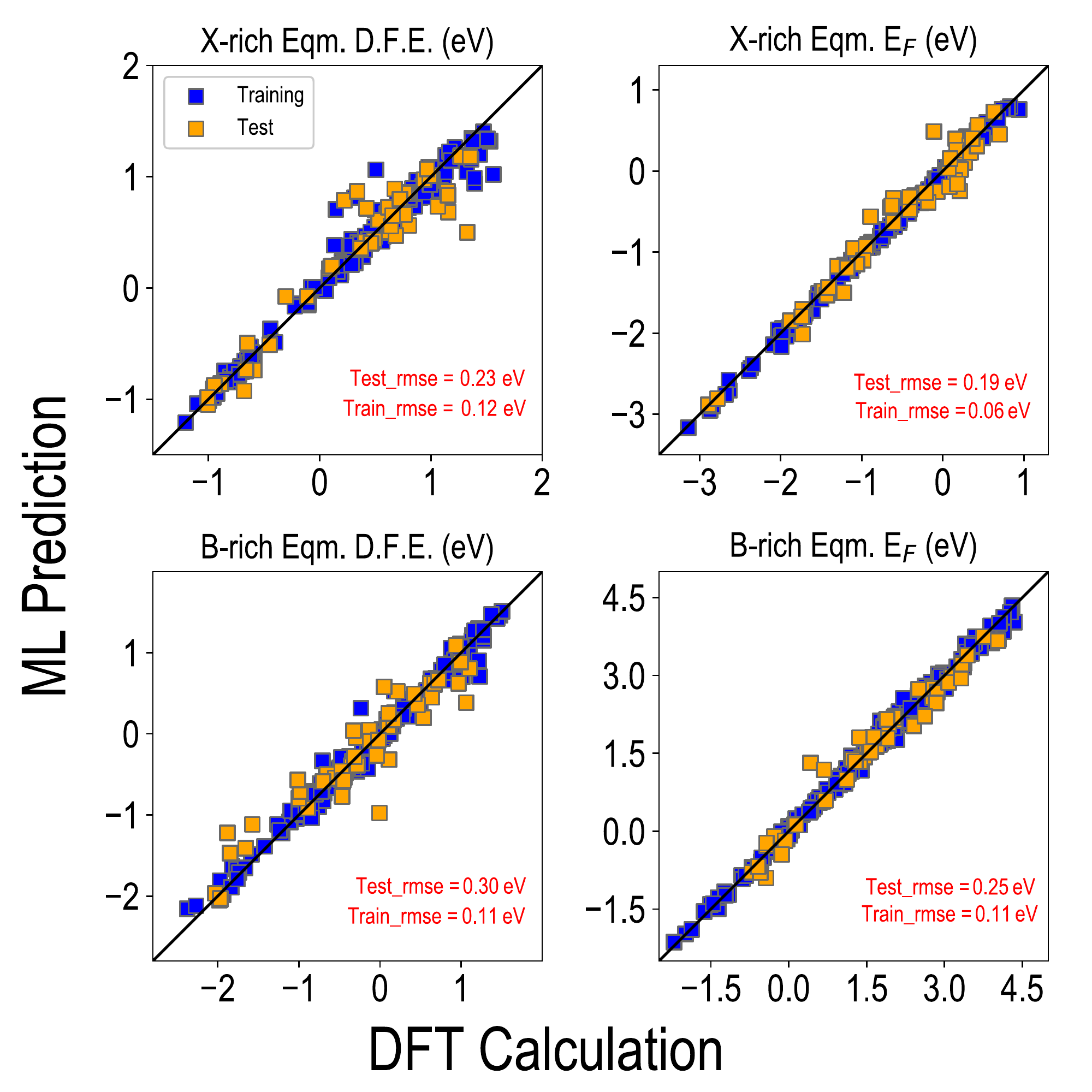}
\caption{\label{Fig:SI_NN_defects} 
Optimized NN predictive models for the equilibrium defect formation energy (D.F.E.) and Fermi level (E$_{F}$) (as determined by the energetics of A-site and X-site vacancies) at X-rich and B-rich chemical potential conditions.}
\end{figure*}

\begin{figure*}[h]
\centering
\includegraphics[width=0.65\linewidth]{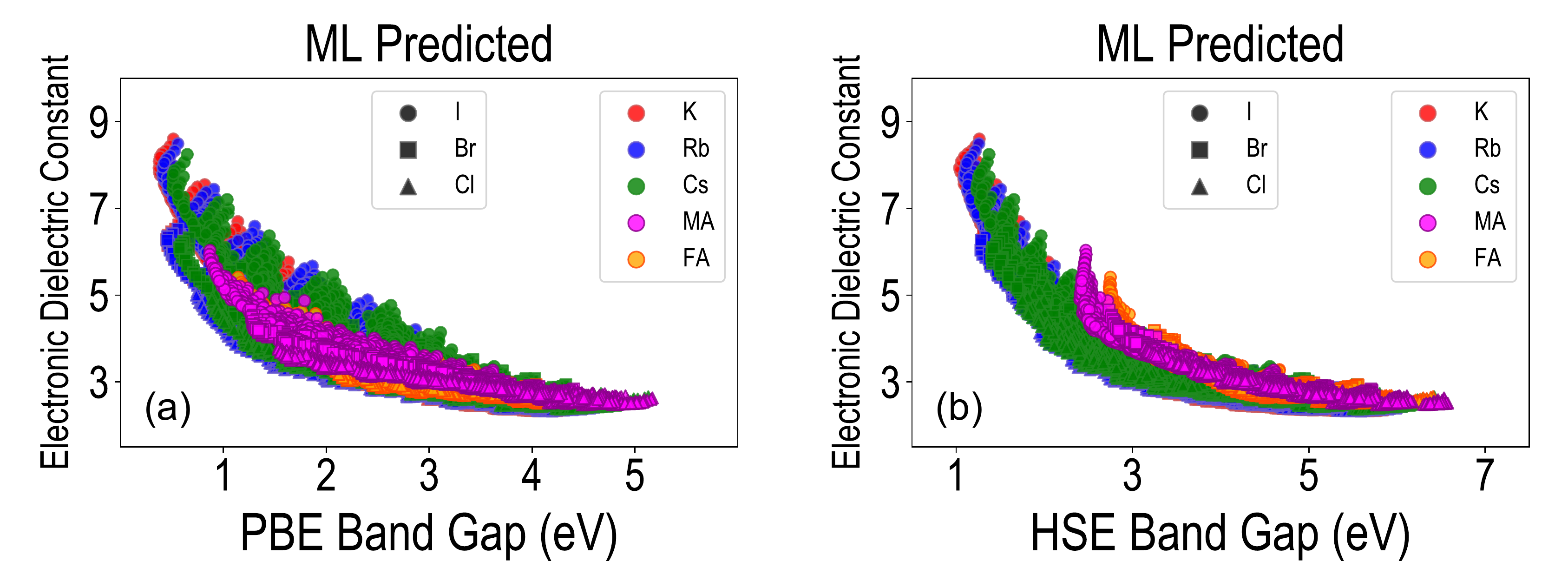}
\caption{\label{Fig:SI_gap_eps_ml} 
ML-predicted electronic component of the dielectric constant plotted against the ML-predicted PBE and HSE band gaps for the entire chemical space of 17,955 compounds.}
\end{figure*}

\begin{table*}[h]
\centering
  \caption{\ New compounds used for testing out-of-sample predictive power.}
  \label{table:SI_outside}
  \begin{tabular}{|c|c|c|c|c|c|}
    \hline
   &  &    &    &    &   \\
\textbf{Data Type}  &  \textbf{Perovskite Formula}  &  \textbf{DFT-PBE $\Delta$H$_{decomp}$}  &  \textbf{DFT-PBE E$_{gap}$}  &  \textbf{ML-PBE $\Delta$H$_{decomp}$}  &  \textbf{ML-PBE E$_{gap}$} \\
   &     &  \textbf{(eV p.f.u.)}  &  \textbf{(eV p.f.u.)}  &  \textbf{(eV)}  &  \textbf{(eV)}  \\
   &  &    &    &    &   \\
\hline
   &  &    &    &    &   \\
New A/X site mixed     &      K$_{0.125}$FA$_{0.875}$SnBr$_{3}$     &      -0.83     &      1.60     &      -0.87     &      1.36   \\
New A/X site mixed     &      MA$_{0.5}$FA$_{0.5}$CaCl$_{3}$     &      -0.32     &      4.31     &      -0.42     &      4.74   \\
New A/X site mixed     &      Rb$_{0.125}$Cs$_{0.875}$CaBr$_{3}$     &      -0.20     &      4.31     &      -0.15     &      4.44   \\
New A/X site mixed     &      K$_{0.5}$MA$_{0.5}$SnBr$_{3}$     &      -0.09     &      1.54     &      -0.13     &      0.51   \\
New A/X site mixed     &      Rb$_{0.375}$Cs$_{0.625}$GeBr$_{3}$     &      -0.22     &      0.72     &      -0.24     &      0.70   \\
New A/X site mixed     &      MACa$_{0.125}$Pb$_{0.875}$Br$_{1.5}$Cl$_{1.5}$     &      -0.13     &      2.43     &      -0.16     &      2.29   \\
New A/X site mixed     &      MASr$_{0.125}$Pb$_{0.875}$Br$_{1.5}$Cl$_{1.5}$     &      -0.13     &      2.47     &      -0.15     &      2.36   \\
New A/X site mixed     &      MABa$_{0.125}$Pb$_{0.875}$Br$_{1.5}$Cl$_{1.5}$     &      -0.10     &      2.50     &      -0.15     &      2.40   \\
New A/X site mixed     &      MAGe$_{0.125}$Pb$_{0.875}$Br$_{1.5}$Cl$_{1.5}$     &      -0.16     &      2.18     &      -0.23     &      1.97   \\
New A/X site mixed     &      MASn$_{0.125}$Pb$_{0.875}$Br$_{1.5}$Cl$_{1.5}$     &      -0.16     &      1.98     &      -0.23     &      1.97   \\
New A/X site mixed     &      CsPbI$_{0.375}$Br$_{2.625}$     &      -0.12     &      1.69     &      -0.18     &      1.70   \\
New A/X site mixed     &      MASrBr$_{1.125}$Cl$_{1.875}$     &      0.02     &      4.73     &      -0.02     &      4.79   \\
New A/X site mixed     &      FAGeBr$_{2.625}$Cl$_{0.375}$     &      -1.01     &      2.91     &      -0.97     &      2.36   \\
New A/X site mixed     &      RbSnBr$_{2.625}$Cl$_{0.375}$     &      -0.05     &      0.58     &      -0.12     &      0.67   \\
New A/X site mixed     &      MASnI$_{2.25}$Br$_{0.75}$     &      -0.13     &      1.09     &      -0.19     &      0.91   \\
   &  &    &    &    &   \\
New B-site Element     &      MAPb$_{0.875}$Be$_{0.125}$I$_{3}$     &      0.15     &      1.82     &      0.06     &      1.51   \\
New B-site Element     &      MAPb$_{0.875}$Mg$_{0.125}$I$_{3}$     &      0.08     &      1.89     &      0.06     &      1.66   \\
New B-site Element     &      MAPb$_{0.875}$Si$_{0.125}$I$_{3}$     &      0.11     &      1.48     &      0.06     &      1.52   \\
New B-site Element     &      MAPb$_{0.875}$Zn$_{0.125}$I$_{3}$     &      0.12     &      1.35     &      0.04     &      1.70   \\
New B-site Element     &      MAPb$_{0.875}$Cd$_{0.125}$I$_{3}$     &      0.05     &      1.41     &      0.04     &      1.80   \\
New B-site Element     &      MAPb$_{0.875}$Be$_{0.125}$Br$_{3}$     &      0.05     &      2.16     &      -0.15     &      1.87   \\
New B-site Element     &      MAPb$_{0.875}$Mg$_{0.125}$Br$_{3}$     &      -0.06     &      2.11     &      -0.13     &      1.99   \\
New B-site Element     &      MAPb$_{0.875}$Si$_{0.125}$Br$_{3}$     &      -0.02     &      1.74     &      -0.17     &      1.85   \\
New B-site Element     &      MAPb$_{0.875}$Zn$_{0.125}$Br$_{3}$     &      -0.03     &      2.19     &      -0.16     &      1.91   \\
New B-site Element     &      MAPb$_{0.875}$Cd$_{0.125}$Br$_{3}$     &      -0.06     &      1.74     &      -0.16     &      1.99   \\
New B-site Element     &      MAPb$_{0.875}$Be$_{0.125}$Cl$_{3}$     &      -0.17     &      2.65     &      -0.18     &      2.44   \\
New B-site Element     &      MAPb$_{0.875}$Mg$_{0.125}$Cl$_{3}$     &      -0.14     &      2.62     &      -0.17     &      2.58   \\
New B-site Element     &      MAPb$_{0.875}$Si$_{0.125}$Cl$_{3}$     &      -0.08     &      2.10     &      -0.23     &      2.42   \\
New B-site Element     &      MAPb$_{0.875}$Zn$_{0.125}$Cl$_{3}$     &      -0.09     &      2.64     &      -0.19     &      2.47   \\
New B-site Element     &      MAPb$_{0.875}$Cd$_{0.125}$Cl$_{3}$     &      -0.17     &      2.24     &      -0.18     &      2.53   \\
   &  &    &    &    &   \\
3$\times$3$\times$3 supercell alloy     &      MAPb$_{0.96}$Ba$_{0.04}$I$_{3}$     &      0.04     &      1.61     &      0.06     &      1.81   \\
3$\times$3$\times$3 supercell alloy     &      MAPb$_{0.93}$Sn$_{0.07}$I$_{3}$     &      0.02     &      1.50     &      0.05     &      1.68   \\
3$\times$3$\times$3 supercell alloy     &      MAPb$_{0.96}$Sn$_{0.04}$Br$_{3}$     &      -0.14     &      1.84     &      -0.16     &      1.95   \\
3$\times$3$\times$3 supercell alloy     &      MAPb$_{0.93}$Sn$_{0.035}$Ba$_{0.035}$Br$_{3}$     &      -0.13     &      1.93     &      -0.14     &      2.02   \\
3$\times$3$\times$3 supercell alloy     &      MAPb$_{0.89}$Sr$_{0.11}$Br$_{3}$     &      -0.12     &      2.18     &      -0.11     &      2.20   \\
   &  &    &    &    &   \\
    \hline
  \end{tabular}
\end{table*}

\begin{figure*}[h]
\centering
\includegraphics[width=0.60\linewidth]{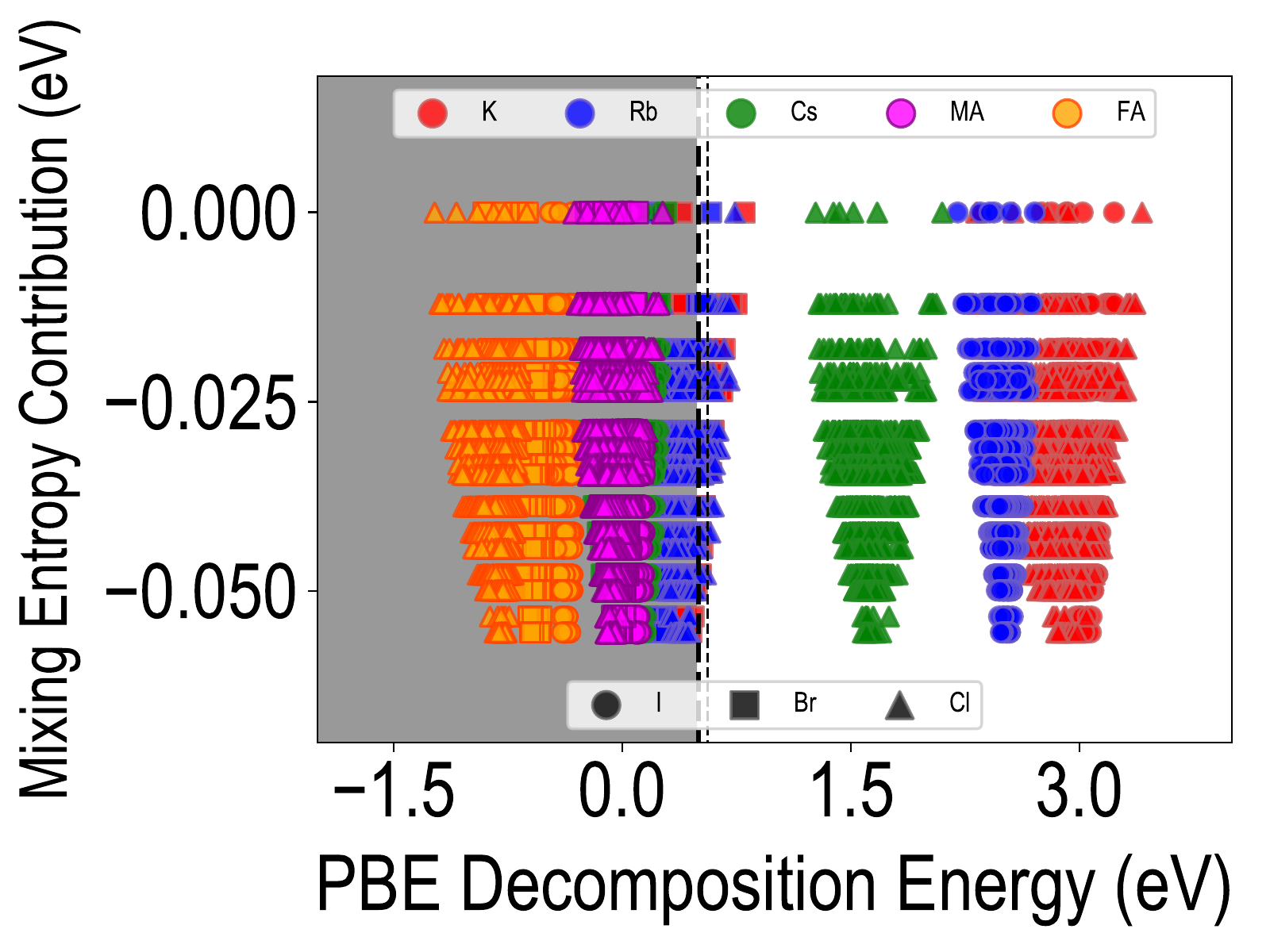}
\caption{\label{Fig:SI_mix_entropy} 
Energy contribution from mixing entropy of halide perovskite alloys plotted against ML-predicted PBE decomposition energies.}
\end{figure*}

\end{document}